\documentclass[fleqn,10pt]{article}
\usepackage{latexsym, graphicx, epsfig, amsmath, amssymb,amsfonts}
\usepackage{natbib,amsthm,version}
\usepackage{amsbsy,bm,multirow,enumerate}
\usepackage[titletoc,page]{appendix}
\usepackage[mathscr]{eucal}
\usepackage{mathtools}
\usepackage{color}
\usepackage[utf8]{inputenc}
\usepackage[english]{babel}
\usepackage{amsthm}
\usepackage{enumerate}
\usepackage[hidelinks]{hyperref}
\usepackage{url}
\usepackage{subfigure}
\usepackage[lined,boxed]{algorithm2e}

\newcommand{\mbs}[1]{\mathbf{#1}}
\newcommand{\mbb}[1]{\mathbb{#1}}

\newtheorem{theorem}{Theorem}[section]

\newtheorem{remark}{Remark}[section]

\theoremstyle{definition}

\title{
  A gPAV-Based Unconditionally Energy-Stable Scheme 
  for Incompressible Flows with Outflow/Open Boundaries
} 
\author{
  Lianlei Lin$^{1*}$, \ Xiaoyu Liu$^{2,3}$, \
  Suchuan Dong$^2$\thanks{Authors of correspondence.
    Email: sdong@purdue.edu, linlianlei@hit.edu.cn} \\
  $^1$School of Electronics and Information Engineering \\
  Harbin Institute of Technology, China \\
  $^2$Center for Computational and Applied Mathematics \\
  Department of Mathematics \\
  Purdue University, USA \\
  $^3$Collge of Infomation Science and Engineering \\
  Northeastern University, China
 } 

\date{(\today)}
\begin{document}
\maketitle



\begin{abstract}

  We present an unconditionally energy-stable scheme for approximating
  the incompressible Navier-Stokes equations on domains
  with outflow/open boundaries. The scheme combines the generalized
  Positive Auxiliary Variable (gPAV) approach and a rotational velocity-correction
  type strategy, and the adoption of the auxiliary variable simplifies
  the numerical treatment for the open boundary conditions.
  The discrete energy stability of the proposed scheme has been proven,
  irrespective of the time step sizes. Within each time step the scheme
  entails the computation of two velocity fields and two pressure
  fields, by solving an individual de-coupled Helmholtz (including
  Poisson) type equation with a constant pre-computable coefficient matrix
  for each of these field variables. The auxiliary variable,
  being a scalar number, is given by a well-defined explicit formula within
  a time step,  which ensures the positivity of its computed values.
  Extensive numerical experiments with several flows
  involving outflow/open boundaries in regimes
  where the backflow instability becomes severe have been presented
  to test the performance of the proposed method and
  to demonstrate its stability at large time step sizes.

\end{abstract}


\vspace{0.05cm}

\section{Introduction}
\label{sec:intro}



This work concerns the numerical approximation and
computation of incompressible flows on domains
with outflow or open boundaries. The presence of the outflow/open
boundary significantly escalates the challenge for incompressible
flow simulations.
A well-known issue 
is the so-called backflow instability~\cite{Dong2015clesobc,DongS2015},
which refers to the numerical instability associated with strong vortices
or backflows at the outflow/open boundary and can cause the simulation
to instantly blow up at moderate or high Reynolds numbers.
The boundary condition imposed on the outflow/open boundary
plays a critical role in the stability of such simulations.
In the past few years a class of effective methods, so-called
energy-stable open boundary
conditions~\cite{DongKC2014,DongS2015,Dong2015clesobc,NiYD2019},
have been developed and can effectively overcome
the backflow instability; see also related
works in~\cite{BruneauF1994,BruneauF1996,BazilevsGHMZ2009,Moghadametal2011,PorporaZVP2012,GravemeierCYIW2012,BertoglioC2014,IsmailGCW2014,BertoglioC2016},
among others.

In the current work
we focus on the numerical approximation of
the incompressible Navier-Stokes equations together with
the energy-stable open boundary conditions (ESOBC),
and propose an unconditionally energy-stable scheme for
such problems.
%
%
Two important issues are encountered immediately, which
call for some comments here.
First, the inclusion of ESOBC, which
is nonlinear in nature~\cite{Dong2015clesobc}, makes the numerical
approximation of the system and the proof of
discrete energy stability considerably more challenging.
Second, the computational cost per time step
of energy-stable schemes is an issue we are conscious of.
The goal here is to develop discretely energy-stable schemes
with a relatively low computational cost,
so that they can be computationally competitive and efficient
even on a per-time-step basis.


There is a large volume of literature on the numerical
schemes for incompressible Navier-Stokes equations (absence of
open/outflow boundaries); see the reviews~\cite{Gresho1991,GuermondMS2006}.
These schemes can be broadly classified into two
categories: semi-implicit splitting type schemes
and unconditionally energy-stable schemes.
The semi-implicit or fractional-step
schemes (see e.g.~\cite{Chorin1968,Temam1969,KimM1985,KarniadakisIO1991,BrownCM2001,XuP2001,GuermondS2003,LiuLP2007,HyoungsuK2011,SersonMS2016},
among others) typically treat the nonlinear term explicitly
and de-couple the computations for the flow variables (pressure/velocity)
by a splitting strategy. These schemes have a low
computational cost per time step, because the coefficient
matrices involved therein are all constant and can be pre-computed.
The main drawback of these schemes lies in their conditional stability,
and the computation is stable only when the time step size
is sufficiently small. Thanks to their low cost,
such schemes have been widely
used in the simulations of turbulence and
flow physics studies of single- and multi-phase problems;
see e.g.~\cite{KravchenkoM2000,DongKER2006,VargheseFF2007,Dong2009,GhaisasSF2015,SaudRT2017,Dong2017,LeeRA2017}.
In the presence of outflow/open boundaries,
the numerical methods employed
in~\cite{DongKC2014,Dong2014obc,DongS2015,Dong2015clesobc,DongW2016,NiYD2019}
also belong to the semi-implicit type schemes.
The unconditionally energy-stable schemes
(see e.g.~\cite{Shen1992,SimoA1994,VerstappenV2003,GuermondMS2005,LabovskyLMNR2009,DongS2010,Sanderse2013,JiangMRT2016,ChenSZ2018}, among others)
typically treat the nonlinear term in a fully implicit or
linearized fashion, and can alleviate or eliminate the constraint
on the time step size encountered with semi-implicit schemes.
The main drawback of energy-stable schemes lies in that
they typically require the solution of a system of nonlinear
algebraic equations or a system of linear algebraic equations
with a variable and time-dependent coefficient matrix
within a time step~\cite{DongS2010}.
Their computational cost per time step is quite high
due to the Newton type nonlinear
iterations and/or the need for frequent re-computations of the coefficient
matrices (every time step).


Mindful of the strengths and weaknesses of traditional energy-stable
algorithms as discussed above,
in this paper we propose a new unconditionally energy-stable scheme
for the incompressible Navier-Stokes equations together with
the convective-like energy-stable open boundary conditions
from~\cite{Dong2015clesobc}.
This scheme in some sense  combines the strengths of
semi-implicit schemes and the traditional energy-stable schemes.
A prominent feature lies in that, while being unconditionally energy-stable,
within a time step
it requires only the solution of de-coupled linear algebraic systems with
constant coefficient matrices that can be pre-computed.
As a result, the scheme is computationally very competitive and efficient.
The unconditional discrete energy stability has been proven in
the presence of outflow/open boundaries,
regardless of the time step sizes.

These attractive properties of the proposed scheme are achieved by
the use of an auxiliary variable associated with the total
energy of the Navier-Stokes system. Such an auxiliary variable
was introduced in a very recent work~\cite{LinYD2019} for
the incompressible Navier-Stokes equations
(see also~\cite{ShenXY2018,YangLD2019,YangD2019twop} for related problems). 
%
The adoption of the auxiliary variable enables us
to deal with the ESOBC in a relatively simple way.
It should be noted that the auxiliary
variable and the Navier-Stokes
equations are treated in a very different way in the current work
than in~\cite{LinYD2019}.
In this work
the incompressible Navier-Stokes equations,
the dynamic equation for the auxiliary variable,
and the energy-stable open boundary conditions
have been reformulated 
based on
the generalized Positive Auxiliary Variable (gPAV) approach.
The gPAV approach is originally developed in~\cite{YangD2019diss}
for general dissipative systems, and provides a systematic procedure
for treating dissipative partial differential equations.

We  treat the gPAV-reformulated system of equations
numerically  in a judicious  way
to arrive at a discrete scheme for simulating incompressible
flows with outflow/open boundaries.
The scheme incorporates features of the rotational
velocity-correction type strategy that is reminiscent of
semi-implicit type algorithms (see
e.g.~\cite{GuermondMS2006,DongKC2014,DongS2012,Dong2014obc,Dong2014}).
The unconditional energy stability of
this scheme is proven. We show that within each time step
the scheme entails the computation of two velocity fields and
two pressure fields, by solving an individual de-coupled
linear equation involving a constant coefficient matrix that can
be pre-computed for each of these field variables.
On the other hand, the auxiliary variable (a scalar number) is computed by
a well-defined explicit formula.
No nonlinear algebraic solver is involved in the current
scheme, and furthermore the existence and positivity of
the computed auxiliary variable
are guaranteed (or preserved).
Note that
this is in sharp contrast with 
the method of~\cite{LinYD2019}, in which Newton-type nonlinear
solvers are required for computing the auxiliary variable and neither
the existence nor the positivity of the computed auxiliary
variable is guaranteed.


The contribution of this paper lies in the unconditionally energy-stable
scheme developed herein for simulating incompressible
flows with outflow/open boundaries.
The discrete formulation of the current algorithm, barring the auxiliary variable,
resembles the conventional rotational velocity-correction scheme
to a certain degree. In such a sense, the current algorithm
can be considered as a modified velocity-correction type scheme,
which turns out to be unconditionally energy-stable.
By contrast, the conventional velocity-correction scheme is
only conditionally stable.
To the best of the authors' knowledge, this is the first time
when a rotational ``velocity-correction'' scheme has been proven
to be unconditionally stable.

The proposed scheme is implemented using the high-order spectral element
method~\cite{SherwinK1995,KarniadakisS2005,ZhengD2011,ChenSX2012} in the
current paper.
It should be noted that the use of spectral elements is not essential
to the current scheme, and other spatial discertization methods can equally
be used in the implementation.
A number of flow problems involving outflow/open boundaries,
and in regimes where the backflow instability becomes
a severe issue for conventional methods,
have been used to demonstrate the performance of the method and its stability
at large time step sizes.


The rest of this paper is structured as follows.
In Section \ref{sec:method} we introduce an auxiliary variable
defined based on the sum of the total system energy and the energy integral
on the outflow/open boundary and introduce its dynamic equation. We
then reformulate the
governing equations together with the energy-stable open boundary condition
into an equivalent system 
utilizing the gPAV approach.
The algorithmic formulation of the scheme is then presented, and we prove
its discrete energy stability property.
The implementation of the scheme is also discussed in some detail.
In Section \ref{sec:tests} we use manufactured analytic solutions
to demonstrate the convergence rates of the proposed scheme, and
use several flow problems involving outflow/open boundaries
to test the performance and demonstrate the stability of the presented
method.
Section \ref{sec:summary} then concludes the discussions
by some closing remarks.


\section{Discretely Energy-Stable Scheme for Incompressible Flows with
Open Boundaries}
\label{sec:method}

\subsection{Incompressible Navier-Stokes Equations and
Energy-Stable Open Boundary Condition}

Consider a domain $\Omega$ in two or three dimensions,
whose boundary is denoted by $\partial\Omega$,
and an incompressible flow contained in this domain.
The dynamics is described by the incompressible
Navier-Stokes equations, in non-dimensional form,
given by
\begin{subequations}
  \begin{align}
    &
    \frac{\partial\mbs u}{\partial t} + \mbs N(\mbs u)
    + \nabla p -\nu\nabla^2\mbs u = \mbs f,
    \label{equ:nse} \\
    &
    \nabla\cdot\mbs u = 0,
    \label{equ:div}
  \end{align}
\end{subequations}
where $\mbs u(\mbs x,t)$ and $p(\mbs x,t)$ are
the velocity and pressure, respectively,
$\mbs N(\mbs u)=\mbs u\cdot\nabla\mbs u$,
$\mbs f(\mbs x,t)$ is an external body force,
and $\mbs x$ and $t$ denote the spatial coordinate and
time. $\nu$ is the non-dimensional viscosity (reciprocal
of the Reynolds number $Re$),
\begin{equation}
  \nu = \frac{1}{Re} = \frac{\nu_f}{U_0L},
\end{equation}
where $\nu_f$ is the kinematic viscosity of the fluid,
and $U_0$ and $L$ are the characteristic velocity and
length scales.


We assume that two types of boundaries (non-overlapping)
may exist in the system: Dirichlet boundary $\partial\Omega_d$
and open boundary $\partial\Omega_o$, namely,
$\partial\Omega=\partial\Omega_d \cup \partial\Omega_o$.
On the Dirichlet boundary the velocity distribution is known,
\begin{equation}\label{equ:dbc}
\mbs u = \mbs w(\mbs x, t), \quad \text{on} \ \partial\Omega_d,
\end{equation}
where $\mbs w$ is the boundary velocity.

On the open boundary $\partial\Omega_o$ neither the velocity nor
the pressure is known. However, we assume that
in general an external boundary force, in the form of a pressure head,
denoted by $p_0$,
may be imposed on $\partial\Omega_o$. For domains with
multiple openings (outlets/inlets), it is assumed that
the imposed external pressure heads on these openings may be
different. To fix the boundary condition for $\partial\Omega_o$,
we consider the energy balance equation for
the system consisting of \eqref{equ:nse}--\eqref{equ:div},
\begin{equation}\label{equ:eng}
  \begin{split}
  \frac{\partial}{\partial t}\int_{\Omega}\frac{1}{2}|\mbs u|^2d\Omega
  &= -\nu\int_{\Omega}\|\nabla\mbs u \|^2d\Omega
  + \int_{\Omega}\mbs f\cdot\mbs u d\Omega
  + \int_{\partial\Omega}\left[
    -p\mbs n\cdot\mbs u + \nu\mbs n\cdot\nabla\mbs u\cdot\mbs u
    -\frac{1}{2}(\mbs n\cdot\mbs u)|\mbs u|^2
    \right]dA \\
  &= -\nu\int_{\Omega}\|\nabla\mbs u \|^2d\Omega
  + \int_{\Omega}\mbs f\cdot\mbs u d\Omega
  + \int_{\partial\Omega_d}\left(
    -p\mbs n + \nu\mbs n\cdot\nabla\mbs u
    -\frac{1}{2}(\mbs n\cdot\mbs w)\mbs w
    \right)\cdot\mbs w  dA \\
    &\quad
    +\int_{\partial\Omega_o}\left[
      -(p-p_0(\mbs x,t))\mbs n
      + \nu\mbs n\cdot\nabla\mbs u
      - \frac{1}{2}(\mbs n\cdot\mbs u)\mbs u
      \right]\cdot\mbs u dA
    - \int_{\partial\Omega_o}p_0(\mbs x,t)\mbs n\cdot\mbs u dA,
  \end{split}
\end{equation}
where $\mbs n$ is the outward-pointing unit vector normal to $\partial\Omega$,
$\|\nabla\mbs u \|^2=\sum_{i,j=1}^{d_{im}}\partial_iu_j\partial_i u_j$
($d_{im}$ denoting the spatial dimension),
and $p_0(\mbs x,t)$ is the imposed external pressure
force on $\partial\Omega_o$, which in general can be a distribution.
Following \cite{Dong2015clesobc}, we consider the following
convective-like boundary condition for the open boundary $\partial\Omega_o$
in this work,
\begin{equation} \label{equ:obc}
  \nu D_0\frac{\partial\mbs u}{\partial t}
  - (p-p_0(\mbs x,t))\mbs n + \nu\mbs n\cdot\nabla\mbs u
  - \mbs H(\mbs n,\mbs u) = \mbs f_b(\mbs x,t),
  \quad \text{on} \ \partial\Omega_o,
\end{equation}
where the constant $D_0\geqslant 0$ represents the inverse
of a convection-velocity scale on $\partial\Omega_o$
(see~\cite{Dong2015clesobc} for details),
$\mbs f_b$ is a prescribed source term for the purpose of
numerical testing only and will be set to $\mbs f_b=0$
in actual simulations. $\mbs H(\mbs n,\mbs u)$ is given by~\cite{Dong2015clesobc},
\begin{equation}\label{equ:def_H}
  \mbs H(\mbs n,\mbs u) = \left[\frac{1}{2}|\mbs u|^2\mbs n
    + \frac{1}{2}(\mbs n\cdot\mbs u)\mbs u\right]\Theta_0(\mbs n,\mbs u),
  \quad
  \Theta_0(\mbs n,\mbs u)=\frac{1}{2}\left(
  1-\tanh\frac{\mbs n\cdot\mbs u}{ U_0\delta}
  \right),
\end{equation}
where $\Theta_0$ represents a smoothed step function,
taking essentially the unit value when $\mbs n\cdot\mbs u<0$
and vanishing otherwise. The small constant $\delta >0$
controls the sharpness of the step, and as $\delta \rightarrow 0$
the function becomes sharper and
$\Theta_0$ approaches the step function.
The boundary condition \eqref{equ:obc} (with $\mbs f_b=0$ and $\delta$
sufficiently small)
is an energy-stable boundary condition for $\partial\Omega_o$,
in the sense that in the absence of the
external forces ($\mbs f=0$, $p_0=0$) and with zero
boundary velocity ($\mbs w=0$) on $\partial\Omega_d$
this boundary condition ensures that a modified energy of
the system will not increase over time. This is because in this case
the energy balance equation \eqref{equ:eng} is reduced to
\begin{equation}
  \frac{\partial}{\partial t}\left(
  \int_{\Omega}\frac{1}{2}|\mbs u|^2d\Omega +
  \nu D_0\int_{\partial\Omega_o}\frac{1}{2}|\mbs u|^2dA
  \right)
  = -\nu\int_{\Omega}\|\nabla\mbs u \|^2d\Omega
  - \int_{\partial\Omega_o} \frac{1}{2}|\mbs u|^2|\mbs n\cdot\mbs u|dA,
  \quad \text{as} \ \delta \rightarrow 0.
\end{equation}

\begin{remark}

  The following more general form for $\mbs H(\mbs n,\mbs u)$ is provided
  in \cite{Dong2015clesobc},
  \begin{equation}\label{equ:H_gen}
    \mbs H(\mbs n,\mbs u) = \left[
      (\theta+\alpha_2)\frac{1}{2}(\mbs u\cdot\mbs u)\mbs n
      + (1-\theta+\alpha_1)\frac{1}{2}(\mbs n\cdot\mbs u)\mbs u
      \right]\Theta_0(\mbs n,\mbs u),
  \end{equation}
  where $\theta$, $\alpha_1$ and $\alpha_2$ are constants satisfying
  the conditions $0\leqslant\theta\leqslant 1$, $\alpha_1\geqslant 0$
  and $\alpha_2\geqslant 0$.
  This form also  ensures that the condition given by \eqref{equ:obc}
  is an energy-stable boundary condition for $\partial\Omega_o$.
  
\end{remark}


The system of equations \eqref{equ:nse}--\eqref{equ:div}
is supplemented by the following initial condition for
the velocity,
\begin{equation}\label{equ:ic}
  \mbs u(\mbs x,0) = \mbs u_{in}(\mbs x),
\end{equation}
where $\mbs u_{in}$ denotes the initial velocity distribution
that satisfies \eqref{equ:div} and is compatible
with the boundary condition \eqref{equ:dbc} on $\partial\Omega_d$.

\subsection{Reformulated Equivalent System}

To facilitate the development of numerical algorithms
we will first reformulate the system consisting of equations
\eqref{equ:nse}--\eqref{equ:div},
the boundary conditions \eqref{equ:dbc}
and \eqref{equ:obc}, and the initial condition
\eqref{equ:ic} into an equivalent system.

Define a biased modified energy,
\begin{equation}\label{equ:def_E}
  E(t) = E[\mbs u] = \int_{\Omega}\frac{1}{2}|\mbs u|^2d\Omega
  + \nu D_0\int_{\partial\Omega_o}\frac{1}{2}|\mbs u|^2 dA
  + C_0,
\end{equation}
where $C_0$ is a chosen energy constant that ensures
 $E(t)>0$ for all $t\geqslant 0$.
Define an auxiliary variable $R(t)$ based on $E(t)$,
\begin{equation}\label{equ:def_R}
  \left\{
  \begin{split}
    &
    R(t) = \sqrt{E(t)}, \\
    &
    E(t) = R^2(t).
  \end{split}
  \right.
\end{equation}
Then $R(t)$ satisfies the following dynamic equation,
\begin{equation}\label{equ:R_equ}
  2R\frac{dR}{dt} = \int_{\Omega}\mbs u\cdot\frac{\partial\mbs u}{\partial t}d\Omega
  + \nu D_0\int_{\partial\Omega_o}\mbs u\cdot\frac{\partial\mbs u}{\partial t}dA.
\end{equation}
Note that both $E(t)$ and $R(t)$ are scalar numbers,
not field functions.

We define another function,
\begin{equation}
  g(\chi) = \left\{
  \begin{array}{ll}
    \chi, & \text{if} \ \chi \leqslant 1, \\
    1, & \text{if} \ \chi > 1.
  \end{array}
  \right.
  \label{equ:def_g}
\end{equation}
Note that $\frac{R^2(t)}{E(t)}=1$, and
so $g\left(\frac{R^2}{E}\right)=1$.

With the variables defined above,
we reformulate
equation \eqref{equ:nse} into the following
equivalent form,
\begin{equation}
\frac{\partial\mbs u}{\partial t} + g\left(\frac{R^2}{E}\right)\mbs N(\mbs u)
+ \nabla p -\nu\nabla^2\mbs u = \mbs f.
\label{equ:nse_1}
\end{equation}
We re-write the boundary condition \eqref{equ:obc} into
\begin{equation}\label{equ:obc_1}
  \nu D_0\frac{\partial\mbs u}{\partial t}
  - (p-p_0)\mbs n + \nu\mbs n\cdot\nabla\mbs u
  - g\left(\frac{R^2}{E} \right)\mbs H(\mbs n,\mbs u) = \mbs f_b,
  \quad \text{on} \ \partial\Omega_o.
\end{equation}

Let $\mbs u_1$, $\mbs u_2$, $p_1$, and $p_2$ denote four
field functions that are to be
specifically defined later in Section \ref{sec:soln}
(by equations \eqref{equ:def_p1}--\eqref{equ:def_p2}
and \eqref{equ:u1_1}--\eqref{equ:u2_2}).
They are related to $\mbs u$ and $p$ by the relations,
$\mbs u = \mbs u_1+ g(\frac{R^2}{E})\mbs u_2$
and $p = p_1 + g(\frac{R^2}{E})p_2$;
see equations \eqref{equ:p_soln} and \eqref{equ:u_soln}
later in Section \ref{sec:soln}.
Following the gPAV idea~\cite{YangD2019diss},
we incorporate the following zero terms into
the right hand side (RHS) of equation \eqref{equ:R_equ},
\begin{equation}\label{equ:zero}
  \begin{split}
    &
    \left(\frac{R^2}{E}-1 \right)\int_{\Omega}\left(
    -\nabla p + \nu\nabla^2\mbs u
    \right)\cdot\mbs u d\Omega
    +\left[g\left(\frac{R^2}{E} \right)- \frac{R^2}{E} \right]
    \int_{\Omega}\mbs N(\mbs u)\cdot\mbs u d\Omega \\
    &
    +\left(\int_{\Omega}\mbs f\cdot\mbs u d\Omega
    - \int_{\Omega}\mbs f\cdot\mbs u d\Omega
    \right)
    + \left(\frac{R^2}{E}-1 \right)\int_{\partial\Omega}\left(
    p\mbs n - \nu\mbs n\cdot\nabla\mbs u
    \right)\cdot\mbs u dA \\
    &
    + \left[\frac{R^2}{E} -g\left(\frac{R^2}{E} \right)\right]
    \int_{\partial\Omega_o} \mbs H(\mbs n,\mbs u)\cdot\mbs u dA
    +\left(
    \int_{\partial\Omega_o} p_0\mbs n\cdot\mbs u dA
    - \int_{\partial\Omega_o} p_0\mbs n\cdot\mbs u dA
    \right) \\
    &
    +\left(
    \int_{\partial\Omega_o} \mbs f_b\cdot\mbs u dA
    - \int_{\partial\Omega_o} \mbs f_b\cdot\mbs u dA
    \right)
    + \left(1-\frac{R^2}{E}\right)\int_{\partial\Omega_d}\left[
      -p\mbs n + \nu\mbs n\cdot\nabla\mbs u
      - \frac{1}{2}(\mbs n\cdot\mbs w)\mbs w
      \right]\cdot\mbs w dA \\
    &
    + \left[\frac{R^2}{E} -g\left(\frac{R^2}{E} \right)\right]
    \int_{\Omega}\mbs f\cdot\mbs u_2 d\Omega
    + \left[\frac{R^2}{E} -g\left(\frac{R^2}{E} \right)\right]
    \int_{\partial\Omega_d}\left(
    -p_2\mbs n + \nu\mbs n\cdot\nabla\mbs u_2
    \right)\cdot\mbs w dA \\
    &
    + \left[\frac{R^2}{E} -g\left(\frac{R^2}{E} \right)\right]
    \int_{\partial\Omega_o} \left(
    -p_0\mbs n\cdot\mbs u_2 + \mbs f_b\cdot\mbs u_2
    \right)dA
    + \left(1-\frac{R^2}{E}\right)\left(
    \left|\int_{\Omega}\mbs f\cdot\mbs u_1d\Omega  \right|
    + \left|\int_{\Omega}\mbs f\cdot\mbs u_2d\Omega  \right|
    \right) \\
    &
    + \left(1-\frac{R^2}{E}\right)\left(
    \left|\int_{\Omega_o}\mbs f_b\cdot\mbs u_1dA  \right|
    + \left|\int_{\Omega_o}\mbs f_b\cdot\mbs u_2dA  \right|
    \right) \\
    &
    + \left(1-\frac{R^2}{E}\right)\left(
    \left|\int_{\partial\Omega_o} p_0\mbs n\cdot\mbs u_1dA  \right|
    + \left|\int_{\partial\Omega_o} p_0\mbs n\cdot\mbs u_2dA  \right|
    \right) \\
    &
    + \left(1-\frac{R^2}{E}\right)\left(
    \left|\int_{\partial\Omega_d}\left[
      -p_1\mbs n + \nu\mbs n\cdot\nabla\mbs u_1
      - \frac{1}{2}(\mbs n\cdot\mbs w)\mbs w
      \right]\cdot\mbs w dA\right|
    +\left|\int_{\partial\Omega_d}\left(
    -p_2\mbs n + \nu\mbs n\cdot\nabla\mbs u_2
    \right)\cdot\mbs w dA\right|
    \right).
  \end{split}
\end{equation}
In the above expression $|(\cdot)|$ denotes the absolute value
of the variable $(\cdot)$.
Then equation \eqref{equ:R_equ} is transformed into
\begin{equation}\label{equ:R_equ_1}
  \begin{split}
  2R\frac{dR}{dt} =& \int_{\Omega}\mbs u\cdot\frac{\partial\mbs u}{\partial t}d\Omega
  + \nu D_0\int_{\partial\Omega_o}\mbs u\cdot\frac{\partial\mbs u}{\partial t} dA \\
  &
  + \frac{R^2}{E}\left[
    -\nu\int_{\Omega}\|\nabla\mbs u \|^2d\Omega 
    -\int_{\partial\Omega_o}\left(
    \frac{1}{2}(\mbs n\cdot\mbs u)|\mbs u|^2
    -\mbs H(\mbs n,\mbs u)\cdot\mbs u
    \right)dA
    \right] \\
  &
  -\int_{\Omega}\left[
    -\nabla p + \nu\nabla^2\mbs u - g\left(\frac{R^2}{E} \right)\mbs N(\mbs u)
    + \mbs f
    \right]\cdot\mbs u d\Omega \\
  &
  -\int_{\partial\Omega_o}\left[
    (p-p_0)\mbs n - \nu\mbs n\cdot\nabla\mbs u
    + g\left(\frac{R^2}{E} \right)\mbs H(\mbs n,\mbs u) + \mbs f_b
    \right]\cdot\mbs u dA \\
  &
  + \int_{\partial\Omega_d}\left[
    -\left(p_1 +\frac{R^2}{E}p_2  \right)\mbs n
    + \nu\mbs n\cdot\nabla\left(\mbs u_1 +\frac{R^2}{E} \mbs u_2  \right)
    - \frac12(\mbs n\cdot\mbs w)\mbs w
    \right]\cdot\mbs w dA \\
  &
  + \int_{\Omega}\mbs f\cdot\left(\mbs u_1 +\frac{R^2}{E} \mbs u_2  \right)d\Omega
  + \int_{\partial\Omega_o}(\mbs f_b-p_0\mbs n)\cdot
  \left(\mbs u_1 +\frac{R^2}{E} \mbs u_2  \right) dA \\
  &
  + \left(1-\frac{R^2}{E}\right)\left(
  \left|\int_{\Omega}\mbs f\cdot\mbs u_1d\Omega  \right|
  + \left|\int_{\Omega}\mbs f\cdot\mbs u_2d\Omega  \right|
  +\left|\int_{\partial\Omega_o}\mbs f_b\cdot\mbs u_1dA  \right|
  + \left|\int_{\partial\Omega_o}\mbs f_b\cdot\mbs u_2dA  \right|
  \right) \\
  &
  + \left(1-\frac{R^2}{E}\right)\left(
  \left|\int_{\partial\Omega_o} p_0\mbs n\cdot\mbs u_1dA  \right|
  + \left|\int_{\partial\Omega_o} p_0\mbs n\cdot\mbs u_2dA  \right|
  +\left|\int_{\partial\Omega_d}\left(
  -p_2\mbs n + \nu\mbs n\cdot\nabla\mbs u_2
  \right)\cdot\mbs w dA\right|
  \right) \\
  &
  + \left(1-\frac{R^2}{E}\right)
  \left|\int_{\partial\Omega_d}\left[
    -p_1\mbs n + \nu\mbs n\cdot\nabla\mbs u_1
    - \frac{1}{2}(\mbs n\cdot\mbs w)\mbs w
    \right]\cdot\mbs w dA\right|.
  \end{split}
\end{equation}

The reformulated equivalent system consists of equations
\eqref{equ:nse_1}, \eqref{equ:div} and
\eqref{equ:R_equ_1}, the boundary conditions \eqref{equ:dbc}
and \eqref{equ:obc_1}, the initial condition \eqref{equ:ic}
for $\mbs u$ and the following initial condition for $R(t)$,
\begin{equation}\label{equ:ic_R}
  R(0) = \sqrt{E(0)}, \quad \text{where} \
  E(0)=
    \int_{\Omega}\frac12|{\mbs u}_{in}|^2d\Omega
    + \nu D_0\int_{\partial\Omega_o}\frac12|{\mbs u}_{in}|^2dA
    + C_0.
\end{equation}
In this system the dynamic variables are $\mbs u(\mbs x,t)$,
$p(\mbs x,t)$ and $R(t)$. $E(t)$ is given by
equation \eqref{equ:def_E}. Note that $R(t)$ is obtained
by solving this coupled system of equations, not by using
equation \eqref{equ:def_R}.
So in such a sense
$R^2(t)$ is an approximation of $E(t)$,
rather than $E(t)$ itself.

\subsection{Numerical Scheme and Discrete Energy Stability}

We next present an unconditionally energy-stable scheme
for numerically solving the reformulated
system of equations. While seemingly a bit involved,
this scheme allows for an efficient
solution algorithm and efficient implementation.

Let $n\geqslant 0$ denote the time step index, and
$(\cdot)^n$ denote the variable $(\cdot)$ at time step $n$.
Define
\begin{equation}
  \mbs u^0 = \mbs u_{in}, \quad
  R^0 = R(0).
\end{equation}
Then given ($\mbs u^n$,$R^n$) and these variables
at previous time steps, we compute
$(\mbs u^{n+1},p^{n+1},R^{n+1})$ by the following
scheme: \\
\noindent\underline{For $p^{n+1}$:}
\begin{subequations}
  \begin{align}
    &
    \frac{\frac32 \tilde{\mbs u}^{n+1}-2\mbs u^n + \frac12\mbs u^{n-1}}{\Delta t}
    + g(\xi)\mbs N(\mbs u^{*,n+1}) + \nabla p^{n+1}
    + \nu\nabla\times\nabla\times\mbs u^{*,n+1}
    =\mbs f^{n+1};
    \label{equ:peq_1} \\
    &
    \xi = \frac{\left(R^{n+3/2}\right)^2}{E[\bar{\mbs u}^{n+3/2}]};
    \label{equ:peq_2} \\
    &
    E[\bar{\mbs u}^{n+3/2}] = \int_{\Omega}
    \frac12\left|\bar{\mbs u}^{n+3/2} \right|^2d\Omega
    + \nu D_0\int_{\partial\Omega_o}
    \frac12\left|\bar{\mbs u}^{n+3/2} \right|^2dA + C_0;
    \label{equ:peq_3} \\
    &
    \nabla\cdot\tilde{\mbs u}^{n+1} = 0; \label{equ:peq_4} \\
    &
    \mbs n\cdot\tilde{\mbs u}^{n+1} = \mbs n\cdot\mbs w^{n+1},
    \quad \text{on} \ \partial\Omega_d;
    \label{equ:peq_5} \\
    \begin{split}
      &
      \nu D_0\frac{\frac32\tilde{\mbs u}^{n+1}-2\mbs u^n+\frac12\mbs u^{n-1}}{\Delta t}
    \cdot\mbs n
     -\left(p^{n+1} -p_0^{n+1} \right)
    + g(\xi) \nu\mbs n\cdot\nabla\mbs u^{*,n+1}\cdot\mbs n
    - g(\xi)\mbs H(\mbs n,\mbs u^{*,n+1})\cdot\mbs n \\
    & \qquad\qquad
    - g(\xi)\nu\nabla\cdot\mbs u^{*,n+1} = \mbs f_b^{n+1}\cdot\mbs n,
    \quad \text{on} \ \partial\Omega_o.
    \end{split}
    \label{equ:peq_6}
  \end{align}
\end{subequations}
\noindent\underline{For $\mbs u^{n+1}$:}
\begin{subequations}
  \begin{align}
    &
    \frac{\frac32\mbs u^{n+1}-\frac32\tilde{\mbs u}^{n+1}}{\Delta t}
    - \nu\nabla^2\mbs u^{n+1} = \nu\nabla\times\nabla\times\mbs u^{*,n+1};
    \label{equ:veq_1} \\
    &
    \mbs u^{n+1} = \mbs w^{n+1} \quad \text{on} \ \partial\Omega_d;
    \label{equ:veq_2} \\
    \begin{split}
    &
    \nu D_0\frac{\frac32\mbs u^{n+1}-2\mbs u^n + \frac12\mbs u^{n-1}}{\Delta t}
    - \left(p^{n+1}-p_0^{n+1} \right)\mbs n
    + \nu\mbs n\cdot\nabla\mbs u^{n+1}
    - g(\xi)\mbs H(\mbs n,\mbs u^{*,n+1}) 
    = \mbs f_b^{n+1}, \\
    & \qquad\qquad
    \text{on} \ \partial\Omega_o.
    \end{split}
    \label{equ:veq_3}
  \end{align}
\end{subequations}
\noindent\underline{For $R^{n+1}$:}
\begin{align}\label{equ:Req_1}
  \begin{split}
    &
    \left(\frac32 R^{n+1}+R^n-\frac12 R^{n-1} \right)
    \frac{\frac32 R^{n+1}-2R^n+\frac12 R^{n-1}}{\Delta t} \\
    &
    =\int_{\Omega}\mbs u^{n+1}\cdot
    \frac{\frac32\mbs u^{n+1}-2\mbs u^n + \frac12\mbs u^{n-1}}{\Delta t}d\Omega 
    + \nu D_0\int_{\partial\Omega_o} \mbs u^{n+1}\cdot
    \frac{\frac32\mbs u^{n+1}-2\mbs u^n + \frac12\mbs u^{n-1}}{\Delta t}dA \\
    &\quad
    + \xi\left[
      -\nu\int_{\Omega}\left\|\nabla\bar{\mbs u}^{n+1} \right\|^2d\Omega
      -\int_{\partial\Omega_o}\left(
      \frac12(\mbs n\cdot\bar{\mbs u}^{n+1})\left|\bar{\mbs u}^{n+1} \right|^2
      -\mbs H(\mbs n,\bar{\mbs u}^{n+1})\cdot\bar{\mbs u}^{n+1}
      \right)dA
      \right] \\
    &\quad
    -\int_{\Omega}\left[
      -\nabla p^{n+1} + \nu\nabla^2\mbs u^{n+1} + \mbs f^{n+1}
      - g(\xi)\mbs N(\mbs u^{*,n+1})
      \right] \cdot\mbs u^{n+1} d\Omega \\
    &\quad
    -\int_{\partial\Omega_o}\left[
      \left(p^{n+1}-p_0^{n+1} \right)\mbs n
      -\nu\mbs n\cdot\nabla\mbs u^{n+1}
      + g(\xi)\mbs H(\mbs n,\mbs u^{*,n+1}) + \mbs f_b^{n+1}
      \right] \cdot\mbs u^{n+1} dA \\
    &\quad
    + \int_{\partial\Omega_d}\left[
      -\left(p_1^{n+1}+\xi p_2^{n+1} \right)\mbs n
      + \nu\mbs n\cdot\nabla\left(\mbs u_1^{n+1}+\xi\mbs u_2^{n+1} \right)
      -\frac12(\mbs n\cdot\mbs w^{n+1})\mbs w^{n+1}
      \right]\cdot\mbs w^{n+1} dA \\
    &\quad
    + \int_{\Omega}\mbs f^{n+1}\cdot\left(\mbs u_1^{n+1}+\xi\mbs u_2^{n+1}  \right)d\Omega
    + \int_{\partial\Omega_o}\left(\mbs f_b^{n+1}-p_0^{n+1}\mbs n \right)\cdot\left(
    \mbs u_1^{n+1}+\xi\mbs u_2^{n+1} \right) dA \\
    &\quad
    + (1-\xi)\left(
    \left|\int_{\Omega}\mbs f^{n+1}\cdot\mbs u_1^{n+1}d\Omega  \right|
    + \left|\int_{\Omega}\mbs f^{n+1}\cdot\mbs u_2^{n+1}d\Omega  \right|
    + \left|\int_{\partial\Omega_o}\mbs f_b^{n+1}\cdot\mbs u_1^{n+1}dA  \right| \right. \\
    &\qquad\qquad\quad
    \left.
    + \left|\int_{\partial\Omega_o}\mbs f_b^{n+1}\cdot\mbs u_2^{n+1}dA  \right|
    + \left|\int_{\partial\Omega_o}p_0^{n+1}\mbs n\cdot\mbs u_1^{n+1}  \right|
    + \left|\int_{\partial\Omega_o}p_0^{n+1}\mbs n\cdot\mbs u_2^{n+1}  \right| \right. \\
    &\qquad\qquad\quad
    \left.
    + \left|\int_{\partial\Omega_d}\left[
      -p_1^{n+1}\mbs n + \nu\mbs n\cdot\nabla\mbs u_1^{n+1}
      -\frac12(\mbs n\cdot\mbs w^{n+1})\mbs w^{n+1}
      \right]\cdot\mbs w^{n+1} dA \right| \right. \\
    &\qquad\qquad\quad
    \left.
    + \left|\int_{\partial\Omega_d}\left[
      -p_2^{n+1}\mbs n + \nu\mbs n\cdot\nabla\mbs u_2^{n+1}
      \right]\cdot\mbs w^{n+1}\right|
    \right).
  \end{split}
\end{align}


The symbols in the above equations are defined as follows.
$\tilde{\mbs u}^{n+1}$ is an auxiliary
field variable approximating $\mbs u^{n+1}$.
$\mbs u^{*,n+1}$ is a second-order explicit approximation of
$\mbs u^{n+1}$, defined by
\begin{equation} \label{equ:ustar}
\mbs u^{*,n+1} = 2\mbs u^{n} - \mbs u^{n-1}.
\end{equation}
$\bar{\mbs u}^{n+1}$ and $\bar{\mbs u}^{n+3/2}$ are second-order approximations
of $\mbs u^{n+1}$ and $\mbs u^{n+3/2}$ to be specified later
in equations \eqref{equ:ubar} and \eqref{equ:un32bar}.
$R^{n+3/2}$ and $R^{n+1/2}$ are defined by
\begin{equation}\label{equ:Rn32}
  R^{n+3/2} = \frac32 R^{n+1} - \frac12 R^n, \quad
  R^{n+1/2} = \frac32 R^n - \frac12 R^{n-1}.
\end{equation}
The following relation will be useful subsequently when dealing with
equation \eqref{equ:Req_1},
\begin{equation}\label{equ:R_rela}
  \begin{split}
  &\left(\frac32 R^{n+1}+R^n-\frac12 R^{n-1} \right)
  \left(\frac32 R^{n+1}-2R^n+\frac12 R^{n-1}  \right) \\
  &= \left(R^{n+3/2} + R^{n+1/2}  \right)\left(R^{n+3/2} - R^{n+1/2}  \right)
  = \left|R^{n+3/2} \right|^2 - \left|R^{n+1/2}  \right|^2.
  \end{split}
\end{equation}
$p_1^{n+1}$, $p_2^{n+1}$, $\mbs u_1^{n+1}$ and $\mbs u_2^{n+1}$ are
field variables related to $p^{n+1}$ and $\mbs u^{n+1}$ that will be
specifically defined later in equations \eqref{equ:def_p1}--\eqref{equ:def_p2}
and \eqref{equ:u1_1}--\eqref{equ:u2_2}.

It is crucial to note that in this scheme all terms are approximated
at time step ($n+1$), except for the term $\frac{R^2(t)}{E(t)}$,
which is approximated at time step $(n+3/2)$ as given in
equation \eqref{equ:peq_2}.
Note that $\xi=\frac{\left(R^{n+3/2} \right)^2}{E[\bar{\mbs u}^{n+3/2}]}$
is a second-order approximation of $\frac{R^2(t)}{E(t)}=1$,
because $R^{n+3/2}$ and $E[\bar{\mbs u}^{n+3/2}]$ are second-order
approximations of $R(t)$ and $E(t)$ at step ($n+3/2$)
and $\frac{R^2(t)}{E(t)}$ is the unit value.
Therefore this treatment does not affect the temporal second-order
accuracy of the scheme.
This treatment was originally used in~\cite{YangD2019diss} for general
dissipative systems.
It allows the auxiliary variable $R^{n+1}$ to be computed explicitly by
a well-defined  formula, and guarantees that the computed values
for $R^{n+1}$ are always positive.

The scheme represented by equations \eqref{equ:peq_1}--\eqref{equ:Req_1}
is unconditionally energy stable because of the following
stability property.
\begin{theorem}
  \label{thm:thm_1}
  In the absence of the external forces and source terms
  ($\mbs f=0$, $p_0=0$, $\mbs f_b=0$) and with
  homogeneous condition on the Dirichlet boundary ($\mbs w=0$),
  and as $\delta \rightarrow 0$ in the open boundary condition \eqref{equ:obc},
  the scheme given by equations \eqref{equ:peq_1}--\eqref{equ:Req_1}
  satisfies the relation
  \begin{equation}
    \left(R^{n+3/2} \right)^2 - \left(R^{n+1/2} \right)^2
    = -\xi\Delta t\left[
      \nu\int_{\Omega}\left\|\nabla\bar{\mbs u}^{n+1}  \right\|^2d\Omega
      + \nu D_0\int_{\partial\Omega_o}\frac{1}{2}\left|\bar{\mbs u}^{n+1}\right|^2
      \left|\mbs n\cdot\bar{\mbs u}^{n+1}\right| dA
      \right]
    \leqslant 0,
    \label{equ:eng_law}
  \end{equation}
  where $R^{n+3/2}$ and $R^{n+1/2}$ are defined in \eqref{equ:Rn32}.
\end{theorem}
\begin{proof}
  Take the $L^2$ inner product between $\mbs u^{n+1}$ and equation \eqref{equ:peq_1}.
  Take the $L^2$ inner produce between $\mbs u^{n+1}$ and
  equation \eqref{equ:veq_1}. Take the inner product
  between $\mbs u^{n+1}$ and equation \eqref{equ:veq_3} and integrate over
  $\partial\Omega_o$. Summing up these equations together with
  equation \eqref{equ:Req_1} leads to
  \begin{equation}\label{equ:eng_1}
    \begin{split}
    &\left(R^{n+3/2} \right)^2 - \left(R^{n+1/2} \right)^2
    = -\xi\Delta t\left[
      \nu\int_{\Omega}\left\|\nabla\bar{\mbs u}^{n+1}  \right\|^2d\Omega \right.\\
      &\quad
      \left.
      + \int_{\partial\Omega_o}\left(\frac{1}{2}\left|\bar{\mbs u}^{n+1}\right|^2
      \left(\mbs n\cdot\bar{\mbs u}^{n+1}\right)
      -\mbs H(\mbs n,\bar{\mbs u}^{n+1})\cdot\bar{\mbs u}^{n+1}
      \right)dA
      \right]
    -\xi S_0\Delta t  + S_1\Delta t,
    \end{split}
  \end{equation}
  where
  \begin{equation}\label{equ:def_S}
    \left\{
    \begin{split}
      &
      S_0 = |A_1| + (|B_1|-B_1) + |A_2| + (|B_2|-B_2)
      + |A_3| + (|B_3|-B_3) + |A_4| + (|B_4|-B_4), \\
      &
      S_1 = (|A_1|+A_1) + |B_1| + (|A_2|+A_2) + |B_2|
      + (|A_3| + A_3) + |B_3| + (|A_4|+A_4) + |A_4|, \\
      &
      A_1 = \int_{\Omega}\mbs f^{n+1}\cdot\mbs u_1^{n+1}d\Omega, \quad
      B_1 = \int_{\Omega}\mbs f^{n+1}\cdot\mbs u_2^{n+1}d\Omega, \\
      &
      A_2 = \int_{\partial\Omega_d}\left[
        -p_1^{n+1}\mbs n + \nu\mbs n\cdot\nabla\mbs u_1^{n+1}
        -\frac12(\mbs n\cdot\mbs w^{n+1})\mbs w^{n+1}
        \right]\cdot\mbs w^{n+1} dA, \\
      &
      B_2 = \int_{\partial\Omega_d}\left[
        -p_2^{n+1}\mbs n + \nu\mbs n\cdot\nabla\mbs u_2^{n+1}
        \right]\cdot\mbs w^{n+1} dA, \\
      &
      A_3 = \int_{\partial\Omega_o}\mbs f_b^{n+1}\cdot\mbs u_1^{n+1}dA, \quad
      B_3 = \int_{\partial\Omega_o} \mbs f_b^{n+1}\cdot\mbs u_2^{n+1}dA, \\
      &
      A_4 = -\int_{\partial\Omega_o}p_0^{n+1}\mbs n\cdot\mbs u_1^{n+1} dA, \quad
      B_4 = -\int_{\partial\Omega_o} p_0^{n+1}\mbs n\cdot\mbs u_2^{n+1}dA.
    \end{split}
    \right.
  \end{equation}
  In light of \eqref{equ:def_H}, one notes that
  \begin{equation}\label{equ:eng_lim}
    \frac12(\mbs n\cdot\bar{\mbs u}^{n+1})\left|\bar{\mbs u}^{n+1} \right|^2
    - \mbs H(\mbs n,\bar{\mbs u}^{n+1})\cdot\bar{\mbs u}^{n+1}
    \rightarrow
    \frac12 \left|\bar{\mbs u}^{n+1} \right|^2\left|\mbs n\cdot\bar{\mbs u}^{n+1} \right|,
    \quad
    \text{as} \ \delta \rightarrow 0.
  \end{equation}
  If $\mbs f=0$, $p_0=0$, $\mbs f_b=0$ and $\mbs w=0$, then
  $S_0=0$ and $S_1=0$.
  Therefore equation \eqref{equ:eng_1} leads to \eqref{equ:eng_law}.
  Note that $E[\bar{\mbs u}^{n+1}]>0$ and $\xi \geqslant 0$
  in light of \eqref{equ:peq_2} and \eqref{equ:peq_3}.
  We conclude that the inequality in \eqref{equ:eng_law} holds.
\end{proof}


\begin{remark}
  \label{rem:rem_0}
  Barring the terms involving the unknown $g(\xi)$, equations
  \eqref{equ:peq_1}--\eqref{equ:veq_3} resemble a rotational
  velocity correction-type scheme for the incompressible
  Navier-Stokes equations
  (see e.g.~\cite{Dong2015clesobc,GuermondMS2006,DongKC2014,DongS2010}).
  Such a velocity correction scheme alone
  is known to be only conditionally stable.
  The current numerical scheme builds upon the
  velocity correction strategy. Because of the auxiliary variable $R(t)$
  introduced
  here and the coupling terms, 
  the overall scheme becomes
  unconditionally energy-stable thanks to Theorem \ref{thm:thm_1}.
\end{remark}

\begin{remark} \label{rem:rem_1}
  Note that the numerical scheme from \cite{LinYD2019} also employs
  an auxiliary variable.  Several major differences distinguish
  the scheme herein from the one from \cite{LinYD2019}:
  \begin{itemize}
  \item
    In the scheme of \cite{LinYD2019}, the pressure $p^{n+1}$ and
    the velocity $\mbs u^{n+1}$ are fully coupled, and the energy stability therein
    is proven in this fully coupled setting. In the implementation
    in \cite{LinYD2019}, a further approximation is made to decouple
    the computations for the pressure and the velocity. The discrete
    energy stability of \cite{LinYD2019}, however, 
    breaks down mathematically with that approximation.
    In contrast, in the current scheme the pressure and the velocity
    are de-coupled, except for the $g(\xi)$ term, which can be 
    dealt with in a straightforward way (see later discussions). The discrete
    energy stability presented here holds in the de-coupled setting.

  \item
    The dynamic equations for the auxiliary variables,
    and their numerical discretizations, in
    the current work and in \cite{LinYD2019}
    are completely different. In \cite{LinYD2019} a nonlinear algebraic
    equation  needs to be solved based on the
    Newton's method when computing
    the auxiliary variable. In contrast, the auxiliary variable
    in the current work is computed by a well-defined explicit
    formula, and no nonlinear algebraic solver is involved. Furthermore,
    the computed values for the auxiliary variable here are guaranteed
    to be positive, and this positivity property is unavailable
    in the method of \cite{LinYD2019}. These points will become
    clear in subsequent discussions.

  \item
    The scheme developed herein is energy stable for flow problems
    involving open/outflow boundaries and Dirichlet boundaries.
    The scheme developed in \cite{LinYD2019} works only with Dirichlet
    boundaries.
        
  \end{itemize}
  
\end{remark}


\subsection{Solution Algorithm and Implementation
  with High-Order Spectral Elements}
\label{sec:soln}

Let us next consider how to implement the scheme
represented by equations \eqref{equ:peq_1}--\eqref{equ:Req_1},
which are seemingly all coupled with one another.
It is critical to realize 
the fact that the variables  $R(t)$, $E[\mbs u]$, $\xi$
and $g(\xi)$ in these equations are but
scalar-valued numbers, not field functions.
By exploiting this fact, we can implement the scheme and
compute different variables in a decoupled and efficient fashion.

Let
\begin{equation}\label{equ:uhat}
  \hat{\mbs u} = 2\mbs u^n - \frac{1}{2}\mbs u^{n-1}, \quad
  \gamma_0 = \frac32.
\end{equation}
We re-write equation \eqref{equ:peq_1} as,
\begin{equation}\label{equ:p_1}
  \frac{\gamma_0}{\Delta t}\tilde{\mbs u}^{n+1}
  +\nabla p^{n+1} = \mbs f^{n+1} + \frac{\hat{\mbs u}}{\Delta t}
  - g(\xi)\mbs N(\mbs u^{*,n+1})
  - \nu\nabla\times\bm{\omega}^{*,n+1},
\end{equation}
where $\bm{\omega}^{*,n+1}=\nabla\times\mbs u^{*,n+1}$ is
the vorticity.
We would like to derive the weak forms of this and subsequent equations so that
certain types of boundary conditions can be incorporated within.
Let $q(\mbs x)$ denote an arbitrary test function, which is
continuous in space.
Its discrete function space will be specified later.
Taking the $L^2$ inner product between $\nabla q$ and
equation \eqref{equ:p_1} yields,
\begin{equation}\label{equ:p_weak}
  \begin{split}
    &
  \int_{\Omega}\nabla p^{n+1}\cdot\nabla q d\Omega
  + \frac{1}{\nu D_0}\int_{\partial\Omega_o}p^{n+1}q dA
  = \int_{\Omega}\left[
    \mbs f^{n+1} + \frac{\hat{\mbs u}}{\Delta t}
    - g(\xi)\mbs N(\mbs u^{*,n+1})
    \right]\cdot\nabla qd\Omega \\
  &\qquad
  -\nu\int_{\partial\Omega}\mbs n\times\bm{\omega}^{*,n+1}\cdot\nabla qdA
  - \frac{\gamma_0}{\Delta t}\int_{\partial\Omega_d}\mbs n\cdot\mbs w^{n+1}qdA
  -\int_{\partial\Omega_o}\left[
     \frac{1}{\nu D_o}\left(
        \mbs f_b^{n+1}\cdot\mbs n -p_0^{n+1}  
     \right. \right.\\
    & \qquad
     - g(\xi)\nu\mbs n\cdot\nabla\mbs u^{*,n+1}\cdot\mbs n 
    \left.\left.
    + g(\xi)\mbs H(\mbs n,\mbs u^{*,n+1})\cdot\mbs n
    + g(\xi)\nu\nabla\cdot\mbs u^{*,n+1}
    \right)
    + \frac{\hat{\mbs u}}{\Delta t}\cdot\mbs n
    \right] qdA,
  \quad \forall q,
  \end{split}
\end{equation}
where we have used integration by part, the divergence theorem,
equations \eqref{equ:peq_4}--\eqref{equ:peq_6}, and the identity
$
\int_{\Omega}\nabla\times\bm{\omega}^{*,n+1}\cdot\nabla q d\Omega
= \int_{\partial\Omega}\mbs n\times\bm{\omega}^{*,n+1}\cdot\nabla q dA.
$

In order to solve equation \eqref{equ:p_weak} for $p^{n+1}$,
we define two field variables $p_1^{n+1}$ and $p_2^{n+1}$
as solutions to the following equations, respectively,
\begin{align}
  \label{equ:def_p1}
  \begin{split}
  &
  \int_{\Omega}\nabla p_1^{n+1}\cdot\nabla q d\Omega
  + \frac{1}{\nu D_0}\int_{\partial\Omega_o}p_1^{n+1}q dA
  = \int_{\Omega}\left(
    \mbs f^{n+1} + \frac{\hat{\mbs u}}{\Delta t}
    \right)\cdot\nabla qd\Omega
    -\nu\int_{\partial\Omega}\mbs n\times\bm{\omega}^{*,n+1}\cdot\nabla qdA \\
    &\qquad\qquad
    - \frac{\gamma_0}{\Delta t}\int_{\partial\Omega_d}\mbs n\cdot\mbs w^{n+1}qdA
    -\int_{\partial\Omega_o}\left[
      \frac{\hat{\mbs u}}{\Delta t}\cdot\mbs n
      +\frac{1}{\nu D_0}\left(-p_0^{n+1}+\mbs f_b^{n+1}\cdot\mbs n  \right)
      \right]q dA,
    \quad \forall q; 
  \end{split}\\
  \label{equ:def_p2}
  \begin{split}
  &
  \int_{\Omega}\nabla p_2^{n+1}\cdot\nabla q d\Omega
  + \frac{1}{\nu D_0}\int_{\partial\Omega_o}p_2^{n+1}q dA
  = -\int_{\Omega}\mbs N(\mbs u^{*,n+1})\cdot\nabla q d\Omega \\
  &\qquad\qquad
  -\frac{1}{\nu D_0}\int_{\partial\Omega_o}\left[
    -\nu\mbs n\cdot\nabla\mbs u^{*,n+1}\cdot\mbs n
    + \mbs H(\mbs n,\mbs u^{*,n+1})\cdot\mbs n
    + \nu\nabla\cdot\mbs u^{*,n+1}
    \right] q dA,
  \quad \forall q.
  \end{split}
\end{align}
Then, noting that $g(\xi)$ is a scalar-valued number, the solution
to \eqref{equ:p_weak} is given by
\begin{equation}\label{equ:p_soln}
  p^{n+1} = p_1^{n+1} + g(\xi) p_2^{n+1},
\end{equation}
where $\xi$ still needs to be determined.

Summation of equations \eqref{equ:peq_1} and \eqref{equ:veq_1}
leads to
\begin{equation}\label{equ:v_1}
  \frac{\gamma_0}{\nu\Delta t}\mbs u^{n+1}
  -\nabla^2\mbs u^{n+1}
  =\frac{1}{\nu}\left[
    \mbs f^{n+1} + \frac{\hat{\mbs u}}{\Delta t}
    -\nabla p^{n+1}
    \right]
  -\frac{1}{\nu}g(\xi)\mbs N(\mbs u^{*,n+1}).
\end{equation}
Let $\varphi(\mbs x)$ denote an arbitrary test function (continuous in space)
that vanishes on $\partial\Omega_d$,
i.e.~$\left.\varphi \right|_{\partial\Omega_d}=0$.
Taking the $L^2$ inner product between $\varphi$ and
equation \eqref{equ:v_1}, we get
\begin{equation}
  \begin{split}
    &
  \int_{\Omega}\nabla\varphi\cdot\nabla\mbs u^{n+1}d\Omega
  + \frac{\gamma_0}{\nu\Delta t}\int_{\Omega}\mbs u^{n+1}\varphi d\Omega
  = \frac{1}{\nu}\int_{\Omega}\left[
    \mbs f^{n+1} + \frac{\hat{\mbs u}}{\Delta t}
    -\nabla p^{n+1}
    -g(\xi)\mbs N(\mbs u^{*,n+1})
    \right] \varphi d\Omega \\
  &\qquad\qquad
  + \int_{\partial\Omega_o}\mbs n\cdot\nabla\mbs u^{n+1}\varphi dA,
  \quad \forall\varphi \ \text{with}\ \varphi|_{\partial\Omega_d}=0,
  \end{split}
  \label{equ:v_2}
\end{equation}
where the divergence theorem has been used.
In light of \eqref{equ:veq_3}, we have
\begin{equation}
  \mbs n\cdot\nabla\mbs u^{n+1}
  =-\frac{\gamma_0 D_0}{\Delta t}\mbs u^{n+1}
  + \frac{D_0}{\Delta t}\hat{\mbs u}
  + \frac{1}{\nu}\left[
    \mbs f_b^{n+1} + (p^{n+1}-p_0^{n+1})\mbs n
    + g(\xi)\mbs H(\mbs n,\mbs u^{*,n+1})
    \right],
  \quad \text{on} \ \partial\Omega_o.
\end{equation}
Equation \eqref{equ:v_2} can then be transformed into
\begin{equation}\label{equ:u_weak}
  \begin{split}
    &
    \int_{\Omega}\nabla\varphi\cdot\nabla\mbs u^{n+1}d\Omega
    + \frac{\gamma_0}{\nu\Delta t}\int_{\Omega}\mbs u^{n+1}\varphi d\Omega
    + \frac{\gamma_0 D_0}{\Delta t}\int_{\partial\Omega_o}\mbs u^{n+1}\varphi dA \\
    &
    = \frac{1}{\nu}\int_{\Omega}\left[
    \mbs f^{n+1} + \frac{\hat{\mbs u}}{\Delta t}
    -\nabla p^{n+1}
    -g(\xi)\mbs N(\mbs u^{*,n+1})
    \right] \varphi d\Omega \\
    &\quad
    + \int_{\partial\Omega_o}\left\{
      \frac{D_0}{\Delta t}\hat{\mbs u}
      + \frac{1}{\nu}\left[
      \mbs f_b^{n+1} + (p^{n+1}-p_0^{n+1})\mbs n
      + g(\xi)\mbs H(\mbs n,\mbs u^{*,n+1})
      \right]
      \right\}\varphi dA,
      \quad \forall\varphi \ \text{with}\ \left.\varphi\right|_{\partial\Omega_d}=0.
  \end{split}
\end{equation}

In order to solve equation \eqref{equ:u_weak} together with \eqref{equ:veq_2}
for $\mbs u^{n+1}$, we define two field variables $\mbs u_1^{n+1}$
and $\mbs u_2^{n+1}$ as solutions to
the following equations: \\
\noindent\underline{For $\mbs u_1^{n+1}$:}
\begin{subequations}
  \begin{align}
    \begin{split}\label{equ:u1_1}
      &
      \int_{\Omega}\nabla\varphi\cdot\nabla\mbs u_1^{n+1}d\Omega
      + \frac{\gamma_0}{\nu\Delta t}\int_{\Omega}\mbs u_1^{n+1}\varphi d\Omega
      + \frac{\gamma_0 D_0}{\Delta t}\int_{\partial\Omega_o}\mbs u_1^{n+1}\varphi dA \\
      &= \frac{1}{\nu}\int_{\Omega}\left(
        \mbs f^{n+1} + \frac{\hat{\mbs u}}{\Delta t}
        -\nabla p_1^{n+1}
        \right) \varphi d\Omega
      + \int_{\partial\Omega_o}\left[
        \frac{D_0}{\Delta t}\hat{\mbs u}
        + \frac{1}{\nu}\left(
        \mbs f_b^{n+1} + (p_1^{n+1}-p_0^{n+1})\mbs n
        \right)
        \right]\varphi dA, \\
      &\qquad \forall\varphi \ \text{with} \ \varphi|_{\partial\Omega_d}=0;
    \end{split} \\
    &
    \mbs u_1^{n+1} = \mbs w^{n+1}, \quad \text{on} \ \partial\Omega_d.
    \label{equ:u1_2}
  \end{align}
\end{subequations}
\noindent\underline{For $\mbs u_2^{n+1}$:}
\begin{subequations}
  \begin{align}
    \begin{split}\label{equ:u2_1}
      &
      \int_{\Omega}\nabla\varphi\cdot\nabla\mbs u_2^{n+1}d\Omega
      + \frac{\gamma_0}{\nu\Delta t}\int_{\Omega}\mbs u_2^{n+1}\varphi d\Omega
      + \frac{\gamma_0 D_0}{\Delta t}\int_{\partial\Omega_o}\mbs u_2^{n+1}\varphi dA \\
      &= -\frac{1}{\nu}\int_{\Omega}\left[
        \nabla p_2^{n+1}+\mbs N(\mbs u^{*,n+1})
        \right] \varphi d\Omega
      + \frac{1}{\nu}\int_{\partial\Omega_o} \left[
        p_2^{n+1}\mbs n + \mbs H(\mbs n,\mbs u^{*,n+1})
        \right]\varphi dA, \\
      &\qquad \forall\varphi \ \text{with} \ \varphi|_{\partial\Omega_d}=0;
    \end{split} \\
    &
    \mbs u_2^{n+1} = 0, \quad \text{on} \ \partial\Omega_d.
    \label{equ:u2_2}
  \end{align}
\end{subequations}
Then, by exploiting the fact that $g(\xi)$ is a scalar-valued number,
the solution to the equations \eqref{equ:u_weak} and \eqref{equ:veq_2}
can be written as
\begin{equation}\label{equ:u_soln}
  \mbs u^{n+1} = \mbs u_1^{n+1} + g(\xi)\mbs u_2^{n+1},
\end{equation}
in which $\xi$ will be determined below.

With $\mbs u_i^{n+1}$ ($i=1,2$) given by equations
 \eqref{equ:u1_1}--\eqref{equ:u2_2},
we define
\begin{align}
  &
  \bar{\mbs u}^{n+1} = \mbs u_1^{n+1} + \mbs u_2^{n+1}, \label{equ:ubar} \\
  &
  \bar{\mbs u}^{n+3/2} = \frac32 \bar{\mbs u}^{n+1} - \frac12\mbs u^n.
  \label{equ:un32bar}
\end{align}
Note that these are second-order approximations of $\mbs u^{n+1}$
and $\mbs u^{n+3/2}$, respectively.

Now we are ready to determine $\xi$. Note that
the combination of equations \eqref{equ:peq_1}, \eqref{equ:veq_1}
and \eqref{equ:Req_1} leads to equation \eqref{equ:eng_1}.
Using equation \eqref{equ:peq_2}, we can compute $\xi$
from \eqref{equ:eng_1} as follows,
\begin{equation}\label{equ:cal_xi}
  \xi = \frac{\left(R^{n+1/2} \right)^2 + S_1\Delta t}{
    E[\bar{\mbs u}^{n+3/2}] + (A_0 + B_0 + S_0)\Delta t
  },
\end{equation}
where $S_0$ and $S_1$ are given in \eqref{equ:def_S}, and
\begin{equation}\label{equ:AB0}
  \left\{
  \begin{split}
    &
    A_0 = \nu\int_{\Omega}\left\| \nabla\bar{\mbs u}^{n+1} \right\|^2d\Omega, \\
    &
    B_0 = \int_{\partial\Omega_o}\left(
    \frac12(\mbs n\cdot\bar{\mbs u}^{n+1})\left|\bar{\mbs u}^{n+1} \right|^2
    - \mbs H(\mbs n,\bar{\mbs u}^{n+1})\cdot\bar{\mbs u}^{n+1}
    \right)dA.
  \end{split}
  \right.
\end{equation}
It can be noted that $A_0\geqslant 0$, $S_0\geqslant 0$ and $S_1\geqslant 0$.
In light of the relation \eqref{equ:eng_lim}, we conclude that
$B_0\geqslant 0$
as $\delta$ in the open boundary condition \eqref{equ:obc} is
chosen to be sufficiently small.
It then follows that $\xi > 0$ if $\delta$ is chosen to be sufficiently
small, regardless of the value $\Delta t$ and the external forces
and source terms.

With $\xi$ known, $p^{n+1}$ and $\mbs u^{n+1}$ can be
computed by equations \eqref{equ:p_soln} and \eqref{equ:u_soln}, respectively.
$R^{n+1}$ is computed as follows,
\begin{equation}\label{equ:Rnp1}
  \left\{
  \begin{split}
    &
    R^{n+3/2} = \sqrt{\xi E[\bar{\mbs u}^{n+3/2}]}, \\
    &
    R^{n+1} = \frac{2}{3}R^{n+3/2} + \frac13 R^n,
  \end{split}
  \right.
\end{equation}
where equations \eqref{equ:peq_2} and \eqref{equ:Rn32} have been used.
It can be noted that these computed values satisfy the
property $R^{n+1}> 0$ and $R^{n+3/2}> 0$.
Algorithm \ref{alg:alg_1} summarizes
the final solution procedure within a time step.

\begin{algorithm}[htb]
  \label{alg:alg_1}
  \SetKwInOut{Input}{input}\SetKwInOut{Output}{output}
  \Input{$\mbs u$ and $R$ of time steps $n$ and $(n-1)$}
  \Output{$\mbs u^{n+1}$, $p^{n+1}$, and $R^{n+1}$}
  \BlankLine\BlankLine
  \Begin{
      Solve equation \eqref{equ:def_p1} for $p_1^{n+1}$\;
      Solve equation \eqref{equ:def_p2} for $p_2^{n+1}$\;
      \BlankLine\BlankLine
      Solve equations \eqref{equ:u1_1}--\eqref{equ:u1_2} for $\mbs u_1^{n+1}$\;
      Solve equations \eqref{equ:u2_1}--\eqref{equ:u2_2} for $\mbs u_2^{n+1}$\;
      \BlankLine\BlankLine
      Compute $\bar{\mbs u}^{n+1}$ and $\bar{\mbs u}^{n+3/2}$ by
      equations \eqref{equ:ubar} and \eqref{equ:un32bar}\;
      Compute $A_0$ and $B_0$ by equation \eqref{equ:AB0}\;
      Compute $S_0$ and $S_1$ by equation \eqref{equ:def_S}\;
      Compute $\xi$ by equation \eqref{equ:cal_xi}\;
      \BlankLine\BlankLine
      Compute $p^{n+1}$ by equation \eqref{equ:p_soln}\;
      Compute $\mbs u^{n+1}$ by equation \eqref{equ:u_soln}\;
      Compute $R^{n+1}$ by equation \eqref{equ:Rnp1}.
      } 
    \caption{Solution algorithm within a time step.}
\end{algorithm}

\begin{remark}
  \label{rem:rem_3}
  Algorithm \ref{alg:alg_1} has several notable properties:
  (i) It is unconditionally energy stable.
  (ii) The computations for different field variables
  ($p_1^{n+1}$, $p_2^{n+1}$, $\mbs u_1^{n+1}$ and $\mbs u_2^{n+1}$)
  are de-coupled, and the resultant linear algebraic systems involve
  only constant coefficient matrices that can be pre-computed.
  (iii) The auxiliary variable is computed by a well-defined
  explicit formula, and its computed values are guaranteed to be
  positive.
  
\end{remark}


Equations \eqref{equ:def_p1}--\eqref{equ:def_p2} and
\eqref{equ:u1_1}--\eqref{equ:u2_2} need to be solved
for the field functions $p_1^{n+1}$, $p_2^{n+1}$, $\mbs u_1^{n+1}$
and $\mbs u_2^{n+1}$. Let us next briefly discuss their
spatial discretizations using $C^0$-type high-order
spectral elements~\cite{KarniadakisS2005}. We discretize the domain $\Omega$
using a mesh consisting of $N_{el}$ conforming elements.
Let $\Omega_h$ denote the discretized domain, $\Omega_{h}^e$ ($e=1,\cdots,N_{el}$)
denote the elements, and $\partial\Omega_h$
denote the discretized domain boundary. The corresponding discretized
Dirichlet and open boundaries are denoted by
$\partial\Omega_{dh}$ and $\partial\Omega_{oh}$
($\partial\Omega_h=\partial\Omega_{dh}\cup\partial\Omega_{oh}$).
Let $K$ (a positive integer) denote a measure of the highest
polynomial degree in field expansions within an element, which will be
called the element order.
Define function spaces,
\begin{equation*}
  \left\{
  \begin{split}
    &
    \mbb{X} = \left\{\ v\in H^1(\Omega_h) \ :\ v\ \text{is a polynomial of degree
      characterized by} \ K \ \text{on}\ \Omega_h^e,\ 1\leqslant e\leqslant N_{el} \
    \right\}; \\
    &
    \mbb X_0 = \left\{\ v\in \mbb X\ :\ v=0\ \text{on}\ \partial\Omega_{dh}  \ \right\}.
  \end{split}
  \right.
\end{equation*}
Let $(\cdot)_h$ denote the discretized version of the variable $(\cdot)$
in what follows.
The fully discretized versions of the equations
\eqref{equ:def_p1}--\eqref{equ:def_p2} are:
Find $p_{1h}^{n+1}\in \mbb X$ and $p_{2h}^{n+1}\in \mbb X$ such that
\begin{align}
  \label{equ:p1disc}
  \begin{split}
  &
  \int_{\Omega_h}\nabla p_{1h}^{n+1}\cdot\nabla q_h d\Omega
  + \frac{1}{\nu D_0}\int_{\partial\Omega_{oh}}p_{1h}^{n+1}q_h dA
  = \int_{\Omega}\left(
    \mbs f_h^{n+1} + \frac{\hat{\mbs u}_h}{\Delta t}
    \right)\cdot\nabla q_hd\Omega \\
    &\qquad
    -\nu\int_{\partial\Omega_h}\mbs n_h\times\bm{\omega}_h^{*,n+1}\cdot\nabla q_hdA 
    - \frac{\gamma_0}{\Delta t}\int_{\partial\Omega_{dh}}\mbs n_h\cdot\mbs w_h^{n+1}q_hdA \\
    &\qquad
    -\int_{\partial\Omega_{oh}}\left[
      \frac{\hat{\mbs u}_h}{\Delta t}\cdot\mbs n_h
      +\frac{1}{\nu D_0}\left(-p_{0h}^{n+1}+\mbs f_{bh}^{n+1}\cdot\mbs n_h  \right)
      \right]q_h dA,
    \quad \forall q_h \in\mbb X; 
  \end{split}\\
  \label{equ:p2disc}
  \begin{split}
  &
  \int_{\Omega}\nabla p_{2h}^{n+1}\cdot\nabla q_h d\Omega
  + \frac{1}{\nu D_0}\int_{\partial\Omega_{oh}}p_{2h}^{n+1}q_h dA
  = -\int_{\Omega_h}\mbs N(\mbs u_h^{*,n+1})\cdot\nabla q_h d\Omega \\
  &\qquad
  -\frac{1}{\nu D_0}\int_{\partial\Omega_{oh}}\left[
    -\nu\mbs n_h\cdot\nabla\mbs u_h^{*,n+1}\cdot\mbs n_h
    + \mbs H(\mbs n_h,\mbs u_h^{*,n+1})\cdot\mbs n_h
    + \nu\nabla\cdot\mbs u_h^{*,n+1}
    \right] q_h dA,
  \quad \forall q_h\in\mbb X.
  \end{split}
\end{align}
The fully discretized versions of the equations
\eqref{equ:u1_1}--\eqref{equ:u2_2} are as follows.
Let $d_{im}$ denote the spatial dimension with $d_{im}=2$ or $3$ below.
\\
\noindent\underline{Find $\mbs u_{1h}^{n+1}\in [\mbb X]^{d_{im}}$ such that:}
\begin{subequations}
  \begin{align}
    \begin{split}\label{equ:u11disc}
      &
      \int_{\Omega_h}\nabla\varphi_h\cdot\nabla\mbs u_{1h}^{n+1}d\Omega
      + \frac{\gamma_0}{\nu\Delta t}\int_{\Omega_h}\mbs u_{1h}^{n+1}\varphi_h d\Omega
      + \frac{\gamma_0 D_0}{\Delta t}\int_{\partial\Omega_{oh}}\mbs u_{1h}^{n+1}\varphi_h dA \\
      &= \frac{1}{\nu}\int_{\Omega_h}\left(
        \mbs f_h^{n+1} + \frac{\hat{\mbs u}_h}{\Delta t}
        -\nabla p_{1h}^{n+1}
        \right) \varphi_h d\Omega
      + \int_{\partial\Omega_{oh}}\left[
        \frac{D_0}{\Delta t}\hat{\mbs u}_h
        + \frac{1}{\nu}\left(
        \mbs f_{bh}^{n+1} + (p_{1h}^{n+1}-p_{0h}^{n+1})\mbs n_h
        \right)
        \right]\varphi_h dA, \\
      &\qquad \forall\varphi_h\in\mbb X_0; 
    \end{split} \\
    &
    \mbs u_{1h}^{n+1} = \mbs w_h^{n+1}, \quad \text{on} \ \partial\Omega_{dh}.
    \label{equ:u12disc}
  \end{align}
\end{subequations}
\noindent\underline{Find $\mbs u_{2h}^{n+1}\in [\mbb X]^{d_{im}}$ such that:}
\begin{subequations}
  \begin{align}
    \begin{split}\label{equ:u21disc}
      &
      \int_{\Omega_h}\nabla\varphi_h\cdot\nabla\mbs u_{2h}^{n+1}d\Omega
      + \frac{\gamma_0}{\nu\Delta t}\int_{\Omega_h}\mbs u_{2h}^{n+1}\varphi_h d\Omega
      + \frac{\gamma_0 D_0}{\Delta t}\int_{\partial\Omega_{oh}}\mbs u_{2h}^{n+1}\varphi_h dA \\
      &= -\frac{1}{\nu}\int_{\Omega_h}\left[
        \nabla p_{2h}^{n+1}+\mbs N(\mbs u_h^{*,n+1})
        \right] \varphi_h d\Omega
      + \frac{1}{\nu}\int_{\partial\Omega_{oh}} \left[
        p_{2h}^{n+1}\mbs n_h + \mbs H(\mbs n_h,\mbs u_h^{*,n+1})
        \right]\varphi_h dA, \\
      &\qquad \forall\varphi_h\in\mbb X_0;
    \end{split} \\
    &
    \mbs u_{2h}^{n+1} = 0, \quad \text{on} \ \partial\Omega_{dh}.
    \label{equ:u22disc}
  \end{align}
\end{subequations}


\section{Representative Numerical Examples}
\label{sec:tests}

In this section we test the  performance
of the algorithm presented above  using
several flow problems in two dimensions involving outflow/open
boundaries. These flows are challenging to
simulate at moderate and high Reynolds numbers
because of the open boundaries and the presence of
strong vortices or backflows on such boundaries.
The effects of the simulation
parameters will be investigated, and in particular
the stability of the method at large
time step sizes will be demonstrated.

\subsection{Convergence Rates}

\begin{figure}
  \centerline{
    \includegraphics[width=3.5in]{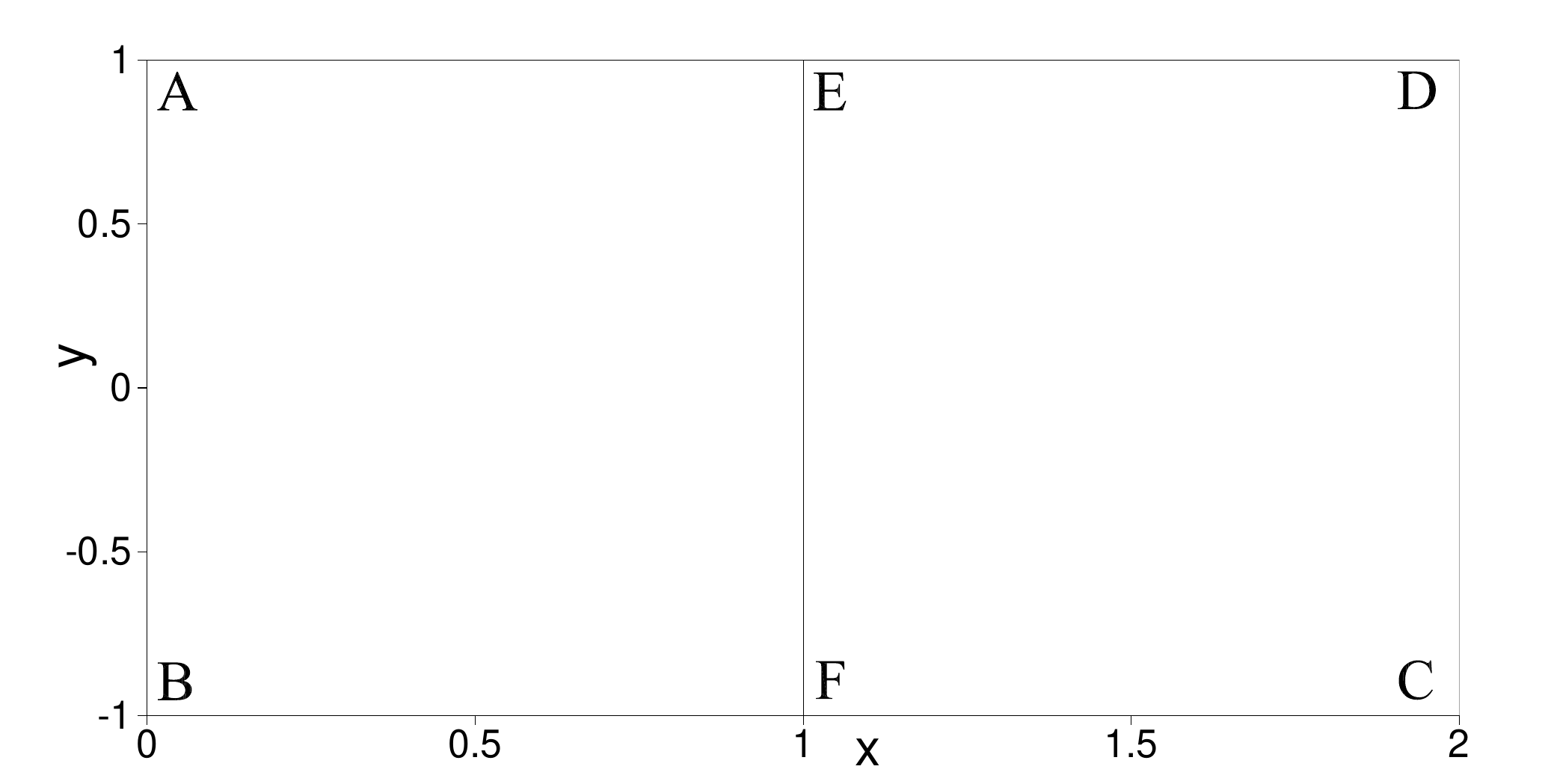}(a)
  }
  \centerline{
    \includegraphics[width=3in]{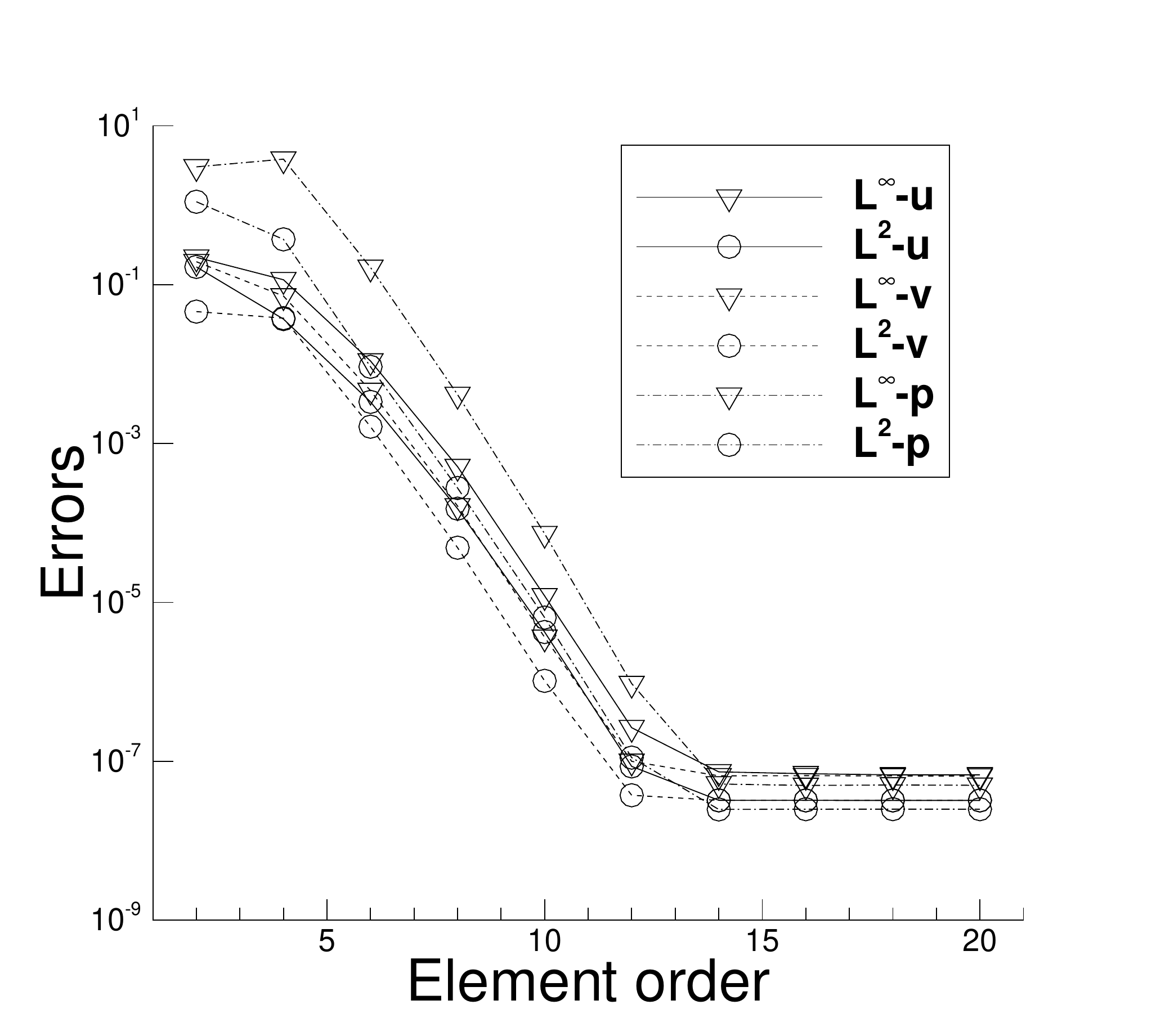}(b)
    \includegraphics[width=3in]{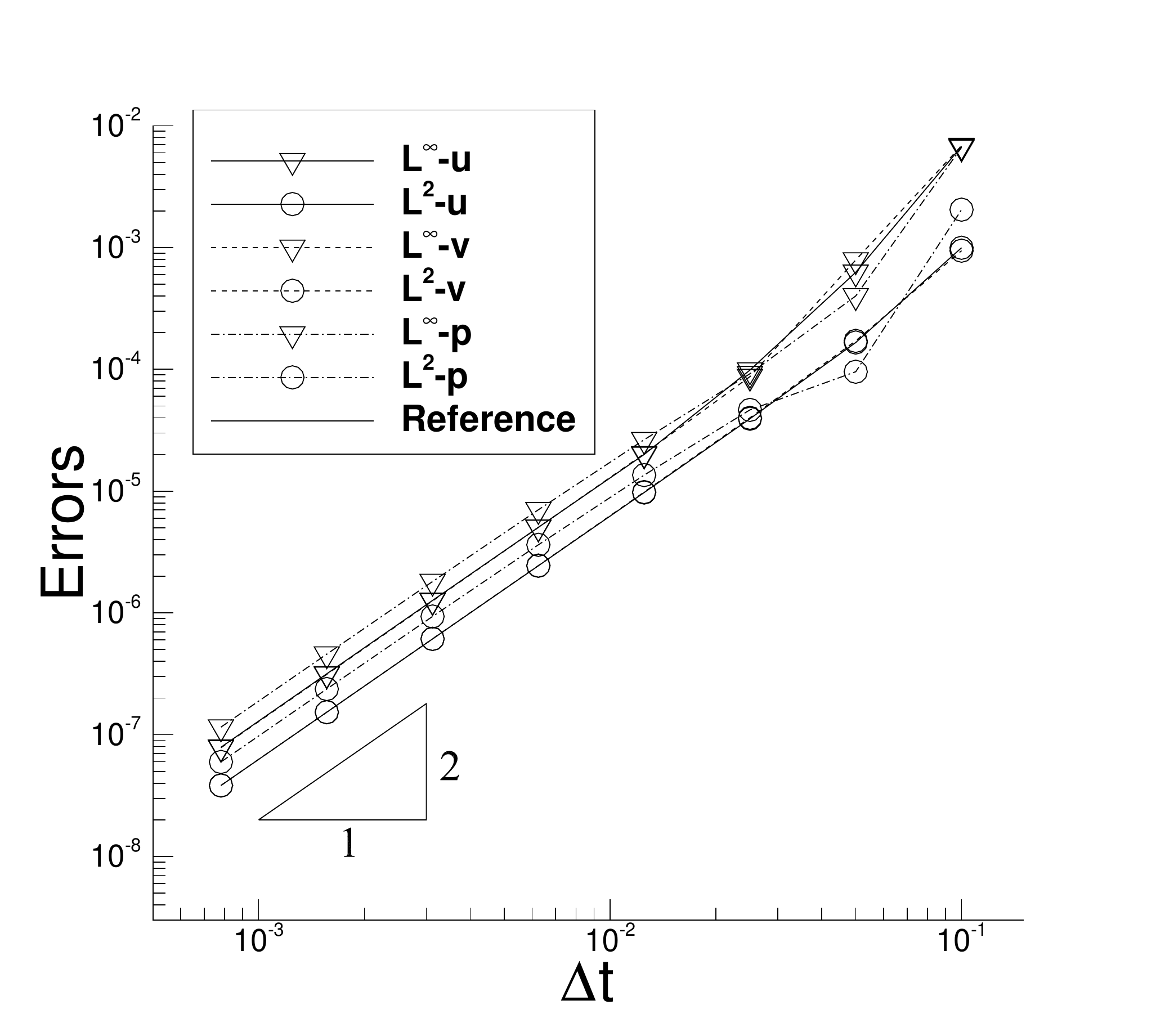}(c)
  }
  \caption{Convergence rates:
    (a) Flow configuration and the mesh.
    (b) Numerical errors versus the
    element order (fixed $t_f=0.1$ and $\Delta t=0.001$).
    (c) Numerical errors versus $\Delta t$ (fixed $t_f=0.2$ and element order $16$).
  }
  \label{fig:conv}
\end{figure}

In this subsection we demonstrate the spatial and temporal
convergence rates of the algorithm from Section \ref{sec:method}
using a manufactured analytical solution to the Navier-Stokes equations.
Consider the computational domain $\overline{ABCD}$ depicted in
Figure \ref{fig:conv}(a),
$0\leqslant x\leqslant 2$ and $-1\leqslant y\leqslant 1$,
which is discretized by two equal-sized quadrilateral
elements $\overline{ABFE}$ and $\overline{EFCD}$.
Consider the following analytic solution to
the equations \eqref{equ:nse}--\eqref{equ:div},
\begin{equation}
  \left\{
  \begin{split}
    &
    u = 2\cos(\pi y)\sin(\pi x)\sin t, \\
    &
    v = -2\sin(\pi y)\cos(\pi x)\sin t, \\
    &
    p = 2\sin(\pi y)\sin(\pi x)\cos t,
  \end{split}
  \right.
  \label{equ:anal}
\end{equation}
where $(u,v)$ are the $x$ and $y$ components of the
velocity $\mbs u$. In equation \eqref{equ:nse}
the non-dimensional viscosity is set to $\nu=0.01$, and
the external force $\mbs f$ is chosen in such a way that
this equation is satisfied by
the expressions in \eqref{equ:anal}.
Dirichlet boundary condition \eqref{equ:dbc} is imposed
on the boundaries $\overline{AB}$, $\overline{BC}$
and $\overline{AE}$, where the boundary
velocity $\mbs w(\mbs x,t)$ is chosen according
to the analytic expressions from \eqref{equ:anal}.
The open boundary condition \eqref{equ:obc},
with $\mbs H(\mbs n,\mbs u)$ given by \eqref{equ:def_H},
is imposed on the boundaries $\overline{CD}$
and $\overline{DE}$, in which the parameters
are set to $D_0=1$, $\delta =0.05$ and $U_0=1$,
and the external boundary pressure force
is set to $p_0=1$ on $\overline{CD}$ and $p_0=10$
on $\overline{DE}$.
The source term $\mbs f_b$ in \eqref{equ:obc}
is chosen such that the analytic expressions from
\eqref{equ:anal} satisfy the equation \eqref{equ:obc}
on these boundaries.
The initial velocity $\mbs u_{in}$ is chosen based on the
analytic expressions from \eqref{equ:anal}
by setting $t=0$.


The method from Section \ref{sec:method} is
employed to integrate the incompressible
Navier-Stokes equations in time from $t=0$ to
$t=t_f$ ($t_f$ to be specified below),
in which we have employed a
constant value $C_0=1$ in equation \eqref{equ:def_E}.
We then compare the numerical solution at $t=t_f$
against the analytic solution of \eqref{equ:anal},
and compute the error in various norms.
The element order and the time step size $\Delta t$
are varied systematically to test the spatial
and temporal convergence behavior of the method.

Figure \ref{fig:conv}(b) illustrates the behavior of
the method for spatial convergence tests.
Here we employ a fixed $t_f=0.1$ and $\Delta t=0.001$,
and vary the element order between $2$ and $20$
in the tests. This figure shows the errors of the numerical
solution at $t=t_f$ in $L^{\infty}$ and $L^2$ norms
as a function of the element order.
The result clearly exhibits an exponential convergence
rate for element orders below  $12$,
and an error saturation for element orders above $12$,
which is due to the dominance of the temporal truncation
error at large element orders.

Figure \ref{fig:conv}(c) illustrates the temporal convergence
behavior of the presented method.
Here we have employed a fixed element order $16$
and $t_f=0.2$, and varied the time step size
systematically between $\Delta t=0.1$ and $\Delta t=7.8125e-4$
in the tests. This figure shows the
$L^{\infty}$ and $L^2$ errors of the numerical solution
at $t=t_f$ as a function of $\Delta t$ for different flow
variables. It is evident that
the method exhibits a second-order convergence rate in time.

\subsection{Flow in a Bifurcation Channel}

\begin{figure}
  \centering
  \includegraphics[width=5in]{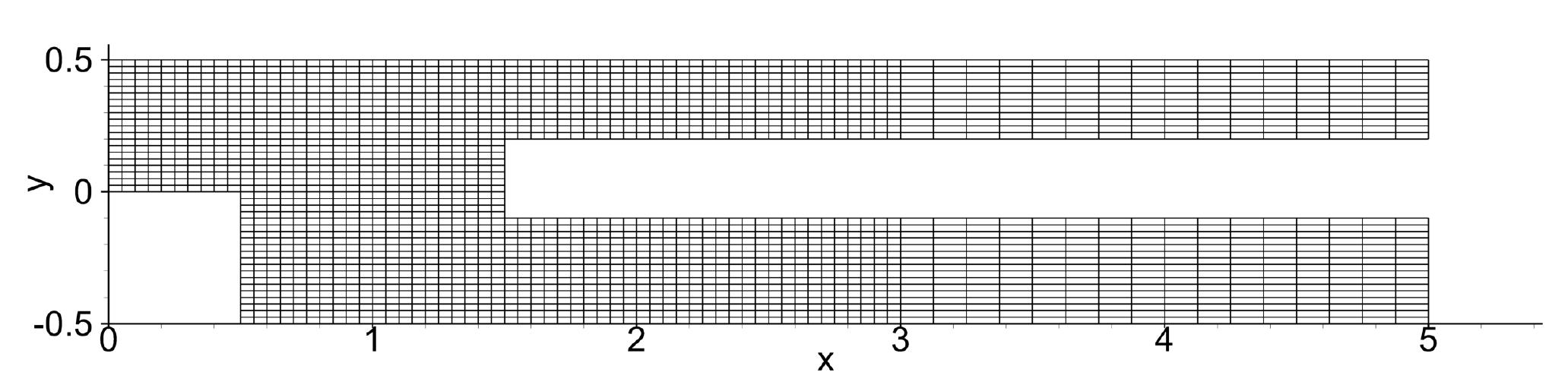}
  \caption{
    Bifurcation channel: flow domain and the mesh of 2288 quadrilateral elements.
  }
  \label{fig:chan_config}
\end{figure}

In this subsection we test our method using
the bifurcation channel problem,
which has been considered by a number of previous
works (see e.g.~\cite{PouxGA2011,DongKC2014}, among others)

Specifically,
we consider an incompressible flow contained in
a bifurcation channel depicted in Figure \ref{fig:chan_config}.
There are three openings in the channel: one on
the left side ($x=0$), and two on the right ($x=5$).
All the rest of the boundaries are walls.
The flow enters the channel through the left opening,
with a velocity profile assumed to be parabolic with
a unit center-line velocity.
External pressure heads are imposed
on the two openings on the right sides,
with $p_{01}$ on the upper right opening and $p_{02}$
on the lower right one. The values for $p_{01}$ and $p_{02}$ will be
specified later.


We discretize the domain using a mesh of $2288$ quadrilateral elements
as shown in Figure \ref{fig:chan_config}, with the element order
 varied in a range of values to be specified below.
The method from Section \ref{sec:method} is employed
to simulate the flow (with no external body force,
i.e.~$\mbs f=0$ in \eqref{equ:nse}).
On the left boundary ($x=0$)
the Dirichlet condition \eqref{equ:dbc} is imposed,
with the boundary velocity given according to the parabolic profile 
specified above. On the wall boundaries the no-clip
condition (i.e.~the Dirichlet condition \eqref{equ:dbc} with
$\mbs w=0$) is imposed. On the right boundaries ($x=5$),
the open boundary condition \eqref{equ:obc} is imposed,
with $\mbs f_b=0$ and $\mbs H(\mbs n,\mbs u)$ given by \eqref{equ:def_H} and
with $p_0=p_{01}$ for the upper right boundary and
$p_0=p_{02}$ for the lower right boundary.

Note that we have used the channel centerline
velocity on the left boundary as the velocity scale ($U_0=1$),
and the channel height in the mid-section ($0.5<x<1.5$)
as the length scale. All the other parameters and
variables are normalized accordingly.
We focus on the Reynolds number $Re=800$ for this problem, chosen
in accordance with~\cite{DongKC2014}, and the flow
is at a steady state.
The other simulation parameters are set to
$D_0=\frac{1}{U_0}=1$ and
  $\delta=0.05$.
The element order,
the time step size $\Delta t$, $C_0$,
$p_{01}$ and $p_{02}$ are varied 
to study their effects on the flow characteristics.


\begin{figure}
  \centering
  \includegraphics[width=5in]{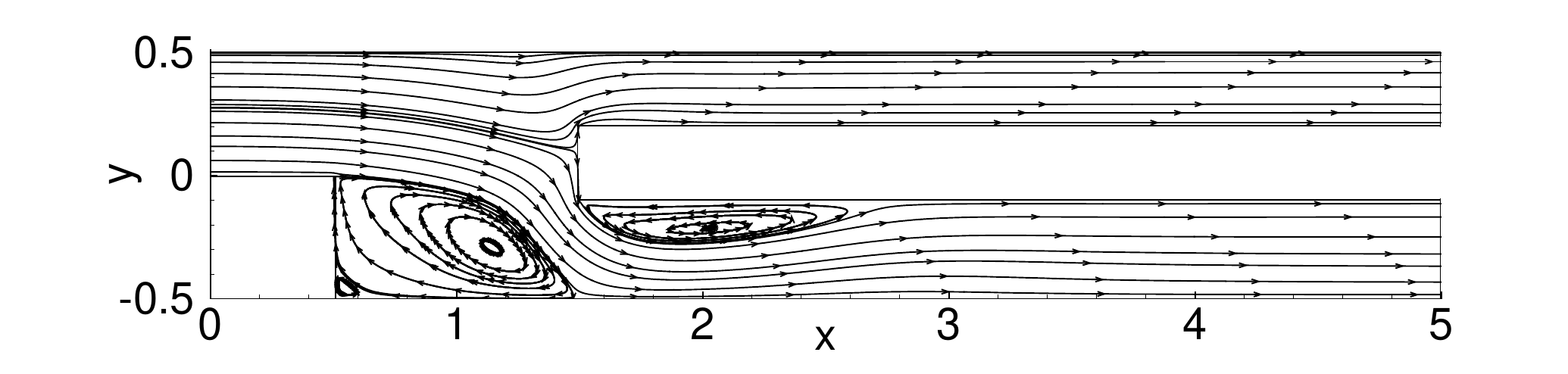}
  \caption{
    Bifurcation channel: flow patterns visualized by streamlines.
  }
  \label{fig:bif_patt}
\end{figure}

We first concentrate on the cases with zero external pressure heads at
the two open boundaries on the right, namely,
$p_{01}=p_{02}=0$.
Figure \ref{fig:bif_patt} shows a visualization of the steady-state
velocity patterns using streamlines, which are obtained
with an element order $8$, time step size $\Delta t=0.001$ and $C_0=1$ in
the simulations.
In this case, the flow enters the domain
through the left boundary and discharges from the domain through the two
open boundaries on the right.
Several re-circulation zones (bubbles) are visible 
from the flow pattern. The most prominent are those behind
the inlet step and the one on the top wall of the lower bifurcation
of the channel.


\begin{table}
  \centering
  \begin{tabular}{lcccc}
    \hline
    Element order & $f_x$ & $f_y$ & $L_1$ & $L_2$ \\
    3 & 0.203 & 2.02e-3 & 0.970 & 1.24 \\
    4 & 0.196 & -8.50e-5 & 0.975 & 1.23 \\
    5 & 0.193 & -2.30e-4 & 0.980 & 1.23 \\
    6 & 0.192 & -6.0e-6 & 0.980 & 1.22 \\
    7 & 0.191 & 1.25e-4 & 0.980 & 1.22 \\
    8 & 0.191 & 1.71e-4 & 0.980 & 1.22 \\
    9 & 0.191 & 1.86e-4 & 0.980 & 1.22 \\
    10 & 0.191 & 1.92e-4 & 0.980 & 1.22 \\
    \hline
  \end{tabular}
  \caption{
    Resolution tests of the bifurcation channel problem.
    $f_x$ and $f_y$ are the $x$ and $y$ components of the total force
    on the channel walls. $L_1$ is the length on the bottom
    wall of the recirculation zone behind the inlet step.
    $L_2$ is the length of the recirculation zone on the
    top wall of the lower bifurcation.
  }
  \label{tab:bif_res}
\end{table}


We have varied the element order systematically in the simulations
to make sure that the numerical results have converged with respect to
the mesh resolution. 
Table \ref{tab:bif_res} provides the $x$ and $y$ components of
the total force ($f_x$ and $f_y$) exerting on the channel walls,
as well as the sizes of the recirculation zones behind the inlet step ($L_1$)
and on the top wall of the lower bifurcation ($L_2$),
corresponding to different element orders. These results are
obtained using a time step size $\Delta t=0.001$, and $C_0=1$
in equation \eqref{equ:def_E}.
The mesh independence of the simulation results is evident for
element orders beyond $6$.


\begin{figure}
  \centerline{
    \includegraphics[width=1.5in]{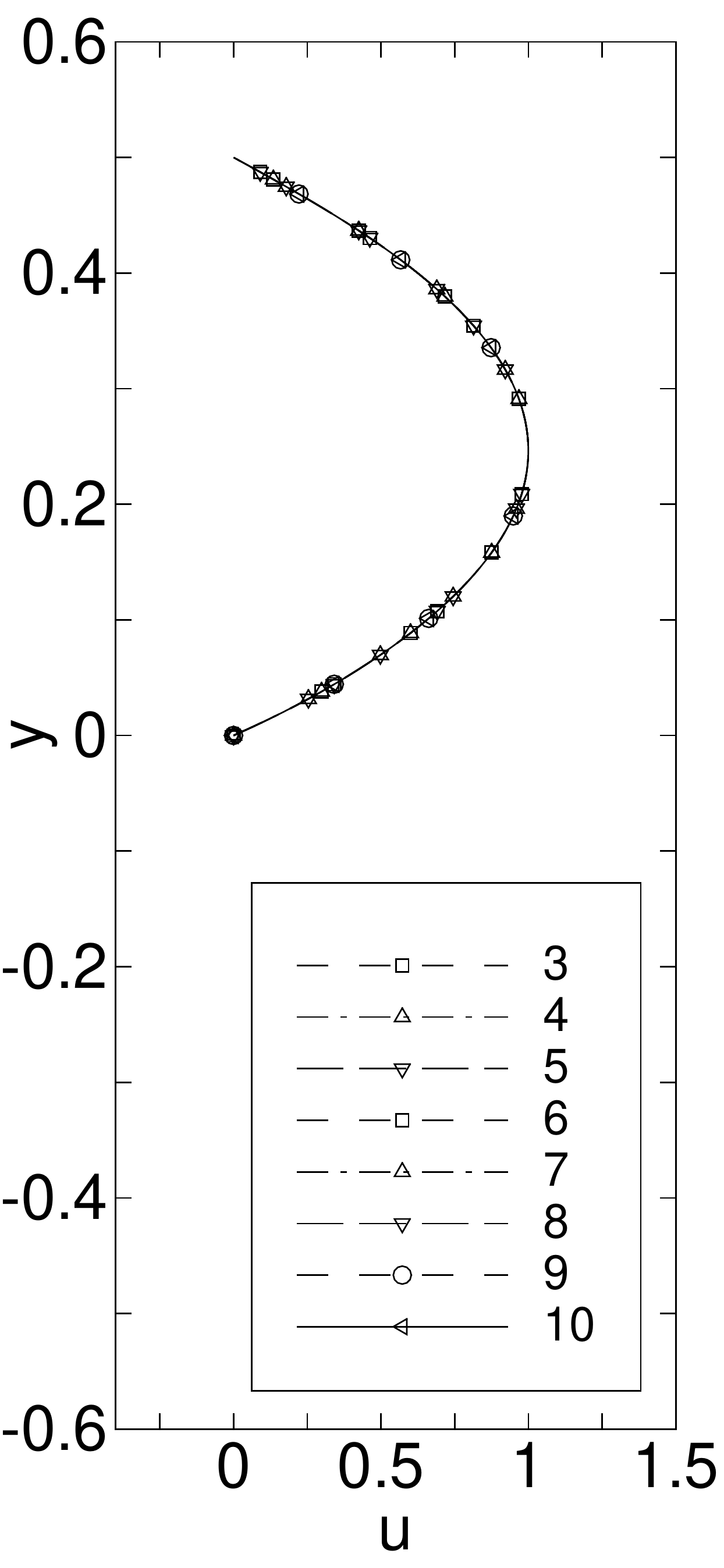}(a)
    \includegraphics[width=1.5in]{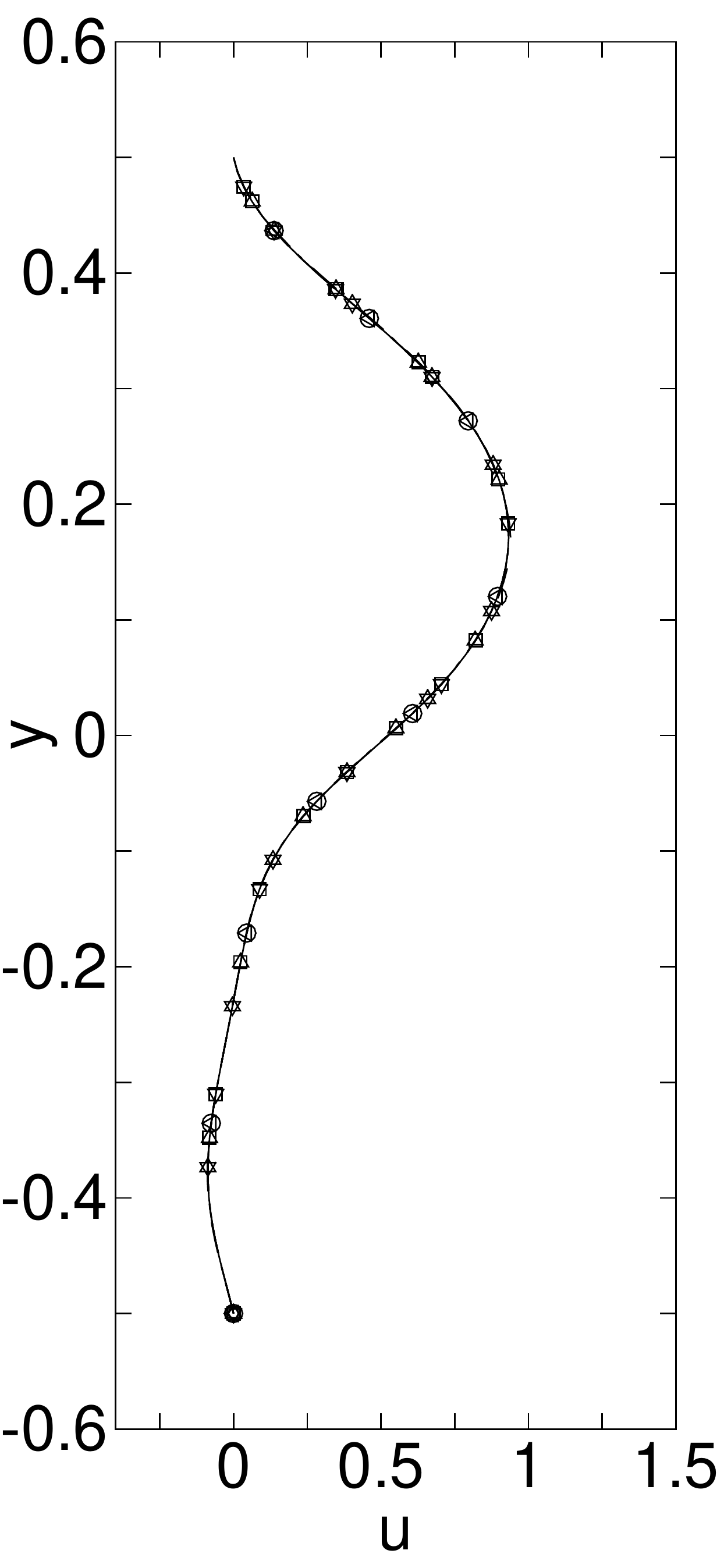}(b)
    \includegraphics[width=1.5in]{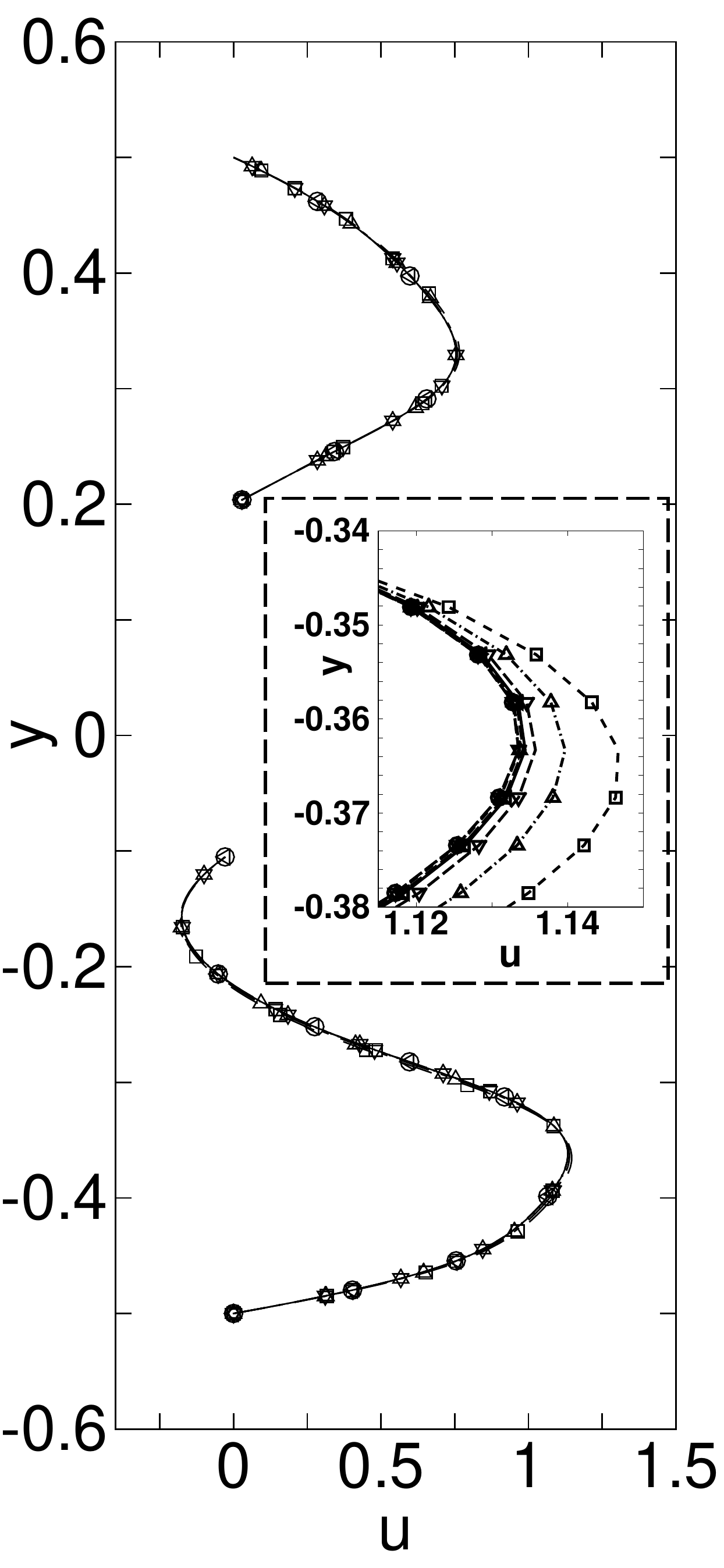}(c)
    \includegraphics[width=1.5in]{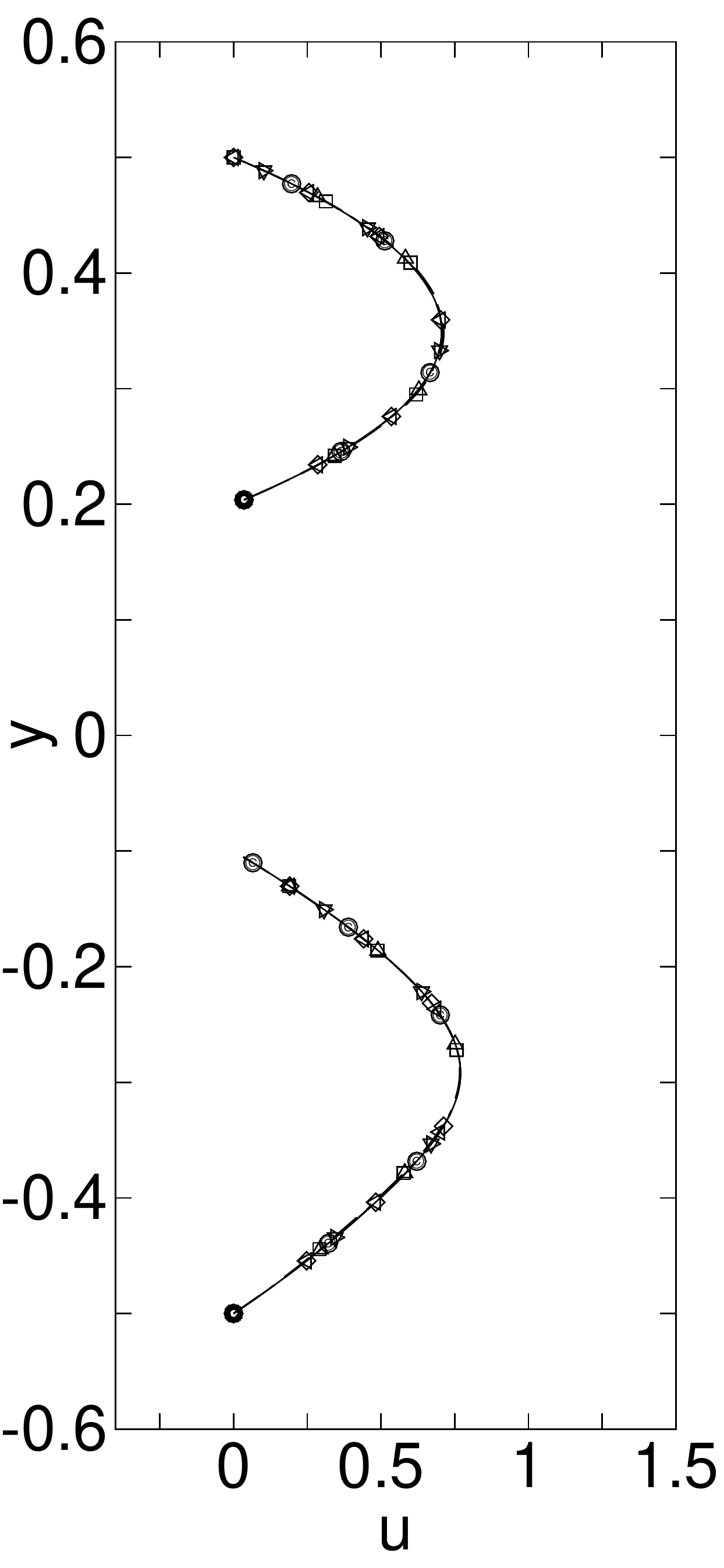}(d)
  }
  \caption{
    Bifurcation channel:
    Comparison of streamwise velocity profiles at several downstream
    locations, $x=0.25$ (a), $1.0$ (b), $2.0$ (c) and $4.0$ (d), computed using
    various element orders.
  }
  \label{fig:bif_profu}
\end{figure}

\begin{figure}
  \centerline{
    \includegraphics[width=1.5in]{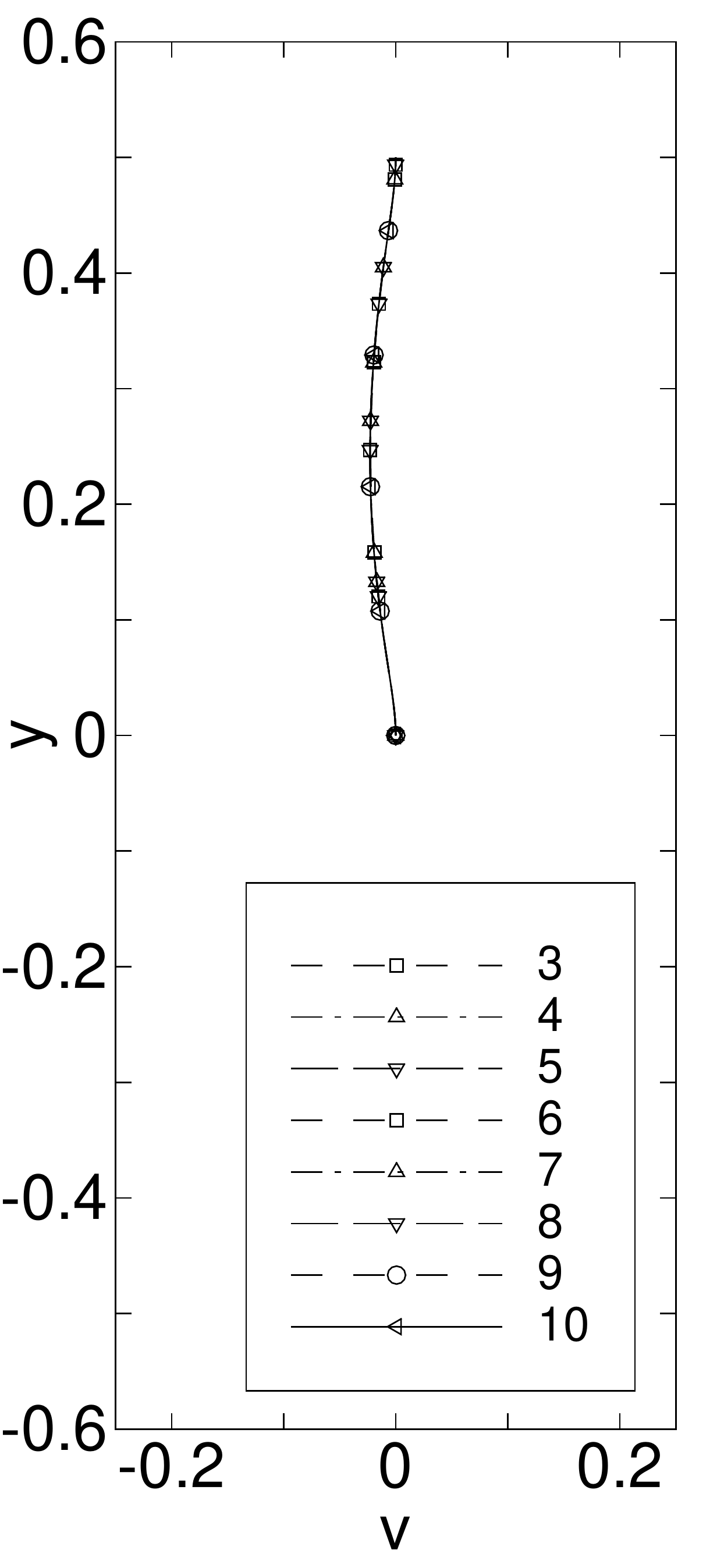}(a)
    \includegraphics[width=1.5in]{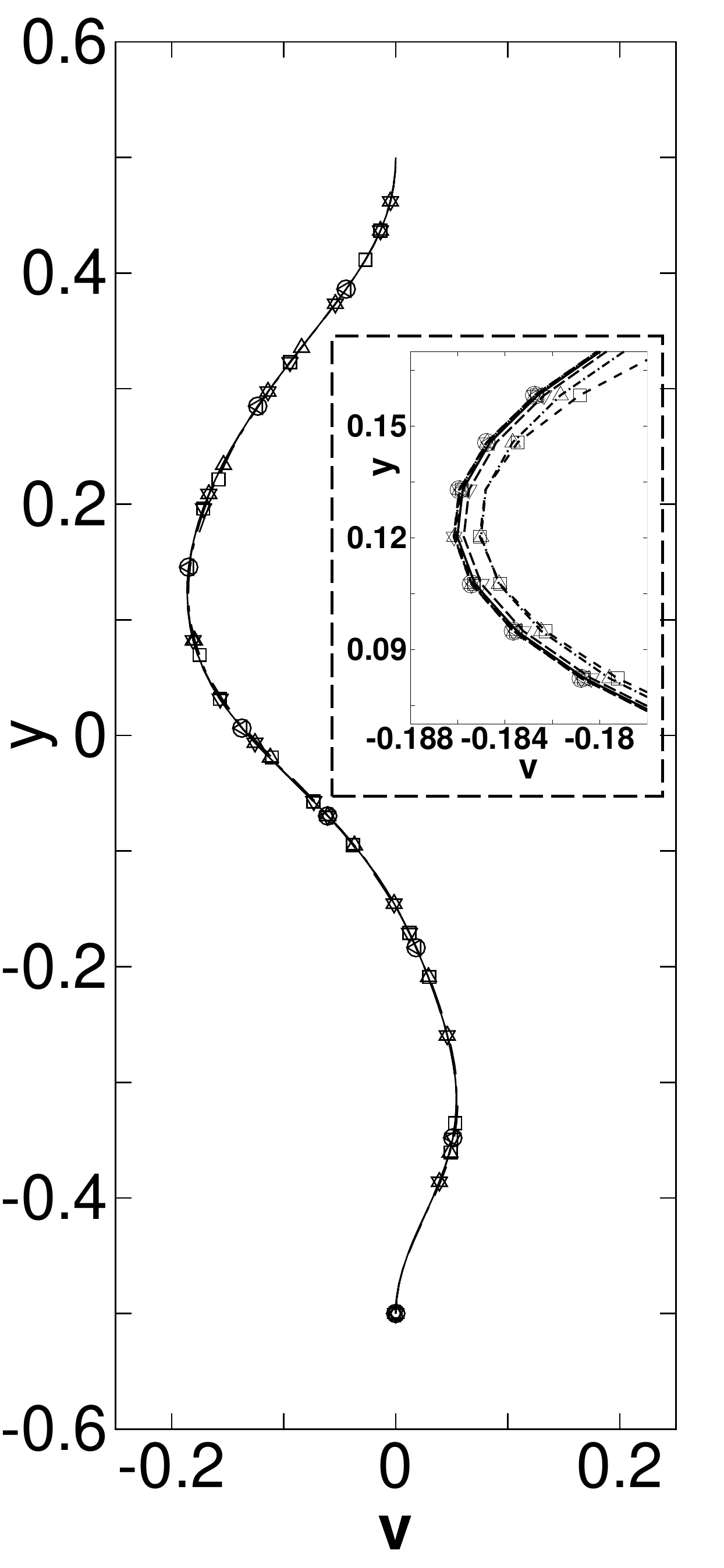}(b)
    \includegraphics[width=1.5in]{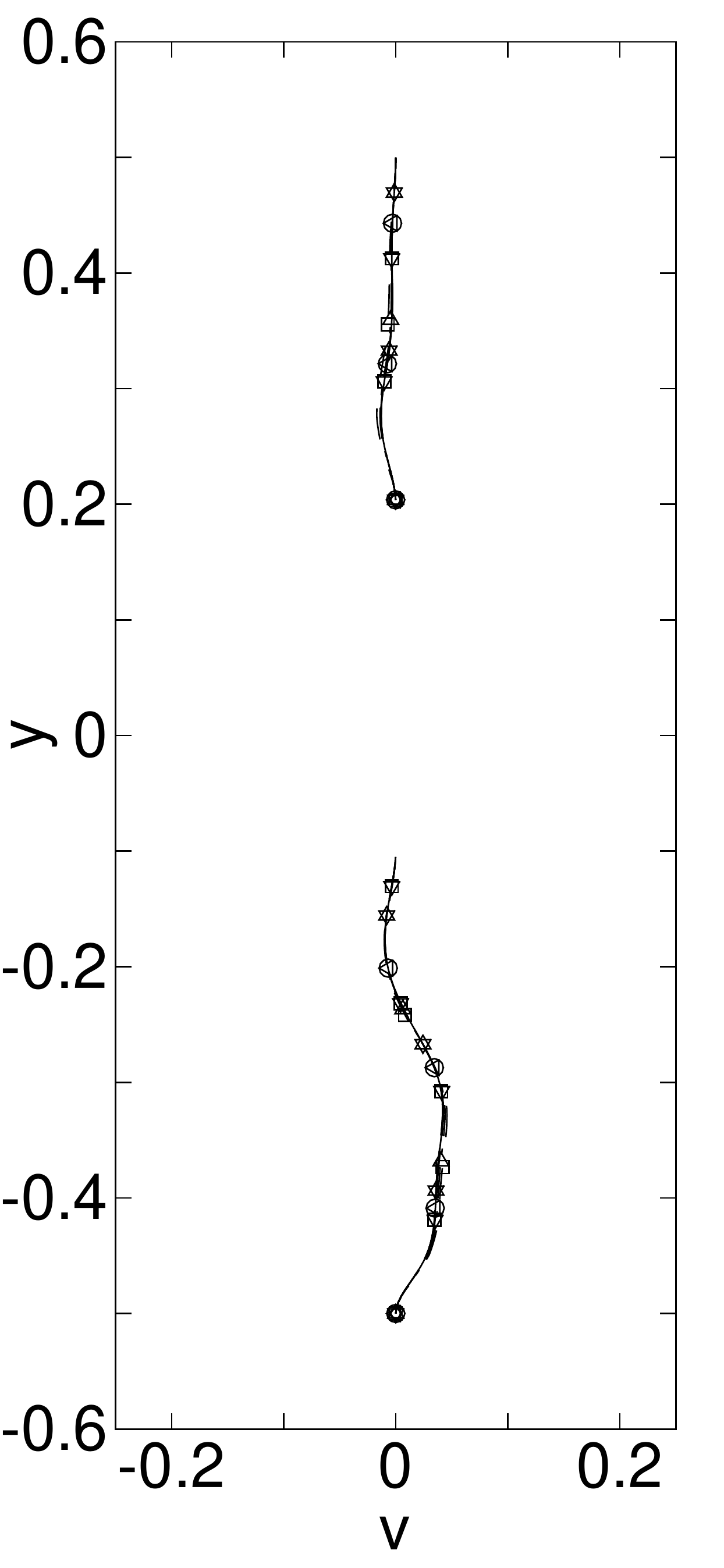}(c)
    \includegraphics[width=1.5in]{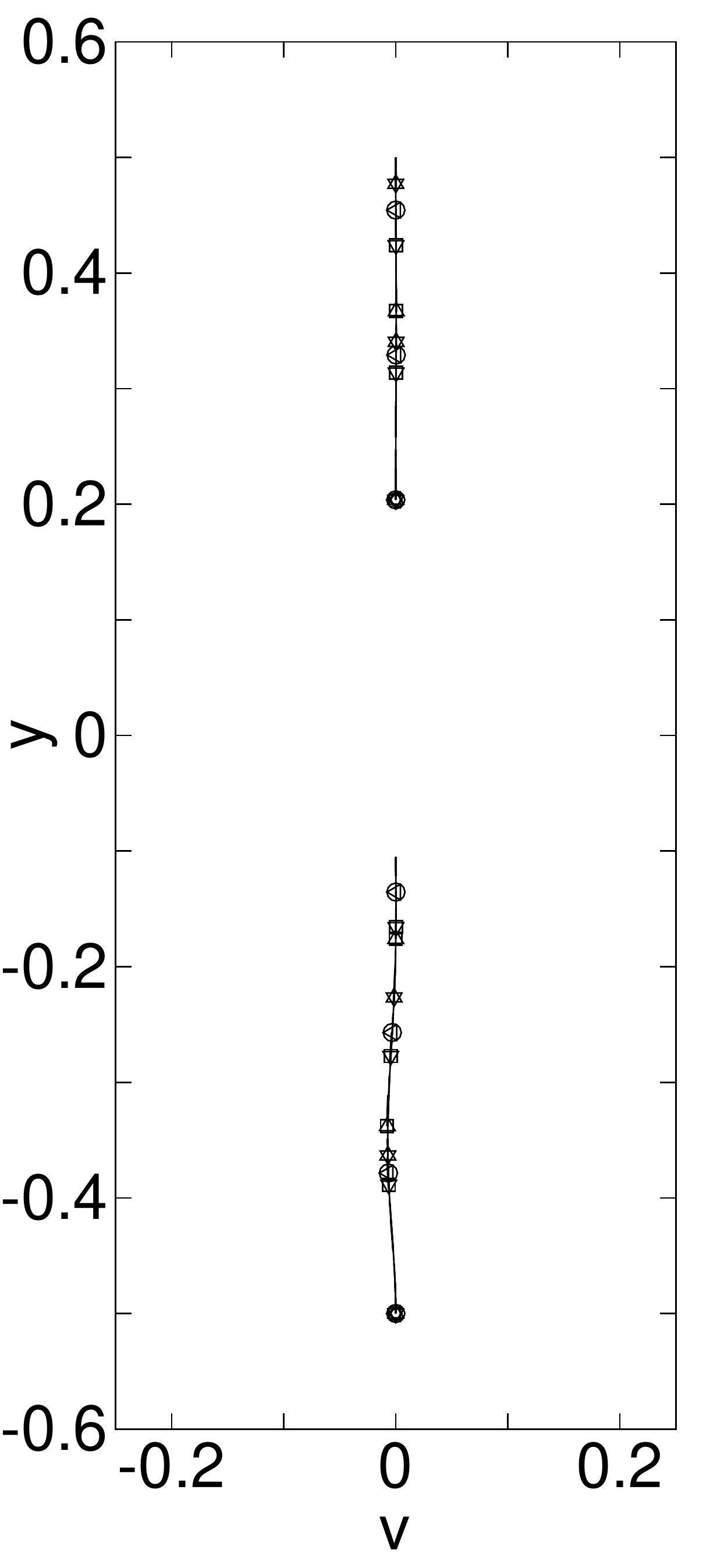}(d)
  }
  \caption{
    Bifurcation channel:
    Comparison of vertical velocity profiles at several downstream
    locations, $x=0.25$ (a), $1.0$ (b), $2.0$ (c) and $4.0$ (d), computed using
    various element orders.
  }
  \label{fig:bif_profv}
\end{figure}

Figure \ref{fig:bif_profu} compares profiles of
the streamwise velocity ($u$) along the $y$ direction
at several downstream locations ($x=0.25$, $1.0$, $2.0$ and $4.0$),
obtained with the range of element orders.
Figure \ref{fig:bif_profv} is a corresponding comparison of the
vertical velocity profiles at the same locations obtained with
different element orders.
Some magnified views of sections of the velocity profiles are provided in
the insets of Figures \ref{fig:bif_profu}(c) and \ref{fig:bif_profv}(b).
These results are computed with $\Delta t=0.001$
and $C_0=1$ in the simulations. 
It can be observed that the velocity profiles corresponding to
element orders $6$ and beyond essentially overlap with one another,
further attesting to the convergence of simulation results.


\begin{table}
  \centering
  \begin{tabular}{lcccc}
    \hline
    $\Delta t$ & $\bar{f}_x$ & $f'_x$ & $\bar{f}_y$ & $f'_y$ \\
    1e-4 & 0.192 & 0 & 1.16e-4 & 0 \\
    5e-4 & 0.191 & 0 & 1.54e-4 & 0 \\
    0.001 & 0.191 & 0 & 1.71e-4 & 0 \\
    0.005 & 0.144 & 1.23e-2 & -6.57e-5 & 3.83e-3 \\
    0.01 & 0.139 & 9.88e-3 & -3.13e-5 & 2.21e-3 \\
    0.05 & 0.126 & 3.99e-3 & -6.46e-5 & 4.72e-4 \\
    0.1 & 0.123 & 1.94e-4 & -6.01e-5 & 7.06e-4 \\
    0.5 & 0.119 & 1.97e-5 & -1.03e-4 & 1.19e-4 \\
    \hline
  \end{tabular}
  \caption{
    Bifurcation channel: forces on the channel walls obtained with
    a range of time step sizes $\Delta t$.
  }
  \label{tab:bif_dt}
\end{table}

\begin{figure}
  \centering
  \includegraphics[width=5in]{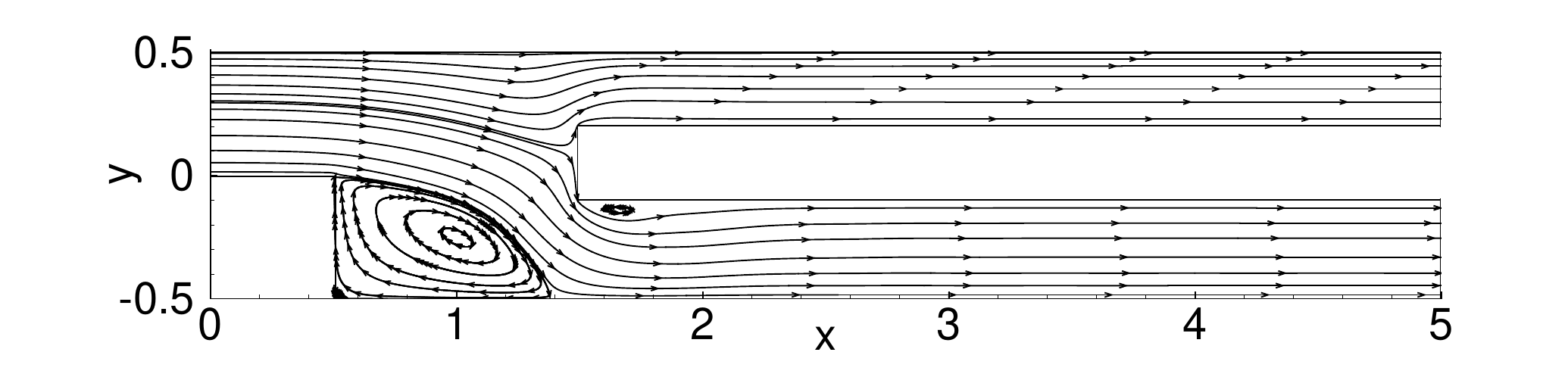}
  \caption{
    Bifurcation channel: flow pattern visualized by streamlines obtained with a
    larger time step size $\Delta t=0.01$.
  }
  \label{fig:bif_ldt}
\end{figure}

Thanks to the unconditional energy stability property (Theorem~\ref{thm:thm_1}),
stable computation results can be obtained using the current
method even with large (or fairly large) time step sizes. 
This point is demonstrated by Table~\ref{tab:bif_dt}, in which
we list the $x$ and $y$ forces on the wall ($f_x$ and $f_y$)
obtained from the simulations
using time step sizes ranging from $\Delta t=10^{-4}$ to $\Delta t=0.5$.
These results correspond to an element order $8$ and
$C_0=1$ in the simulations.
We observe that when $\Delta t$ increases beyond a certain value ($5e-3$),
the computed forces are no longer constant, but  exhibit
a fluctuation in their histories, although these fluctuations are
minuscule. For such cases, $f_x$ and $f_y$ shown in Table \ref{tab:bif_dt}
are the time-averaged forces,  and $f_x'$ and $f'_y$ are the 
root-mean-square (rms) values.
The stability of our method with large $\Delta t$ is evident.
On the other hand,  a deterioration
in accuracy of the simulation result is visible when $\Delta t$
becomes large (or fairly large).
Figure \ref{fig:bif_ldt} shows a visualization of the flow
pattern obtained with a larger time step size $\Delta t=0.01$. 
This figure can be compared with Figure \ref{fig:bif_patt}, which
corresponds to $\Delta t=0.001$.
While the overall flow pattern in Figure \ref{fig:bif_ldt}
seems reasonable, the recirculation zone on the top wall of
the lower bifurcation  is
markedly different in size compared with that of Figure \ref{fig:bif_patt}.

The above simulation results are obtained with
the parameter values $D_0=1$ and $C_0=1$ in the computations.
The parameters $D_0$ and $C_0$ have also been varied
systematically ($D_0$ ranging from $0.1$ to $2.0$;
$C_0$ ranging from $10^{-3}$ to $10^6$),
and we observe no apparent effects of the variation of these parameters
on the simulation results (e.g.~in terms of the forces on the walls).


\begin{figure}
  \centering
  \includegraphics[width=5in]{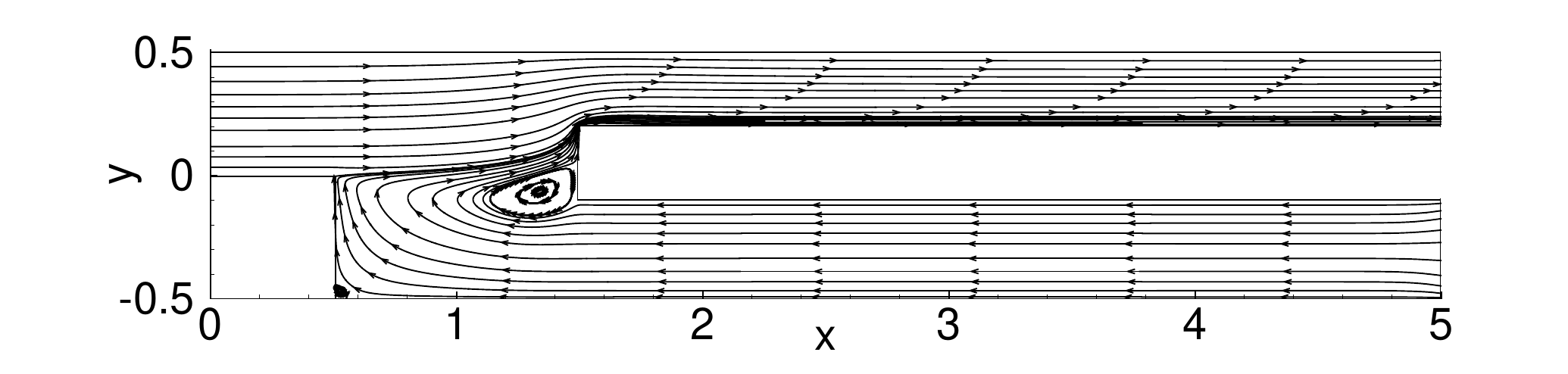}(a)
  \includegraphics[width=5in]{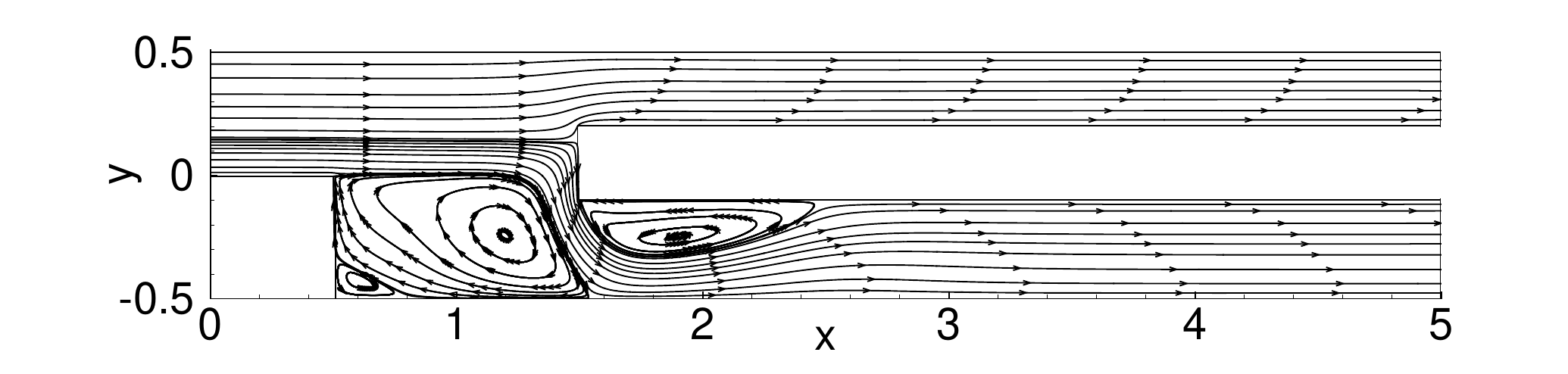}(b)
  \includegraphics[width=5in]{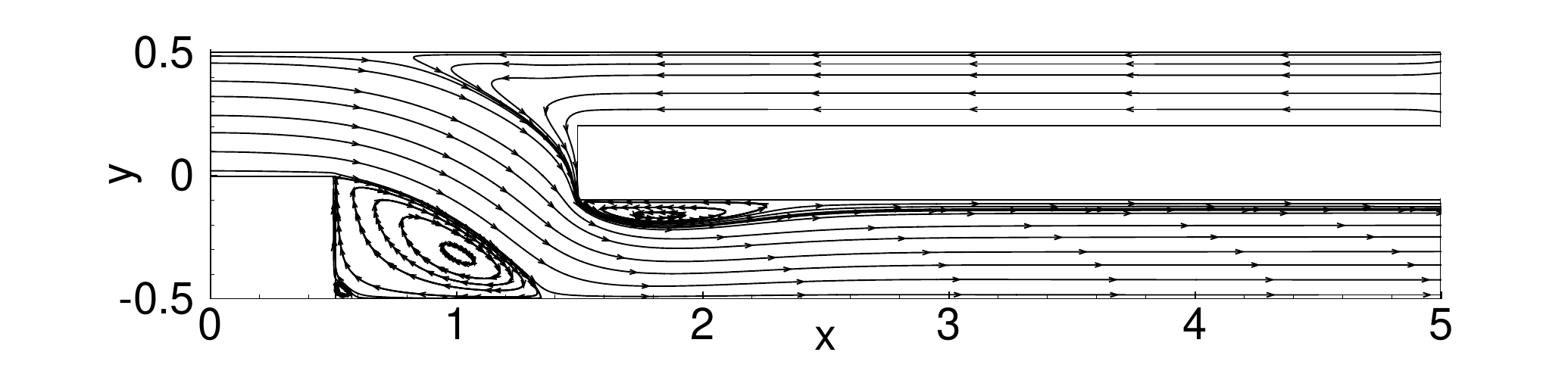}(c)
  \includegraphics[width=5in]{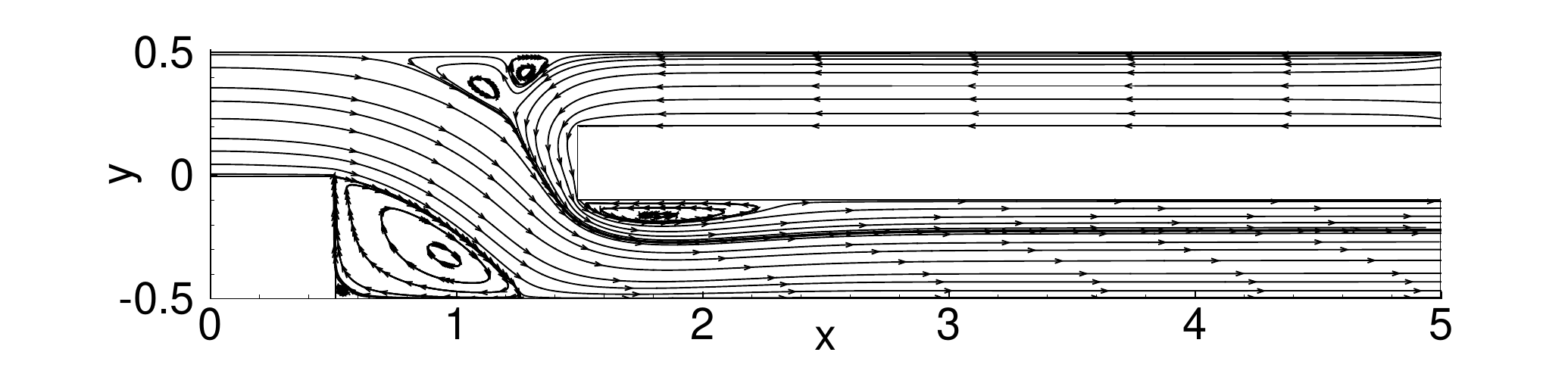}(d)
  \caption{
    Bifurcation channel: 
    flow patterns visualized by streamlines corresponding to
    several external pressure heads imposed on the upper right boundary:
    (a) $p_{01}=-1.5$, (b) $p_{01}=-0.75$, 
    (c) $p_{01}=0.75$, and (d) $p_{01}=1.5$.
    zero pressure head is imposed on the lower right boundary ($p_{02}=0$).
  }
  \label{fig:bif_peff}
\end{figure}

The incorporation of $p_0(\mbs x,t)$ in the open boundary condition~\eqref{equ:obc}
allows one to impose different pressure heads ($p_{01}$ and $p_{02}$)
on the open boundaries of
the bifurcation channel. Depending on the relative values of
$p_{01}$ and $p_{02}$,
the flow pattern in the domain can be modified dramatically. 
Figure \ref{fig:bif_peff} shows a comparison of the flow
patterns visualized by streamlines corresponding to several
$p_{01}$ values (ranging from $-1.5$ to $1.5$) at the upper right opening,
while zero
pressure head is imposed on the lower right one
($p_{02}=0$). These patterns can be compared with
that of Figure \ref{fig:bif_patt}, which corresponds to
$p_{01}=p_{02}=0$. These results are obtained with
an element order $8$, $\Delta t=0.001$, $D_0=1$ and $C_0=1$
in the simulations.
When $p_{01}$ is sufficiently low compared with $p_{02}$,
e.g. with $p_{01}=-1.5$ and $p_{02}=0$ (Figure \ref{fig:bif_peff}(a)),
the flow direction in the lower bifurcation
can be reversed. In this case the lower right boundary
 effectively becomes an inlet, through which the flow
is sucked into the domain.
On the other hand, when $p_{01}$ is sufficiently high compared
with $p_{02}$, e.g. with  
$p_{01}=1.5$ and $p_{02}=0$ (Figure \ref{fig:bif_peff}(d)),
a flow reversal occurs in the upper bifurcation.
In this case, the upper right boundary becomes
an effective inlet, and the flow is pushed into
the domain through that boundary due to the high pressure
head there.

\subsection{Flow past a Circular Cylinder}
\label{sec:cyl}

\begin{figure}
  \centering
  \includegraphics[height=3in]{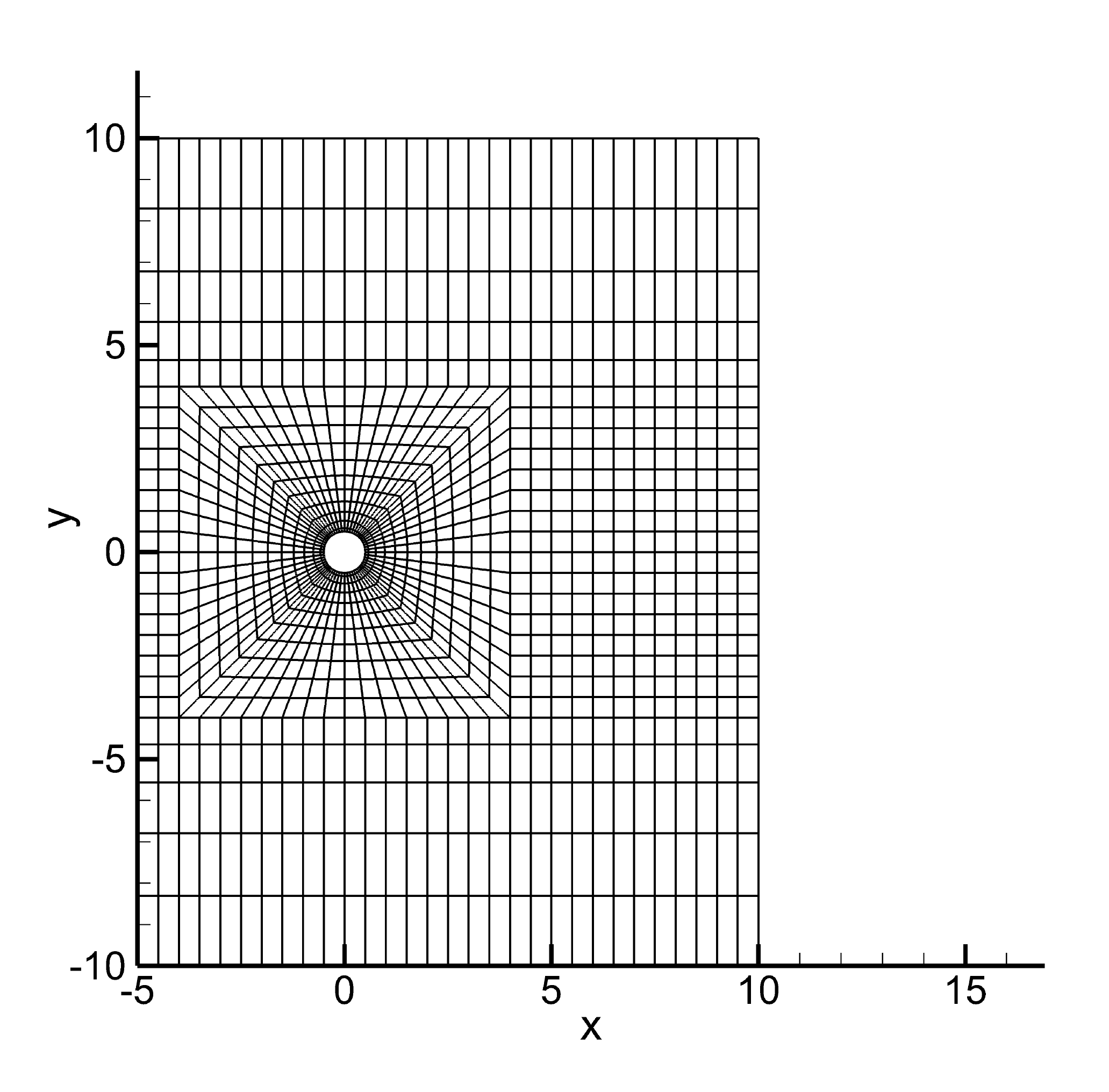}
  \caption{
    Cylinder flow: flow configuration and
    a mesh of 1228 quadrilateral elements.
  }
  \label{fig:cylmesh}
\end{figure}

In this subsection we use a canonical problem,
the flow past a circular cylinder,
to test the performance of the method
developed herein.
Consider the flow domain depicted in Figure \ref{fig:cylmesh}.
The center of the cylinder coincides with
the origin of the coordinate system.
Let $d$ denote the cylinder diameter.
On the left boundary ($x=-5d$), a uniform inflow with
a velocity $U_0=1$ enters the domain in the horizontal
direction. The flow exits the domain through
the right boundary ($x=10d$), which is open.
The top and bottom domain boundaries ($y=\pm 10d$)
are assumed to be periodic.
We choose $U_0$ and $d$ as the characteristic velocity and
length scales, respectively, and all the parameters and
variables are normalized accordingly. 
We would like to study the long-time behavior of this
flow using the method developed here.
As the Reynolds number becomes moderately large (around $Re=2000$
and beyond), vortices shed behind the cylinder can
persist in the entire wake region and pass through the right open boundary,
which can cause a severe issue to
numerical simulations  due to the 
backflow instability~\cite{Dong2015clesobc,DongS2015,NiYD2019,DongKC2014}.

We discretize the domain using a mesh of $1228$ quadrilateral
spectral elements, as shown in Figure \ref{fig:cylmesh},
and the element order is varied in the simulations.
The method from Section \ref{sec:method} is
employed to solve the incompressible Navier-Stokes
equations, with $\mbs f=0$ being set in \eqref{equ:nse}.
Dirichlet boundary condition \eqref{equ:dbc}
is imposed on the left domain boundary, with the
boundary velocity chosen based on the uniform inflow
condition. No slip condition is imposed on
the cylinder surface. Periodic conditions are imposed on the
top/bottom boundaries for all  flow variables.
The open boundary condition~\eqref{equ:obc},
with $\mbs f_b=0$,
$D_0=\frac{1}{U_0}=1$, $p_0=0$,
and $\delta =0.01$ in equation \eqref{equ:def_H},
is imposed on the right boundary.
We employ a fixed $C_0=1$ in equation \eqref{equ:def_E}
in the simulations of this problem.
The element order and the time step size
$\Delta t$ have been varied systematically 
to investigate their effects on the simulation results.
The flow at several Reynolds numbers ranging from
$Re=30$ to $Re=10000$ has been simulated.


\begin{figure}
  \centerline{
    \includegraphics[width=3.1in]{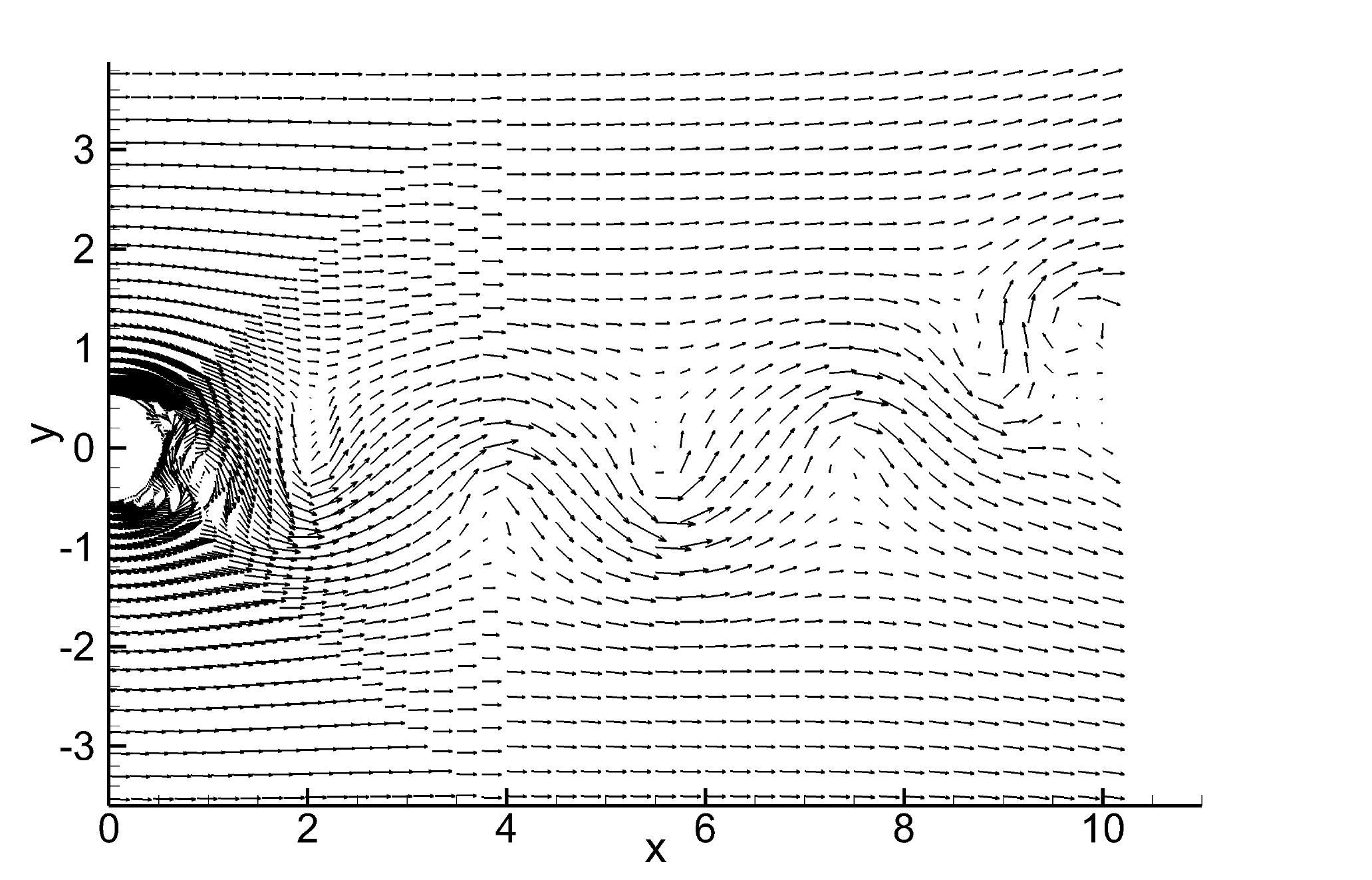}(a)
    \includegraphics[width=3.1in]{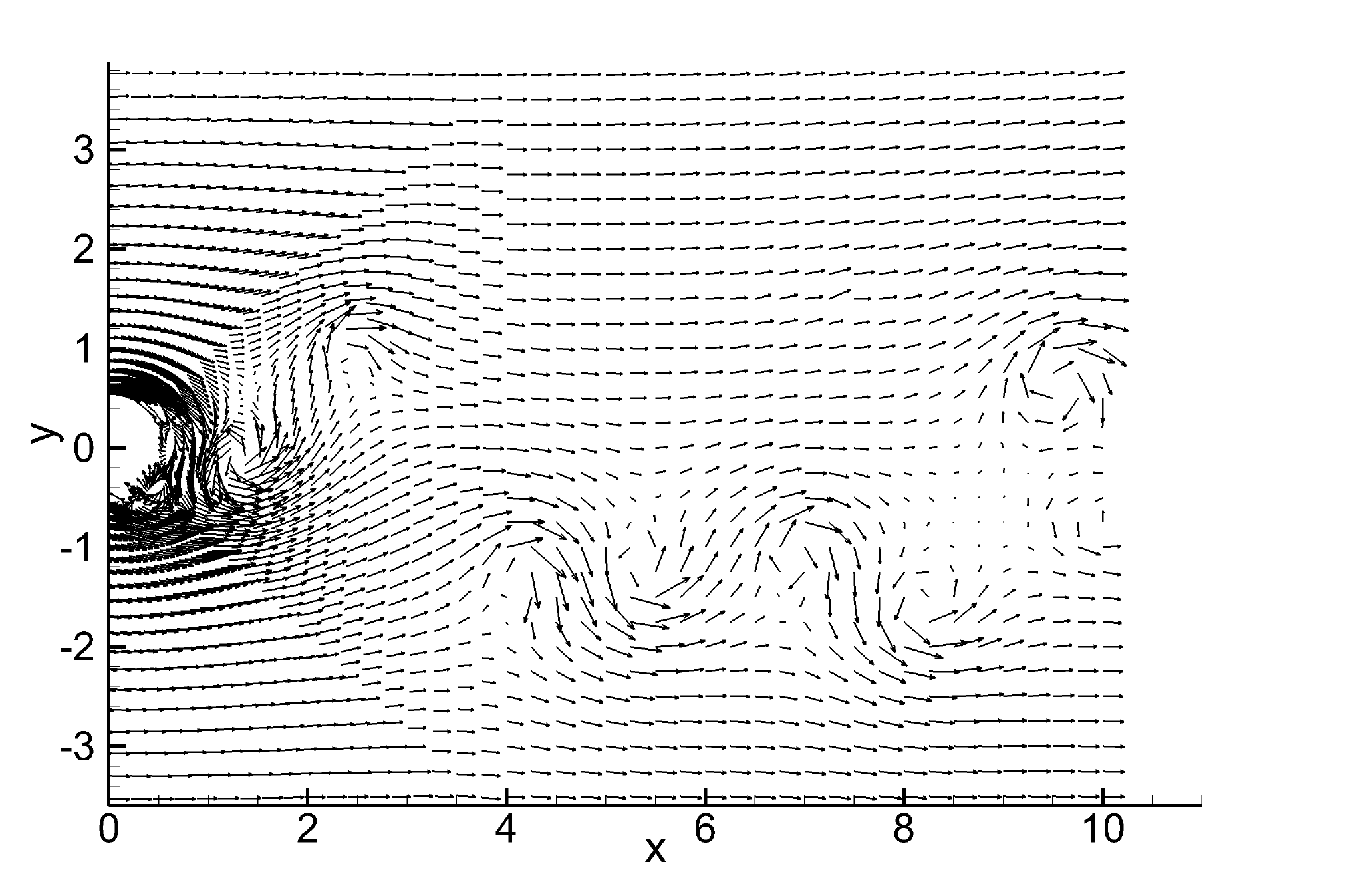}(b)
  }
  \caption{
    Cylinder flow: instantaneous velocity distributions at
    Reynolds numbers (a) $Re=2000$, and (b) $Re=10000$.
  }
  \label{fig:cyl_str}
\end{figure}


\begin{figure}
  \centerline{
    \includegraphics[width=3.1in]{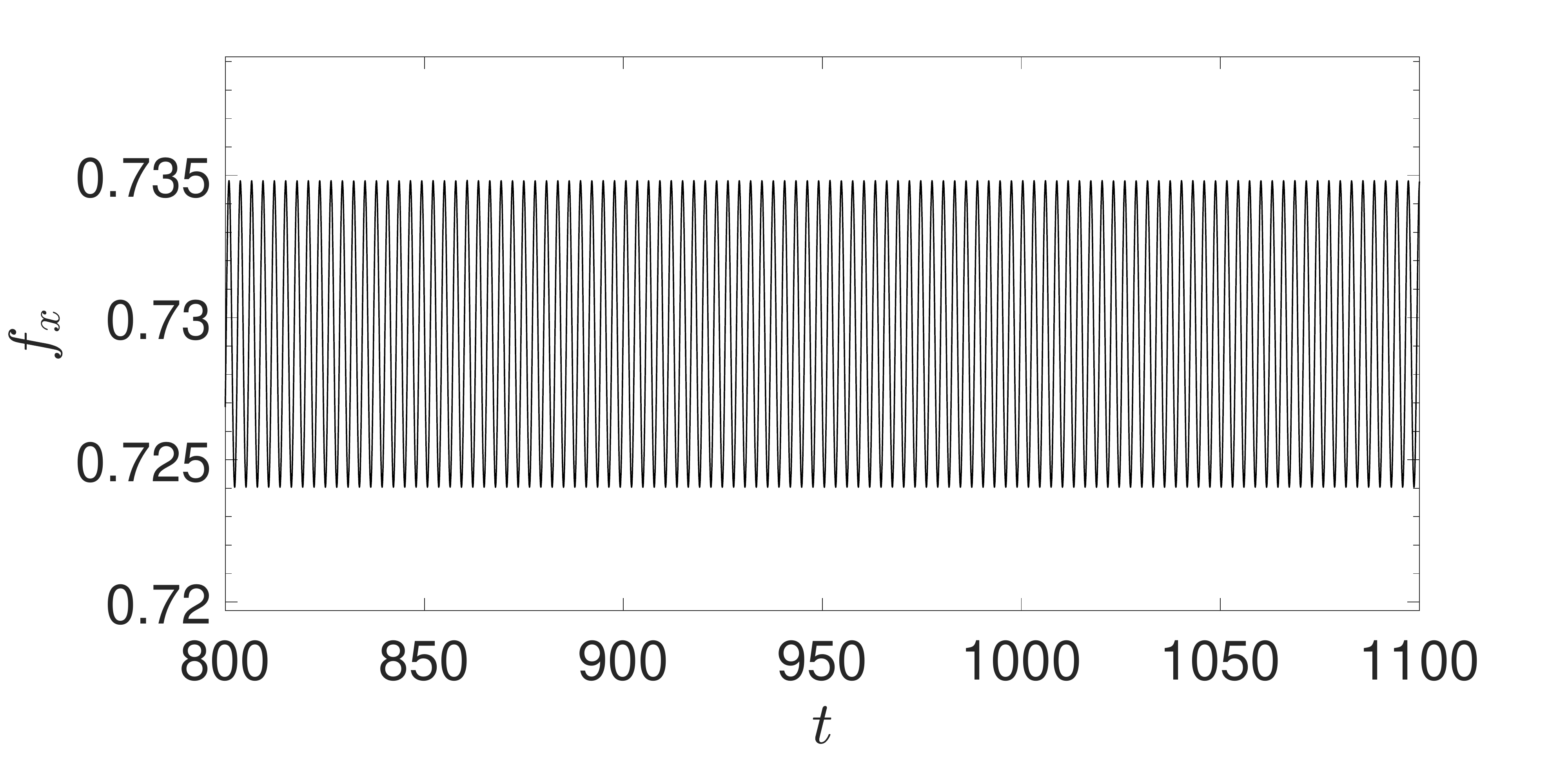}(a)
    \includegraphics[width=3.1in]{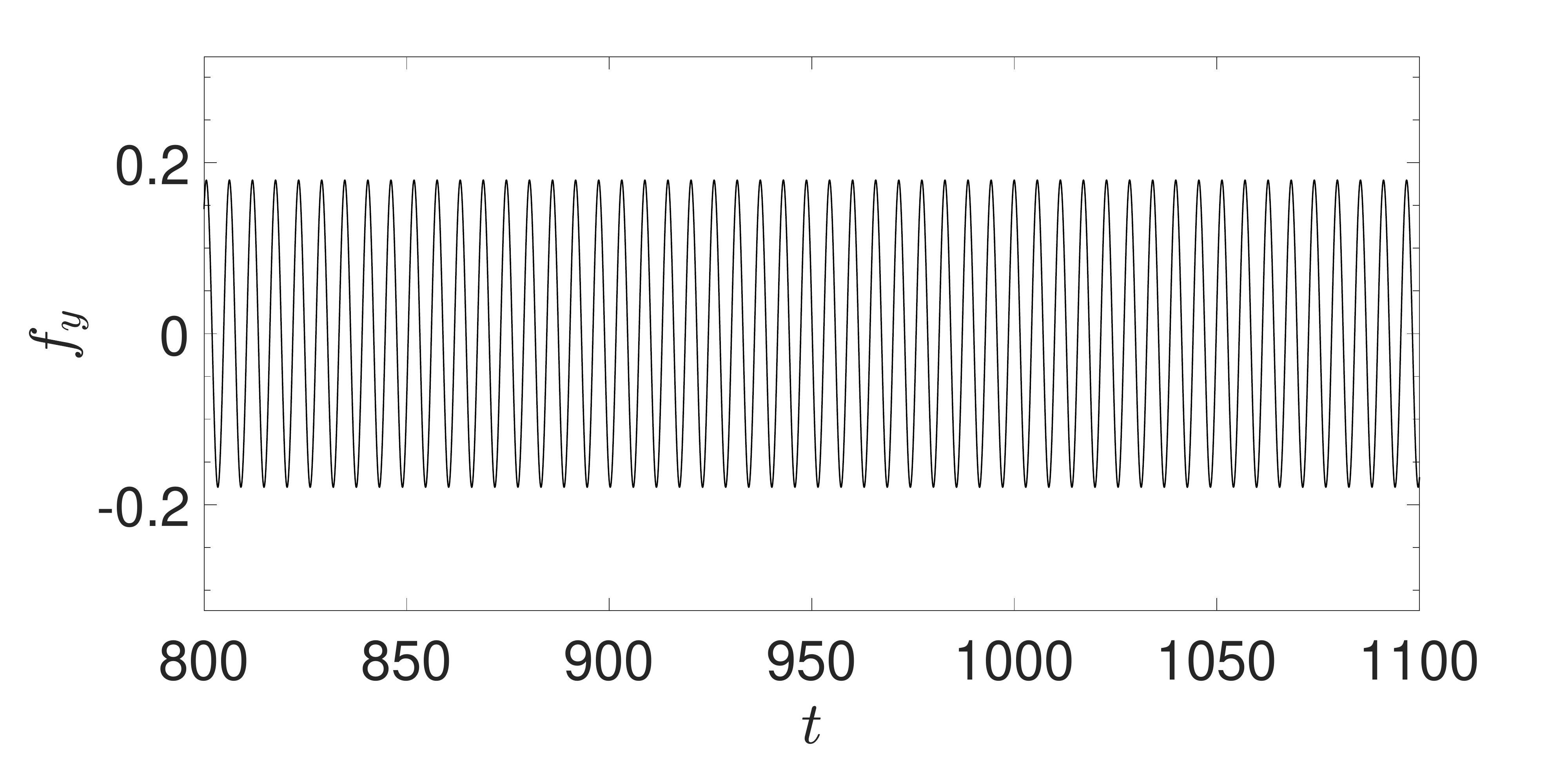}(b)
  }
  \centerline{
    \includegraphics[width=3.1in]{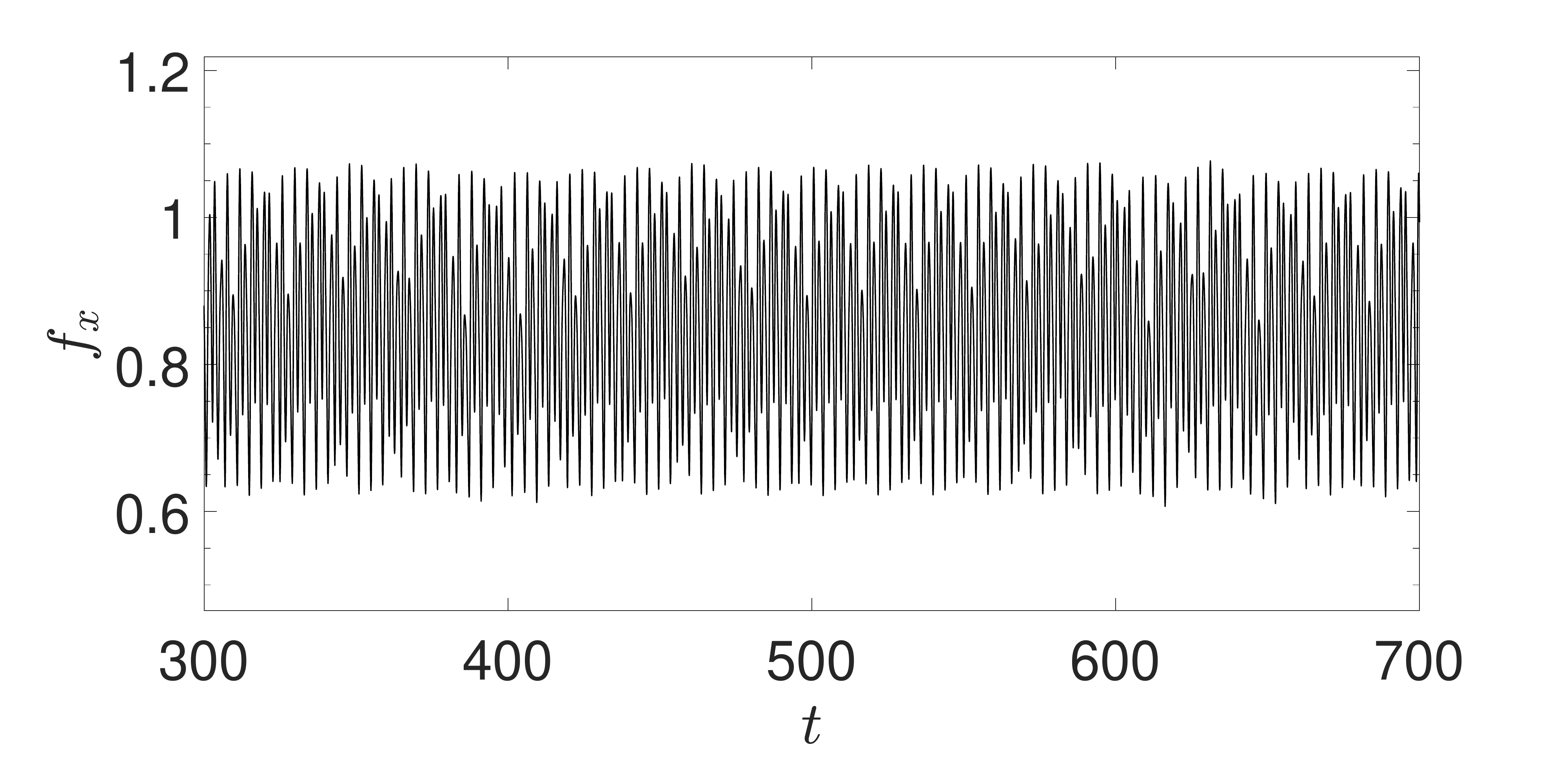}(c)
    \includegraphics[width=3.1in]{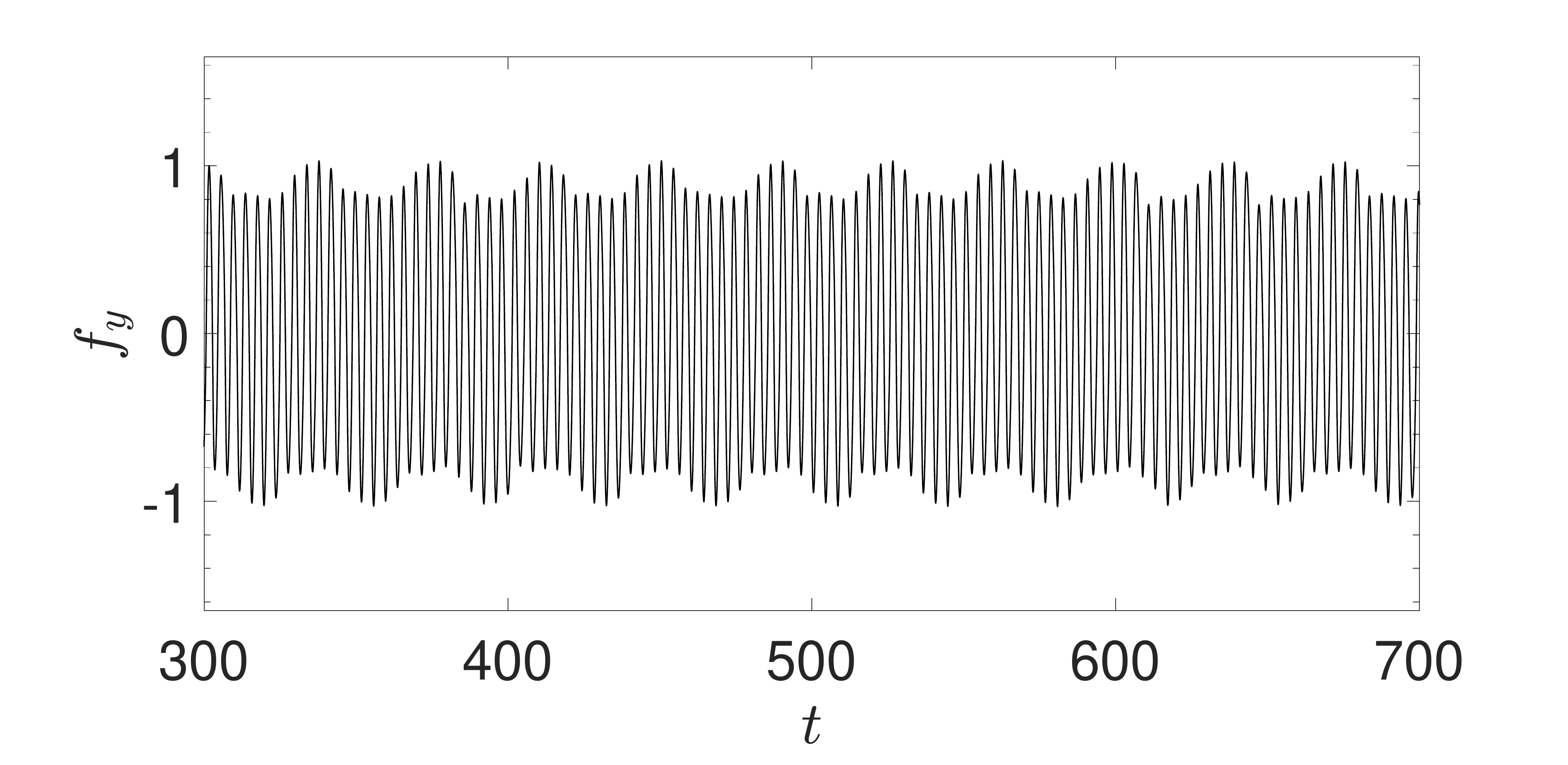}(d)
  }
  \caption{
    Cylinder flow: time histories of the drag (plots (a) and (c)) and 
    lift (plots (b) and (d)).
    Plots (a) and (b) are for $Re=100$, 
    and plots (c) and (d) are for $Re=2000$.
  }
  \label{fig:cyl_for}
\end{figure}

The cylinder flow is at a steady state for low enough Reynolds numbers
(for $Re\lesssim 47$), and it becomes unsteady with vortex
shedding behind the cylinder as the Reynolds number increases.
We refer the reader to the review article~\cite{Williamson1996}
for a discussion of different flow regimes in the cylinder wake.
Figure \ref{fig:cyl_str} shows instantaneous velocity field
distributions at Reynolds numbers $Re=2000$ and $Re=10000$
obtained from current two-dimensional simulations.
A train of irregular vortices can be observed behind the cylinder,
which persist in the entire wake region at these Reynolds numbers
with the current domain.
In particular, vortices and backflows can be clearly observed at
the outflow/open boundary, and they pose no problem to the current method.
This is 
thanks to the energy-stable open boundary condition~\eqref{equ:obc}, which
can effectively overcome the backflow instability issue
(see~\cite{Dong2015clesobc} for details).
Figure \ref{fig:cyl_for} shows a window of time histories of
the drag and lift on the cylinder  from
current simulations for Reynolds numbers
$Re=100$ and $Re=2000$, respectively.
These forces  fluctuate about some constant mean level,
and their overall characteristics stay the same over time.
These results demonstrate the long-term stability of our simulations,
and show that the flow has reached a statistically stationary state.


\begin{table}
  \centering
  \begin{tabular}{lllcccc}
    \hline
    Reynolds number & Method & Element order & mean drag & rms drag & mean lift & rms lift \\
    \hline
    30 & Current & 4 & 0.968 & 0 & 0 & 0 \\
    && 5 & 0.968 & 0 & 0 & 0 \\
    && 6 & 0.968 & 0 & 0 & 0 \\
    && 7 & 0.968 & 0 & 0 & 0 \\
    && 8 & 0.968 & 0 & 0 & 0 \\
    && 9 & 0.968 & 0 & 0 & 0 \\
    && 10 & 0.968 & 0 & 0 & 0 \\ \cline{2-7}
    & Dong (2015) & -- & 0.968 & 0 & 0 & 0 \\
    & Dong \& Shen (2015) & -- & 0.968 & 0 & 0 & 0 \\
    \hline
    100 & Current & 4 & 0.729 & 3.79e-3 & 1.61e-4 & 0.127 \\
    && 5 & 0.730 & 3.82e-3 & 9.16e-5 & 0.127 \\
    && 6 & 0.729 & 3.81e-3 & 1.40e-4 & 0.127 \\
    && 7 & 0.729 & 3.81e-3 & -2.80e-5 & 0.127 \\
    && 8 & 0.729 & 3.81e-3 & 2.15e-4 & 0.127 \\
    && 9 & 0.729 & 3.81e-3 & -6.50e-5 & 0.127 \\ \cline{2-7}
    & Dong (2015) & -- & 0.729 & -- & -- & 0.127 \\
    & Dong \& Shen (2015) & -- & 0.729 & -- & -- & 0.127 \\
    \hline
    2000 & Current & 4 & 0.724 & 0.0996 & 1.46e-3 & 0.537 \\
    && 5 & 0.863 & 0.117 & 3.53e-4 & 0.642 \\
    && 6 & 0.893 & 0.125 & -2.45e-5 & 0.699 \\
    && 7 & 0.865 & 0.123 & -4.66e-4 & 0.671 \\
    && 8 & 0.853 & 0.123 & -5.79e-4 & 0.657 \\
    && 9 & 0.848 & 0.122 & 2.33e-3 & 0.653 \\ \cline{2-7}
    & Dong (2015) & -- & 0.853 & -- & -- & 0.657 \\
    \hline
  \end{tabular}
  \caption{
    Spatial resolution test for cylinder flow: mean and rms
    drag and lift forces on the cylinder computed using
    various element orders for several Reynolds numbers.
    The data from Dong (2015)~\cite{Dong2015clesobc} and
    Dong \& Shen (2015)~\cite{DongS2015} are included for
    comparison.
  }
  \label{tab:cyl_for}
\end{table}


Based on the force history data,
we can obtain the statistical quantities
such as the time-averaged
mean and root-mean-square (rms) forces.
In current simulations we have varied the element order
systematically between $4$ and $10$ to study its effect
on the numerical result.
In Table~\ref{tab:cyl_for} we list the mean and
rms drag and lift forces for several Reynolds numbers
($Re=30$, $100$ and $2000$) corresponding to different element orders.
At $Re=30$ the flow is at a steady state, and so the values in
the table are the steady-state forces and no time-averaging is performed
for this Reynolds number.
In these simulations the time step sizes are
$\Delta t=1e-3$ for $Re=30$ and $100$, and
$\Delta t=5e-4$ for $Re=2000$.
For comparison, the forces obtained from~\cite{Dong2015clesobc,DongS2015}
for these Reynolds numbers
have also been included in this table.
For the lower Reynolds numbers ($Re=30$ and $100$)
the computed forces are basically the same using
all these element orders.
For the higher Reynolds number ($Re=2000$) we can observe a larger discrepancy
between the obtained forces corresponding to the
element orders $4$ and $5$ and those corresponding to
higher element orders. On the other hand, when the element order
increases to $7$ and higher, the obtained forces
are quite close to one another, exhibiting a sense of convergence.
The converged values of the forces from current simulations
are in good agreement with those of~\cite{Dong2015clesobc} and~\cite{DongS2015}.
In subsequent simulations an element order $8$ is employed
for this problem.


\begin{table}
  \centering
  \begin{tabular}{llcccc}
    \hline
    Reynolds number & $\Delta t$ & mean drag & rms drag & mean lift & rms lift \\
    \hline
    30 & 0.001 & 0.968 & 0 & 0 & 0 \\
    & 0.005 & 0.968 & 0 & 0 & 0 \\
    & 0.01 & 0.968 & 0 & 0 & 0 \\
    & 0.05 & 0.782 & 0.128 & -3.21e-5 & 6.80e-3 \\
    & 0.1 & 0.589 & 0.0950 & -4.94e-5 & 4.42e-4 \\
    \hline
    100 & 0.001 & 0.729 & 3.81e-3 & 2.15e-4 & 0.127 \\
    & 0.005 & 0.729 & 3.81e-3 & -1.72e-7 & 0.127 \\
    & 0.01 & 0.550 & 0.128 & 2.15e-4 & 0.0616 \\
    & 0.05 & 0.333 & 0.0823 & 1.95e-4 & 9.70e-3 \\
    & 0.1 & 0.238 & 0.0461 & 1.195e-4 & 4.53e-3 \\
    \hline
    2000 & 2.5e-4 & 0.853 & 0.123 & -5.79e-4 & 0.657 \\
    & 5.0e-4 & 0.855 & 0.123 & 6.71e-4 & 0.658 \\
    & 0.001 & 0.504 & 0.288 & 2.54e-3 & 0.440 \\
    & 0.005 & 0.167 & 0.0896 & 3.90e-4 & 0.179 \\
    & 0.01 & 0.0898 & 0.0578 & -3.2e-4 & 0.0780 \\
    & 0.05 & 0.0251 & 7.16e-3 & -5.83e-3 & 2.08e-3 \\
    & 0.1 & 0.0205 & 1.05e-3 & 1.06e-4 & 9.48e-4 \\
    \hline
  \end{tabular}
  \caption{
    Cylinder flow: mean and rms drag and lift forces obtained
    with a range of time steps sizes $\Delta t$ for several
    Reynolds numbers.
  }
  \label{tab:cyl_ldt}
\end{table}

Thanks to the unconditional energy stability property,
stable simulation results can be obtained using our method
irrespective of the time step size.
We have varied $\Delta t$ systematically and performed
simulations  using these $\Delta t$
values for several Reynolds numbers.
Table~\ref{tab:cyl_ldt} lists the mean and rms forces
on the cylinder computed using different $\Delta t$ values
at three Reynolds numbers $Re=30$, $100$ and $2000$.
The element order is fixed at $8$ in these tests.
Long-time simulations have been performed with each $\Delta t$,
and the statistical quantities shown in the table are computed
based on the drag and lift histories
from these simulations.
%
These results attest to the stability of simulations
using the current method, even with large (or fairly large)
$\Delta t$ values.
It can also be observed that the accuracy of the simulation
results could deteriorate  when $\Delta t$ becomes large.
For example, at $Re=30$ it should physically be
 a steady flow. However, with larger time
step sizes (e.g.~$\Delta t=0.05$ and $0.1$) the obtained
velocity fields actually become unsteady, and there is a notable difference
between the computed mean and rms drag values
when compared with those corresponding to smaller $\Delta t$.
%
At higher Reynolds numbers, the computed values for the
mean and rms drags and lifts with large $\Delta t$
appear smaller than those corresponding to small
$\Delta t$.
%
These tests suggest that, while the  method is
unconditionally energy stable and can produce stable 
results with various time step sizes ranging from small to large values,
the result corresponding to a large $\Delta t$ 
should only serve as a reference solution and should not be
blindly trusted. Convergence tests should be performed
with respect to the simulation parameters (e.g.~$\Delta t$ and
spatial resolution) using the current method, as well as
with any other numerical method for that matter.


\begin{figure}
  \centerline{
    \includegraphics[width=3.1in]{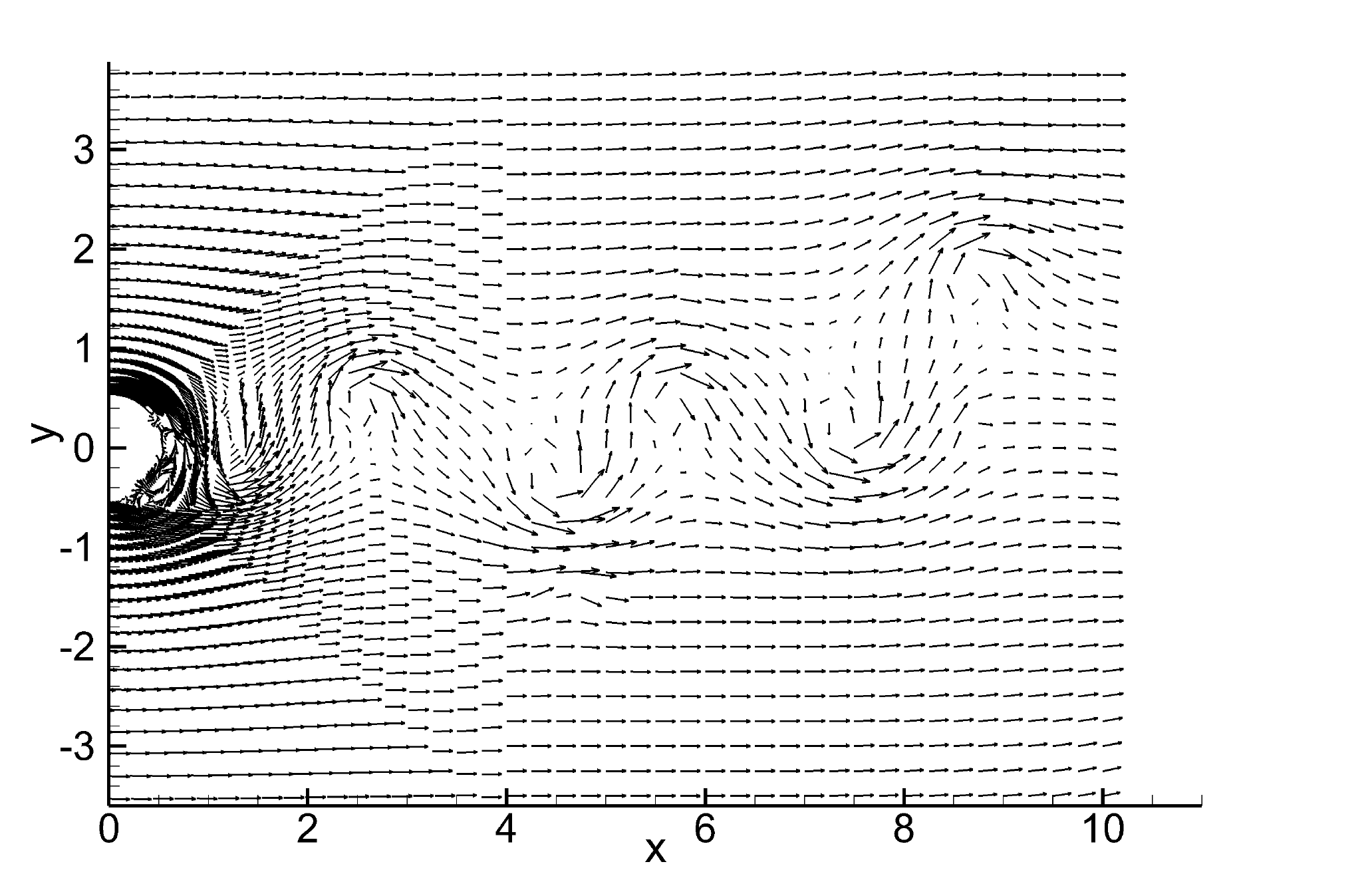}(a)
    \includegraphics[width=3.1in]{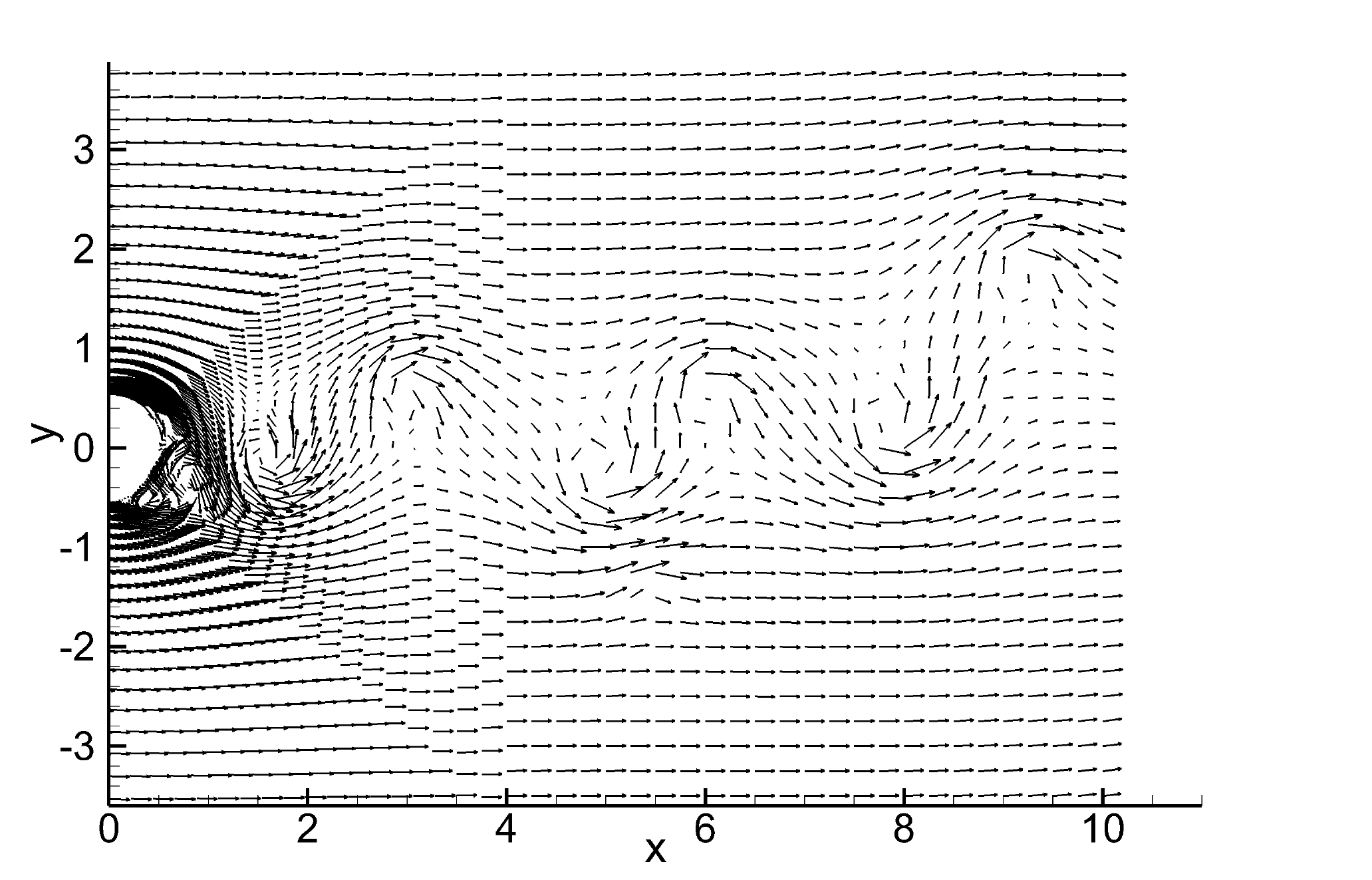}(b)
  }
  \centerline{
    \includegraphics[width=3.1in]{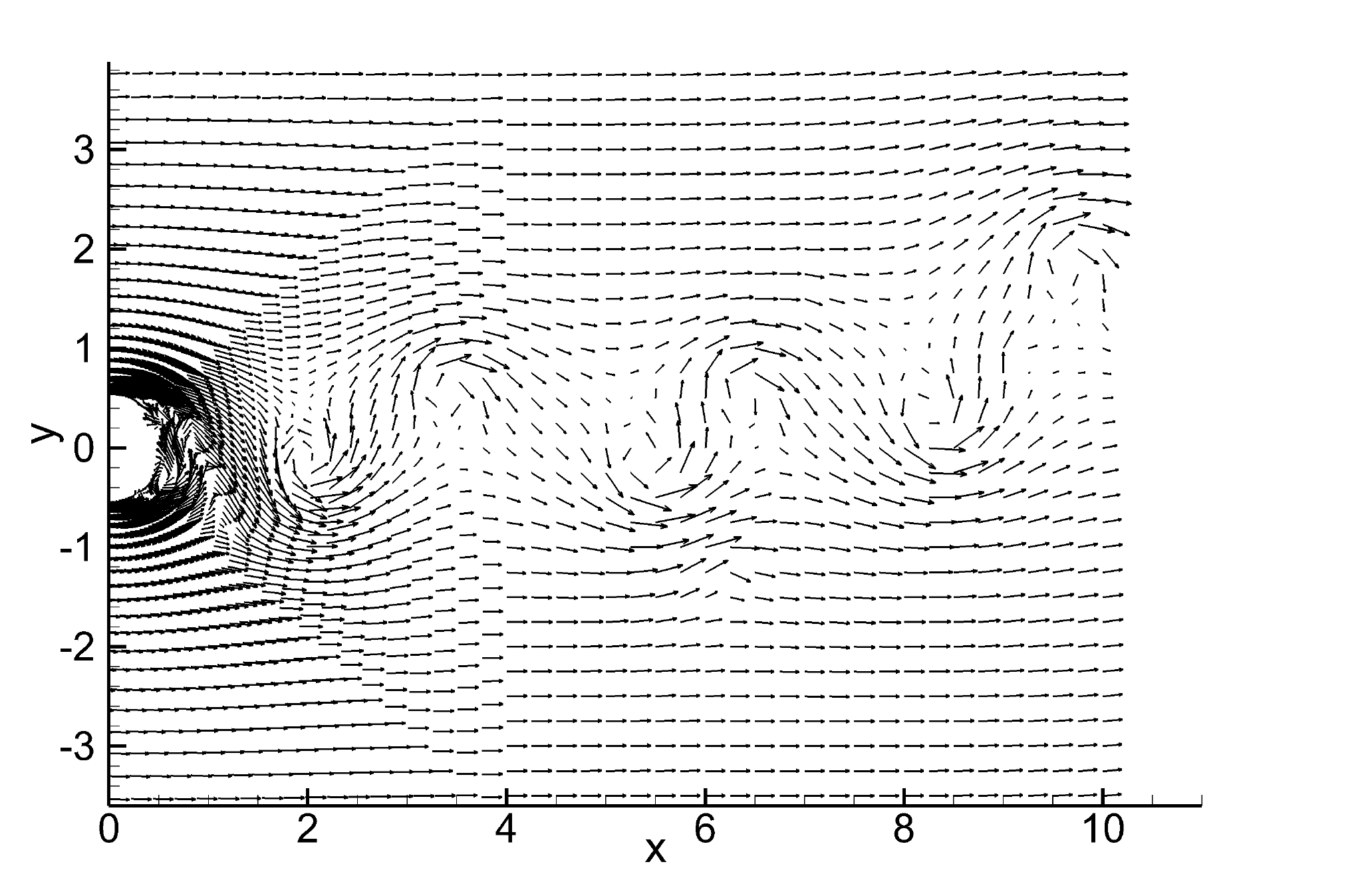}(c)
    \includegraphics[width=3.1in]{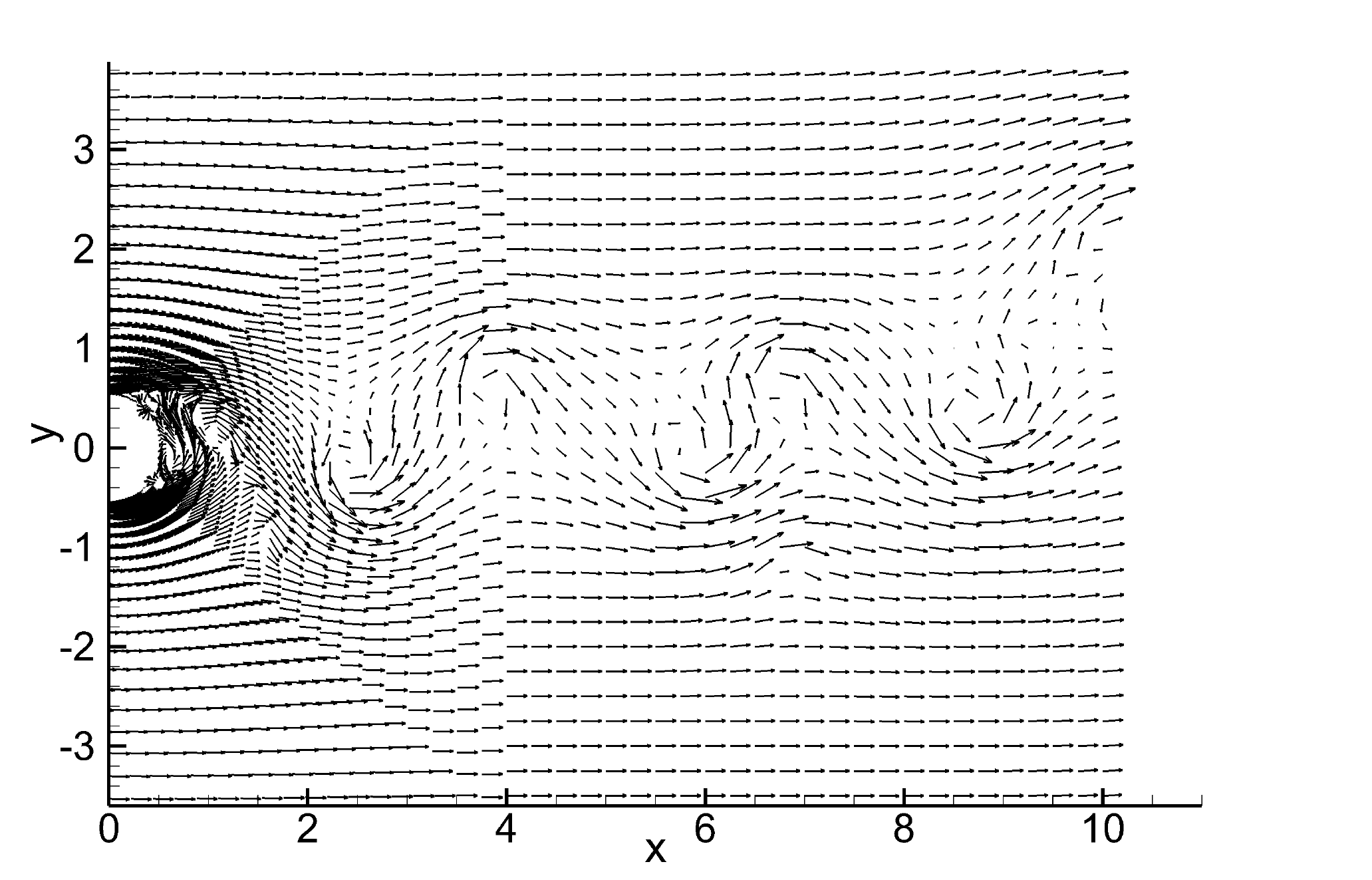}(d)
  }
  \centerline{
    \includegraphics[width=3.1in]{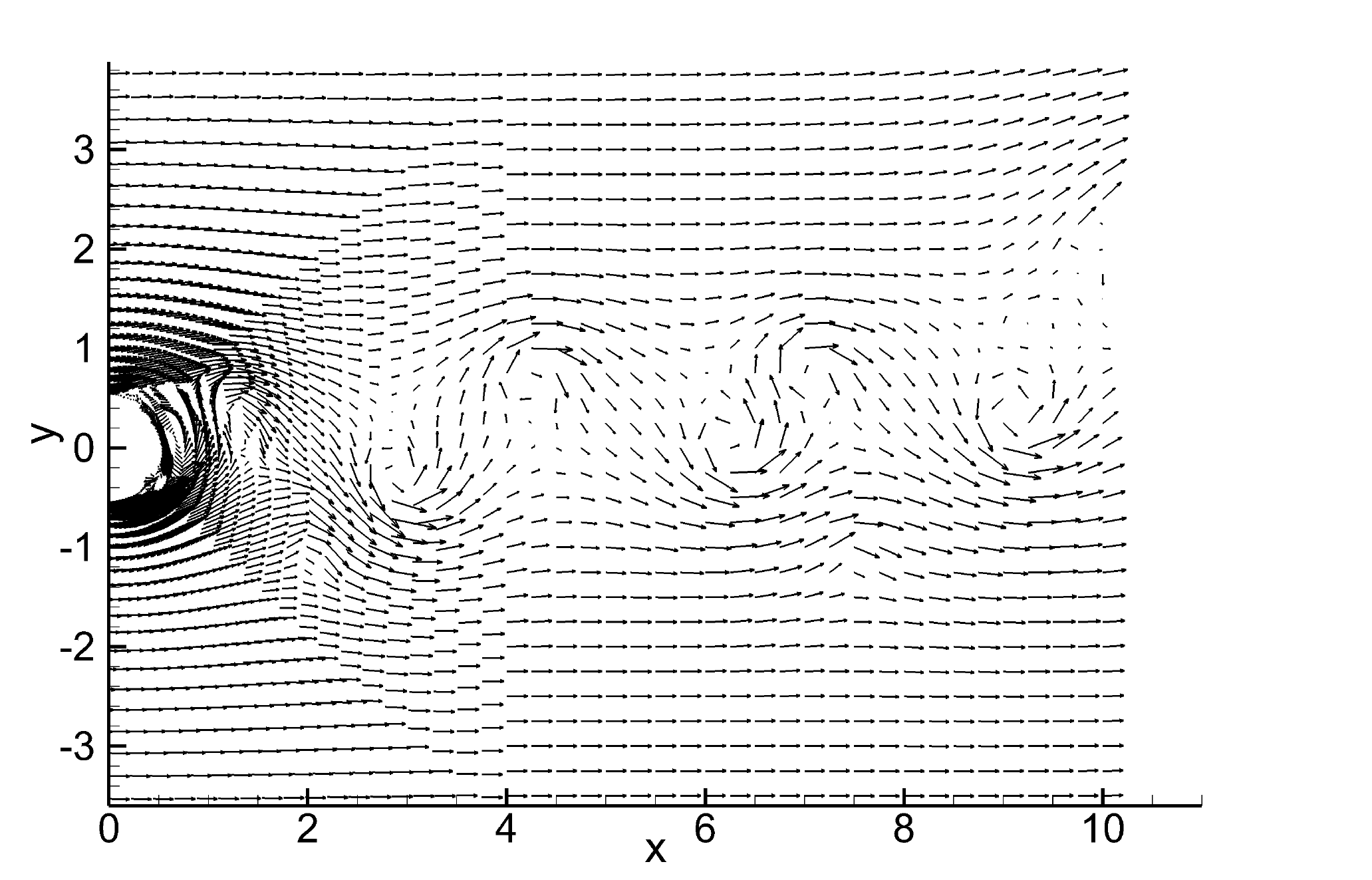}(e)
    \includegraphics[width=3.1in]{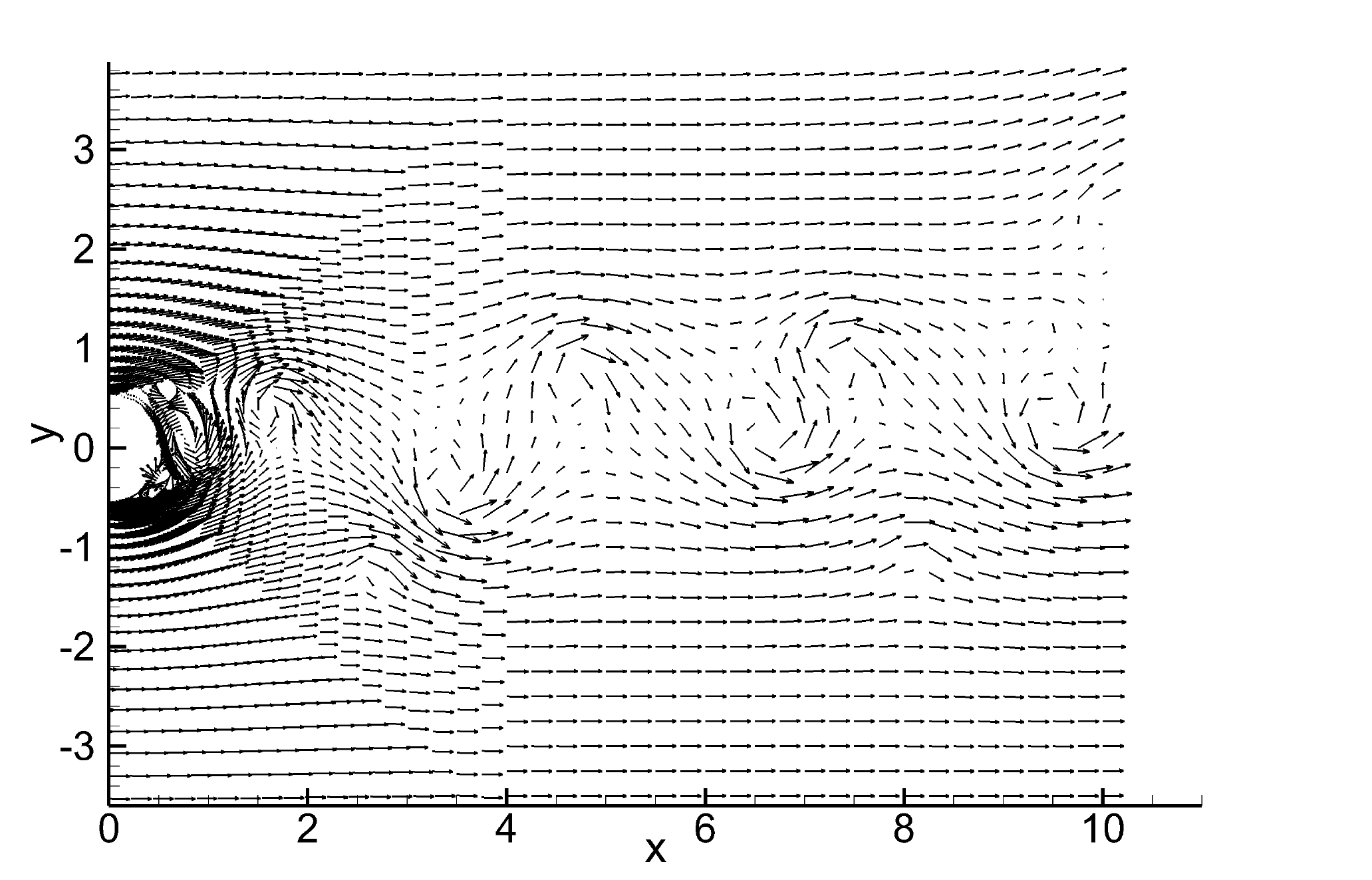}(f)
  }
  \centerline{
    \includegraphics[width=3.1in]{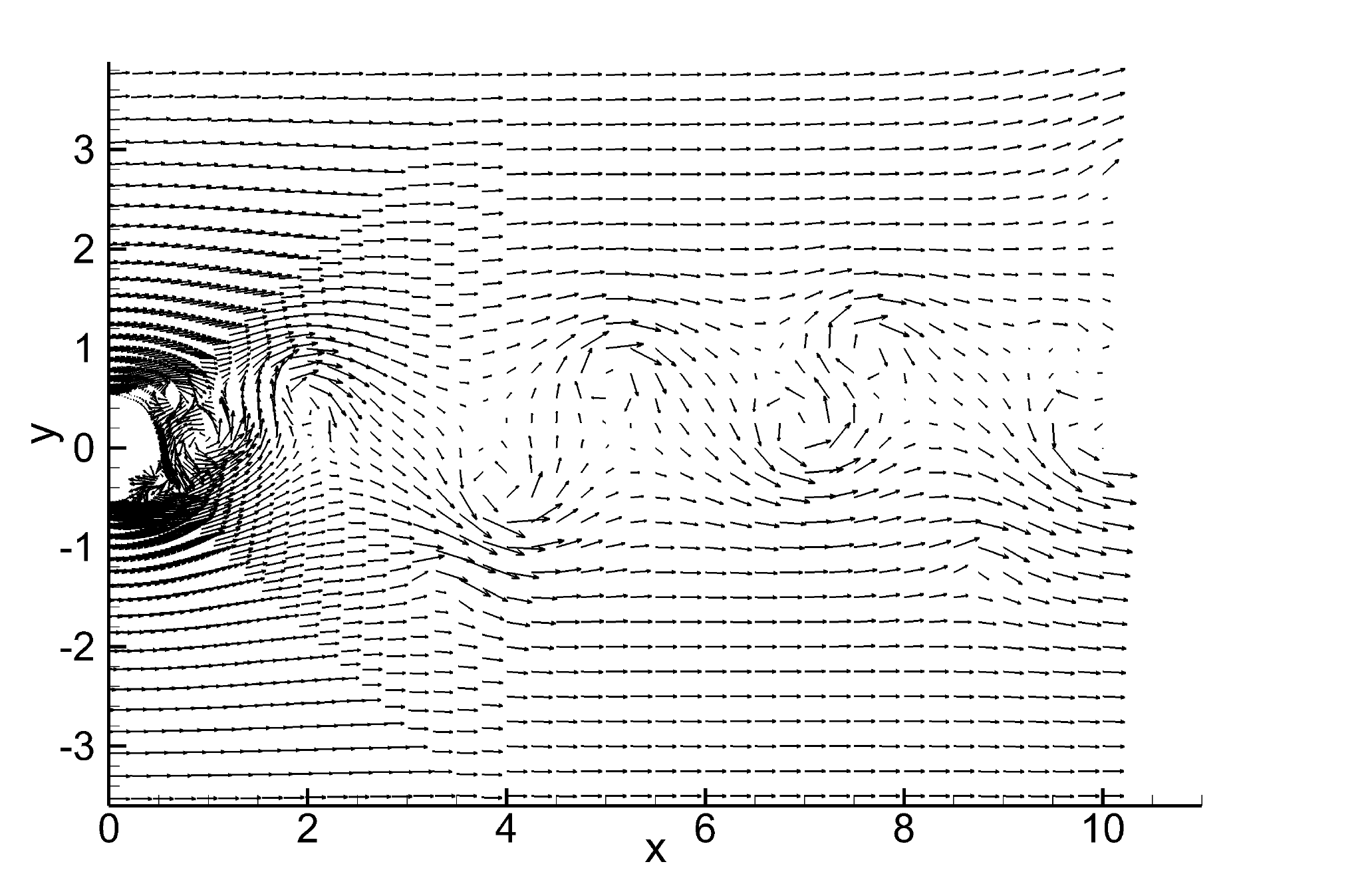}(g)
    \includegraphics[width=3.1in]{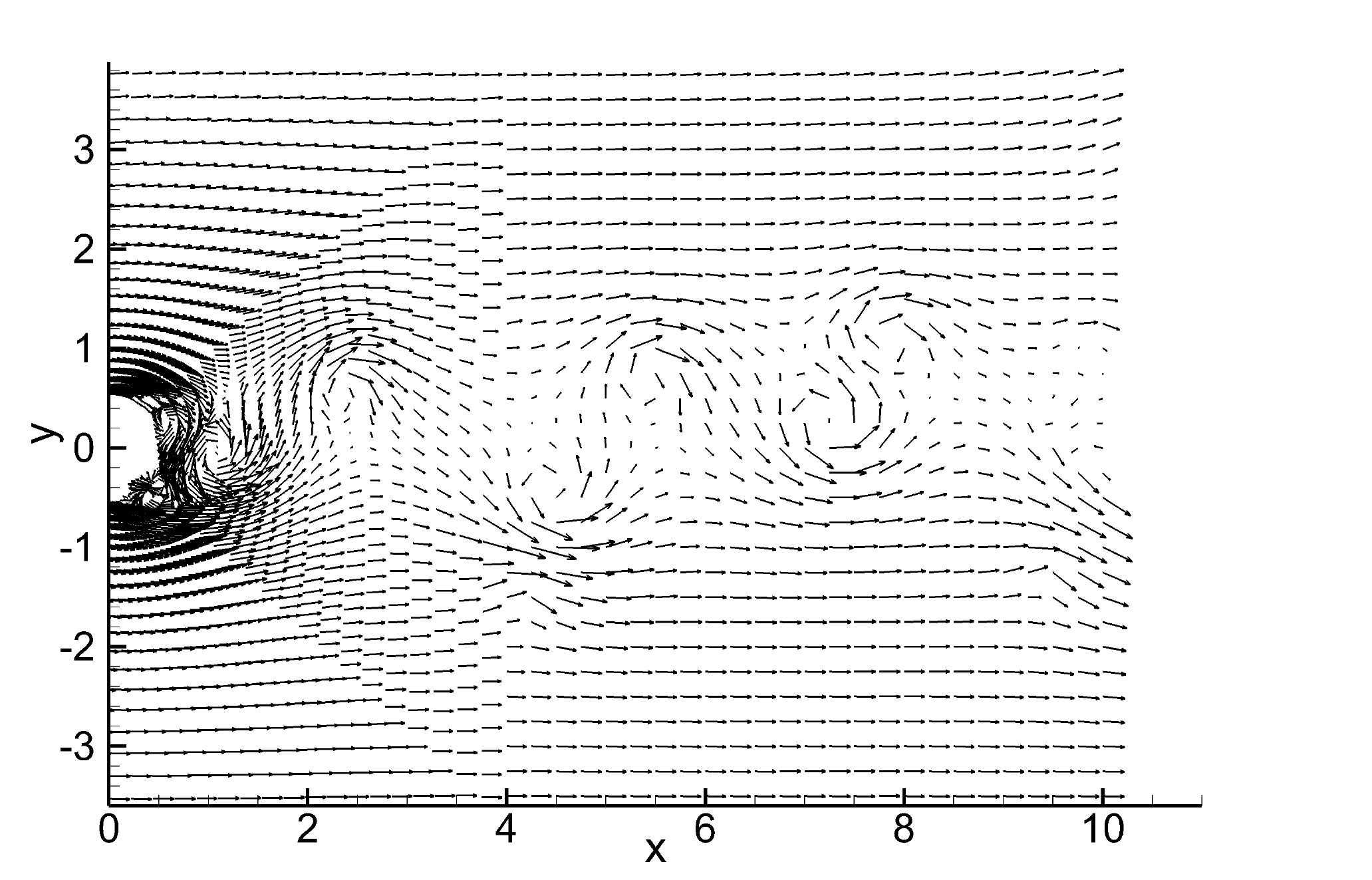}(h)
  }
  \caption{
    Cylinder flow ($Re=5000$): temporal sequence of snapshots of
    velocity distributions:
    (a) $t=1208.9$,
    (b) $t=1209.5$,
    (c) $t=1210.1$,
    (d) $t=1210.7$,
    (e) $t=1211.3$,
    (f) $t=1211.9$,
    (g) $t=1212.5$,
    (h) $t=1213.1$.
    Velocity vectors are plotted on a sparser grid for clarity.
  }
  \label{fig:cyl_snap}
\end{figure}

Finally, Figure \ref{fig:cyl_snap} illustrates
the dynamics of the cylinder flow
with a temporal sequence of snapshots (with a time
interval $0.6$ between consecutive frames) of
the velocity distributions at $Re=5000$.
Vortex shedding behind the cylinder generates a Karman
vortex street in the wake. 
As the vortices exit the domain, backflows can
be observed at the outflow/open boundary;
see e.g.~Figures \ref{fig:cyl_snap}(b)--(e) and (f)--(h).
While some distortions to the vortices are evident, 
it is observed that the current method
can allow the vortices to cross the outflow/open boundary
in a fairly smooth and natural fashion.


\subsection{Jet Impinging on a Wall}


\begin{figure}
  \centerline{
    \includegraphics[width=5in]{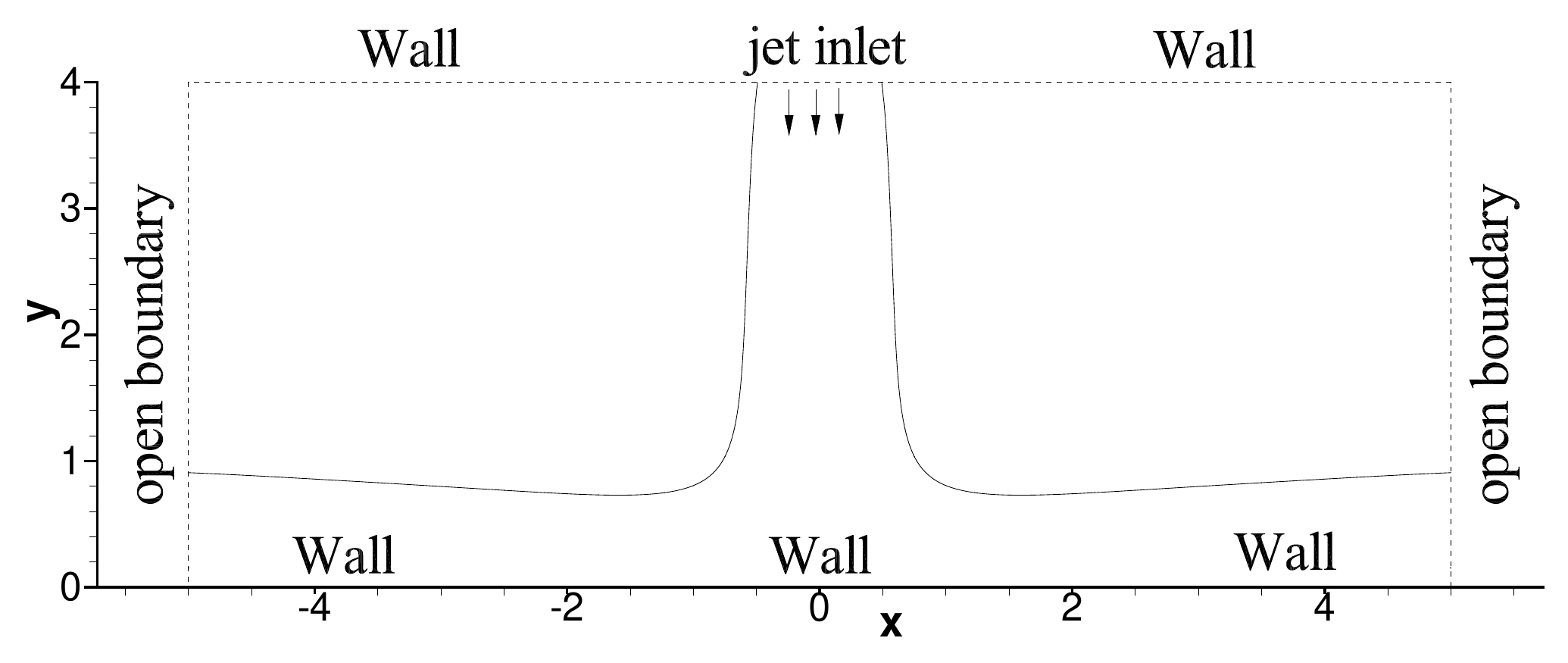}
  }
  \caption{Impinging jet: flow configuration and boundary conditions.}
  \label{fig:jet_conf}
\end{figure}

In the last numerical test we simulate a jet impinging on a wall
using the current method.
A sketch of the problem configuration is shown in Figure \ref{fig:jet_conf}.
We consider a rectangular domain, $-5\leqslant x/d\leqslant 5$
and $0\leqslant y/d\leqslant 4$, where $d=1$ denotes the diameter
of the jet at the inlet.
The top and bottom  of the domain are solid walls. 
A jet stream is introduced into the domain through an opening
in the middle of the top wall, with a diameter $d$.
The jet velocity at the inlet is assumed to have the following distribution:
\begin{equation}
  \left\{
  \begin{split}
    & u = 0 \\
    &
    v = -U_0\left(
    \left[H(x,0) - H(x,R_0) \right]\tanh\frac{1-x/R_0}{\sqrt{2}\epsilon}
    + \left[H(x,-R_0) - H(x,0)  \right]\tanh\frac{1+x/R_0}{\sqrt{2}\epsilon}
    \right)
  \end{split}
  \right.
  \label{equ:jet_vel}
\end{equation}
where $U_0=1$ denotes the velocity scale,
$R_0=d/2$ is the jet radius, and $\epsilon=d/40$.
$H(x,a)$ is the Heaviside step function, taking a unit value if $x\geqslant a$
and vanishing otherwise.
The left and right sides of the domain are open, and the fluids can
freely leave or enter the domain through these boundaries.
A pressure head is imposed on the open boundaries,
with $p_{01}$ on the left boundary and $p_{02}$ on the right one.
The jet stream enters the domain through the inlet and,
depending on the relative levels for $p_{01}$ and $p_{02}$,
may exit the domain through both sides  or
through only one side of the open boundary.


All the parameters and variables are normalized with the velocity scale
$U_0$ and the length scale $d$.
We discretize the domain using a mesh of $640$ equal-sized quadrilateral spectral
elements, with $40$ elements along the horizontal direction and
$16$ elements along the vertical direction.
The method from Section \ref{sec:method} is employed to
simulate the incompressible Navier-Stokes equations,
with $\mbs f=0$ in equation \eqref{equ:nse}.
The Dirichlet condition \eqref{equ:dbc} is imposed
on the wall boundaries, with the boundary velocity set to $\mbs w=0$,
and also at the jet inlet, with the boundary velocity $\mbs w$
set according to equation \eqref{equ:jet_vel}.
The boundary condition \eqref{equ:obc} is imposed on
the left and right open boundaries, with $\mbs f_b=0$,
$D_0=\frac{1}{U_0}=1$ and $\delta =0.01$. The external
pressure force in \eqref{equ:obc} is set to
$p_0=p_{01}$ on the left boundary and
$p_0=p_{02}$ on the right boundary.
The Reynolds number,
the element order, the time step size $\Delta t$,
and the external pressure heads $p_{01}$ and $p_{02}$
have been varied to study their effects on the flow characteristics.
We employ a constant $C_0=1$ in simulations for this problem. 
Long-time simulations have been performed such that
the flow has reached a statistically stationary state.
Therefore the initial velocity has no effect on the 
results reported below.

\begin{figure}
  \centering
  \includegraphics[width=4in]{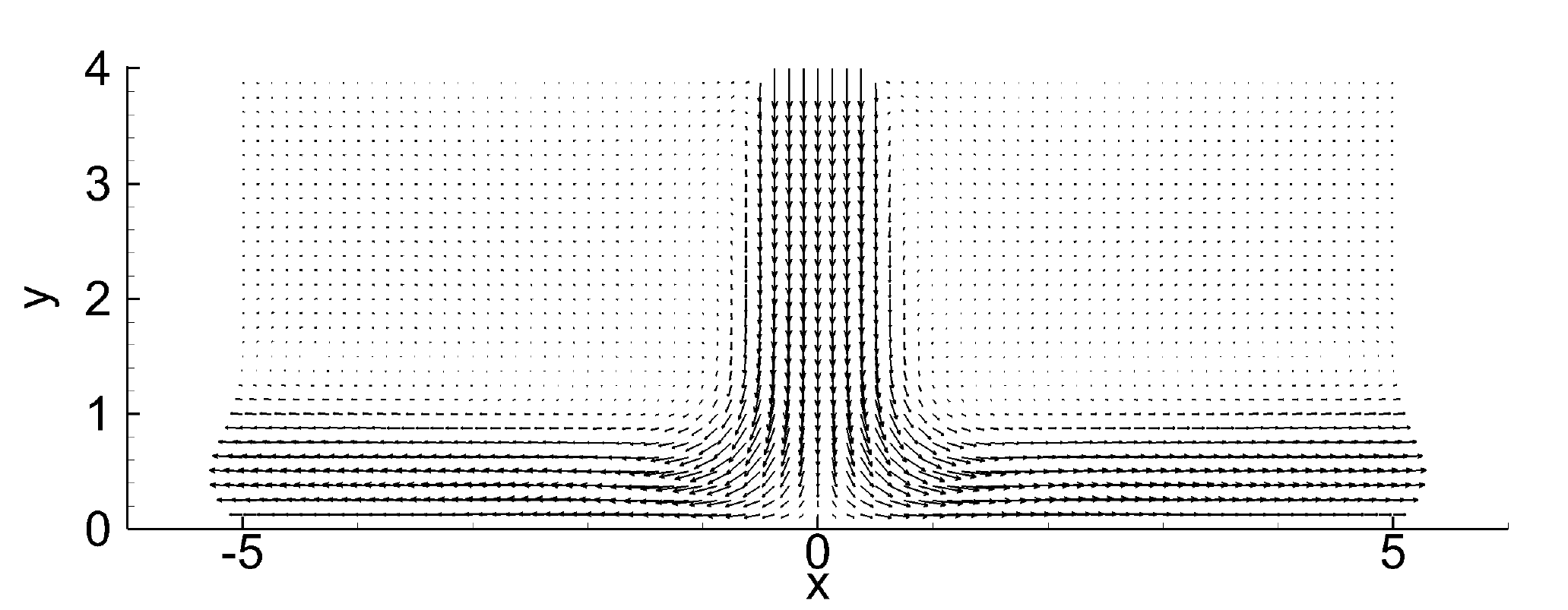}(a)
  \includegraphics[width=4in]{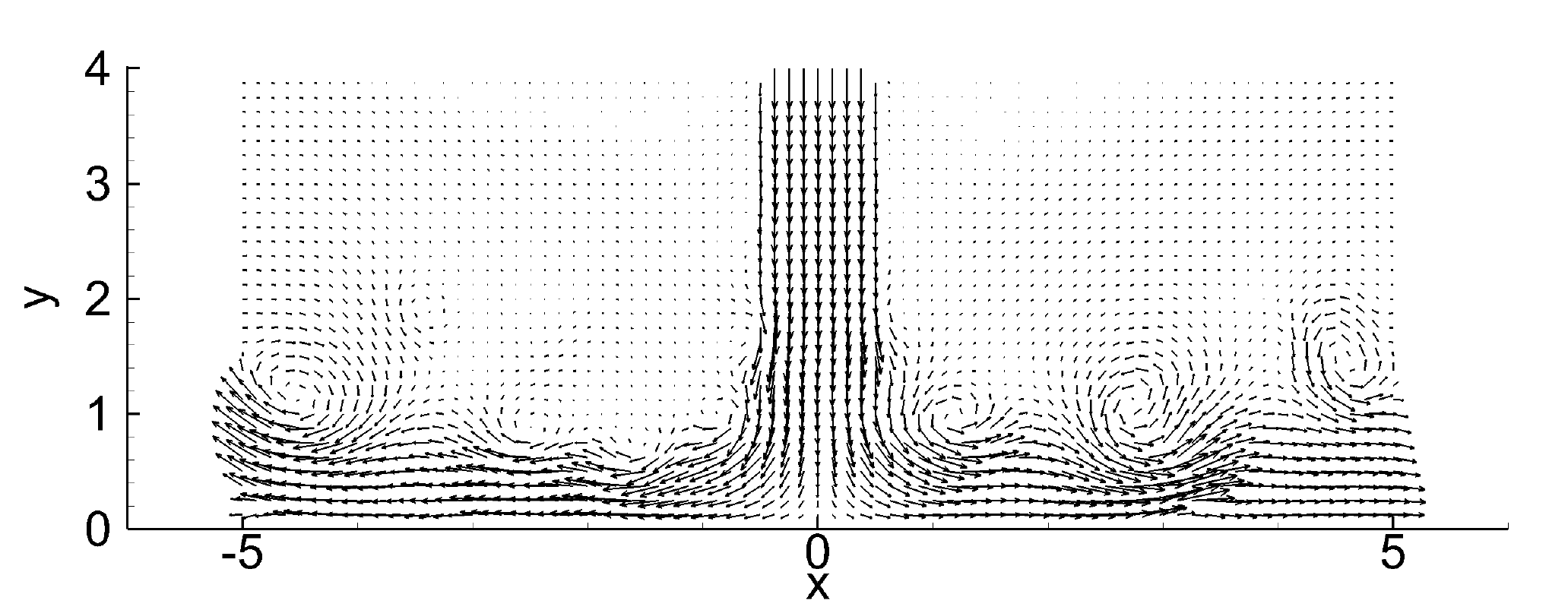}(b)
  \includegraphics[width=4in]{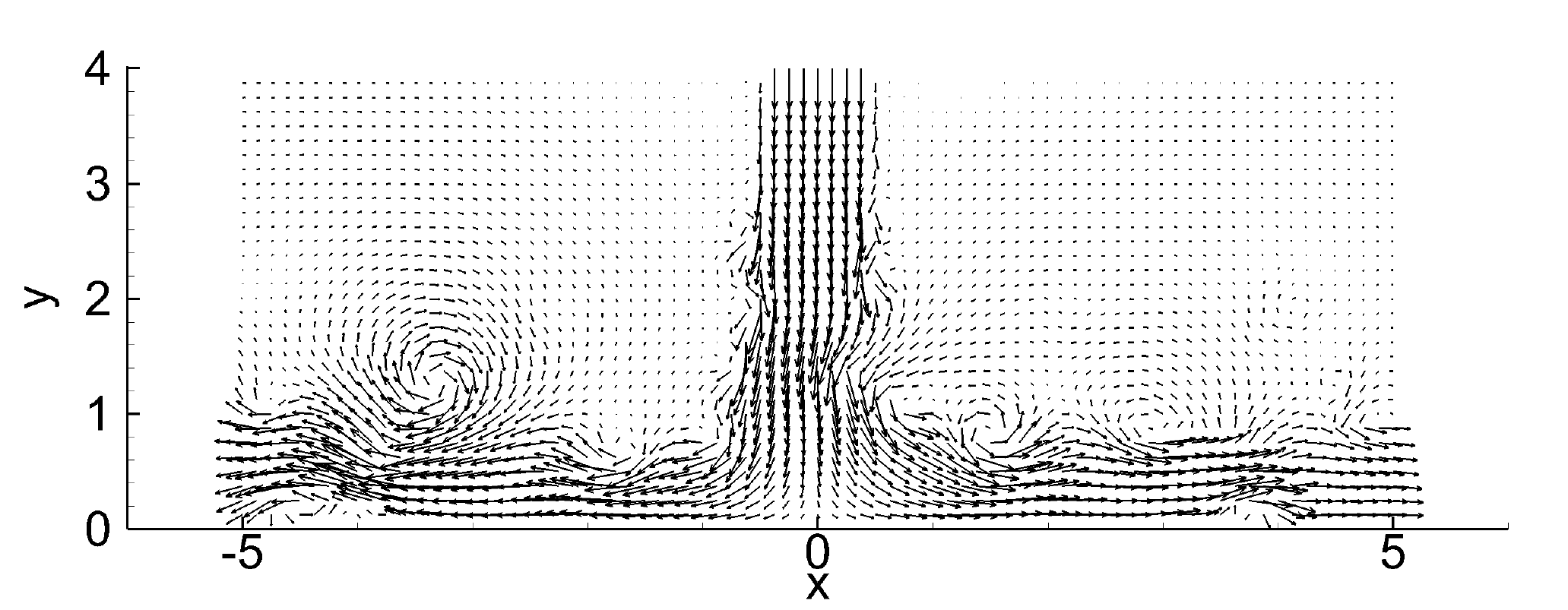}(c)
  \caption{
    Impinging jet: velocity
    distributions at Reynolds numbers (a) $Re=300$,
    (b) $Re=2000$, and (c) $Re=5000$.
    Velocity vectors are plotted on a set of sparser grid points for clarity.
  }
  \label{fig:jet_char}
\end{figure}

Let us first focus on the cases with zero external pressure heads
on the left and right open boundaries ($p_{01}=p_{02}=0$).
Figure \ref{fig:jet_char} provides an overview of the
flow characteristics of this  problem
for several Reynolds numbers.
At $Re=300$ it is a steady flow (Figure \ref{fig:jet_char}(a)).
After impinging on the bottom wall,
the vertical jet splits into two horizontal streams near the wall,
which flow out of the domain through the left and right open
boundaries. It is noted that strong flows mostly occupy
the regions near the bottom wall or near the domain centerline
($x/d=0$), while in the rest of the domain
the flow is quite weak.
With the increase of Reynolds number,
the flow becomes unsteady.
A train of vortices are observed to form along the profile of
the vertical jet or the near-wall horizontal
streams (Figures \ref{fig:jet_char}(b)-(c)), which
travel along with the jet and leave the
domain through the open boundaries.


\begin{figure}
  \centering
  \includegraphics[width=4in]{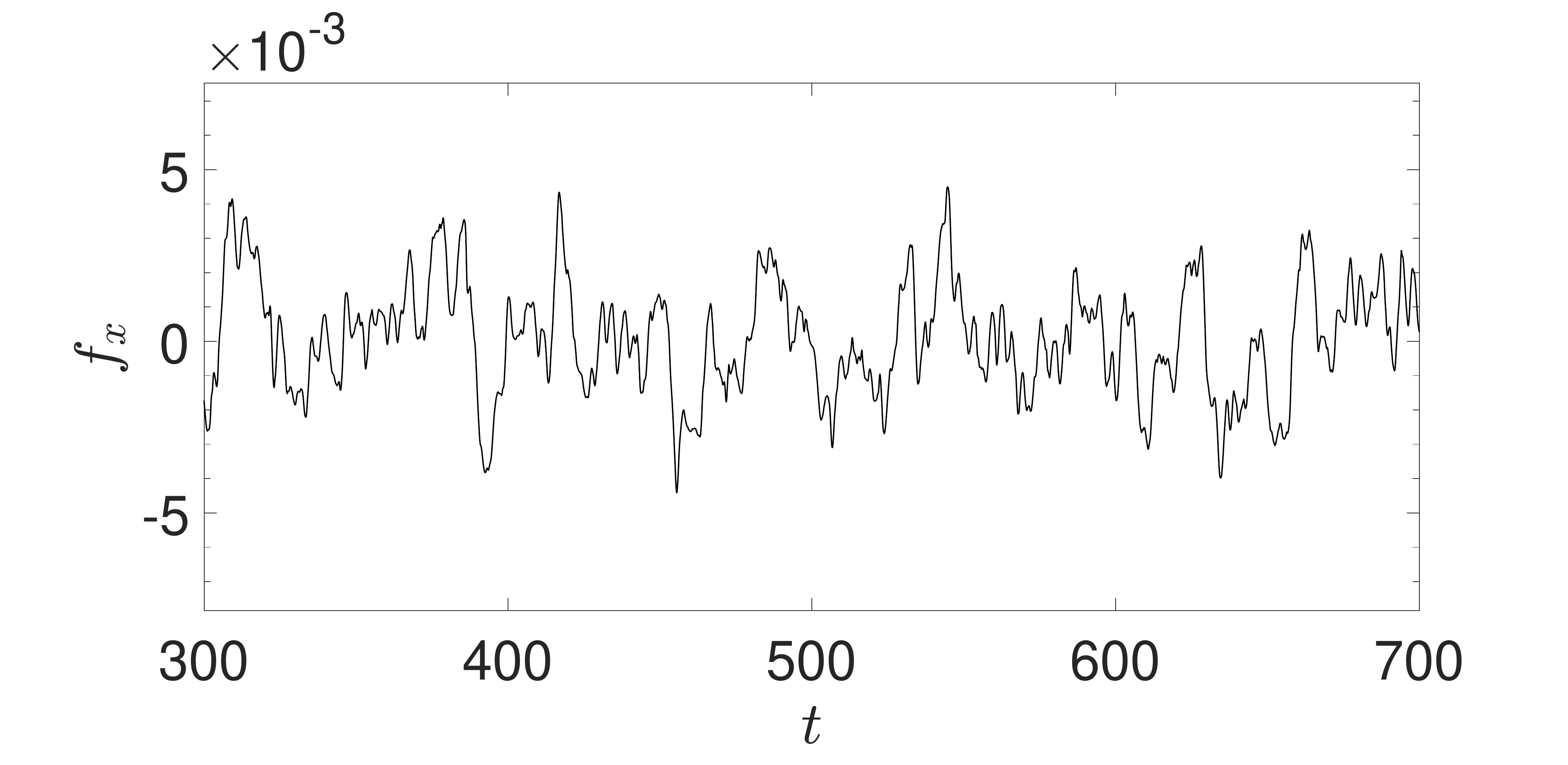}(a)
  \includegraphics[width=4in]{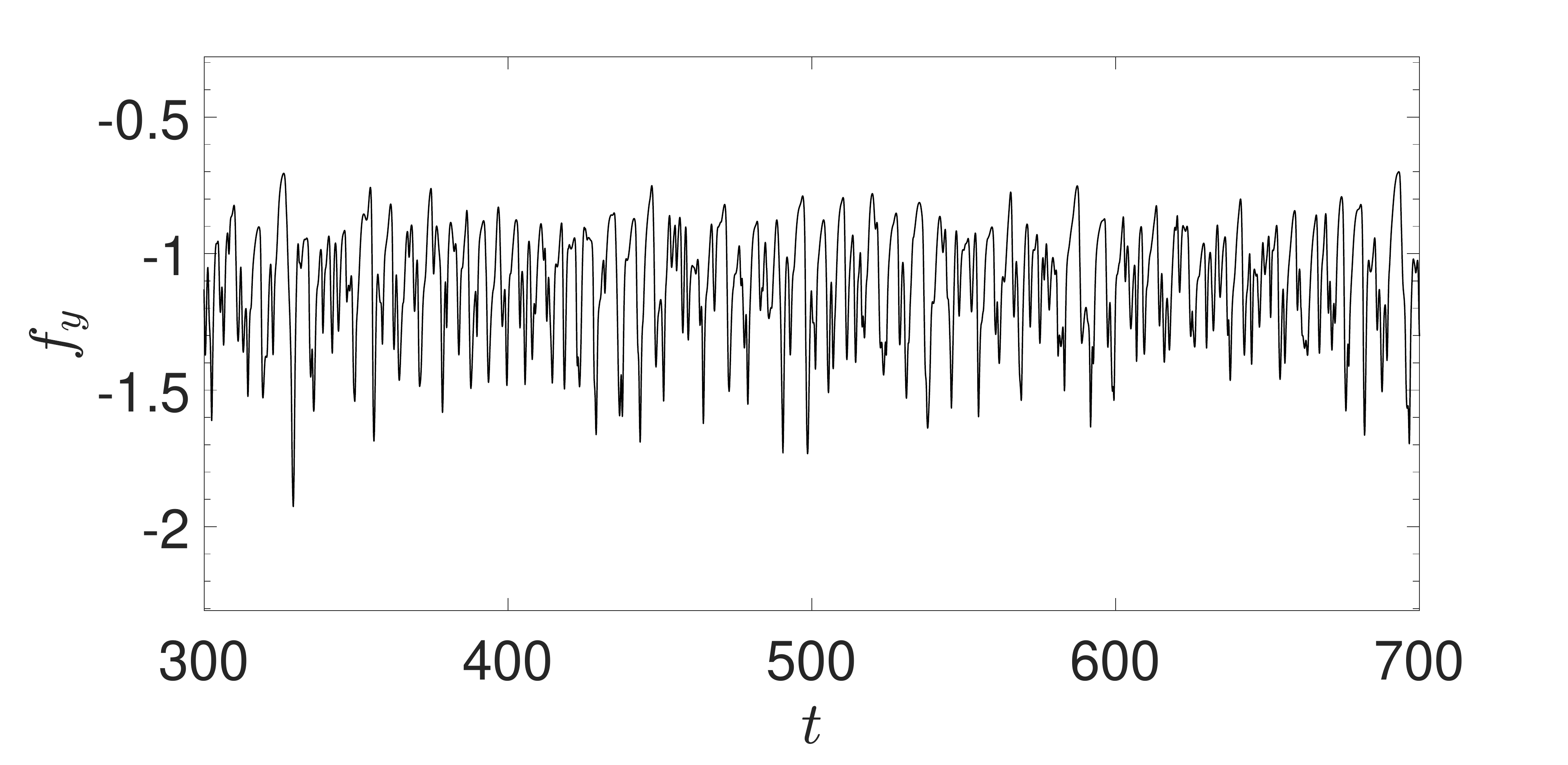}(b)
  \caption{
    Impinging jet (Re=2000): time histories of
    the x-component (a) and y-component (b) of the force
    on the walls.
  }
  \label{fig:jet_for}
\end{figure}

Figure \ref{fig:jet_for} shows a window of the time histories
of the total force ($x$ and $y$ components)
exerting on the domain walls at $Re=2000$,
which corresponds to the case shown in Figure \ref{fig:jet_char}(b).
These results are obtained with an element order $6$ and
a time step size $\Delta t=0.001$.
One can note that the forces are highly unsteady and fluctuational.
The horizontal force  is very weak and
 essentially negligible compared with the vertical force,
because of symmetry and the zero pressure heads imposed on
the left and right boundaries. 
The long time histories indicate that the flow has indeed reached
a statistically stationary state and that the simulation is
long-time stable.


\begin{table}
  \centering
  \begin{tabular}{llllll}
    \hline
    $Re$ & Element order & $\overline{f}_x$ & $f'_x$ & $\overline{f}_y$ & $f'_y$ \\
    300 & 4 & 0 & 0 & -0.965 & 0 \\
    & 5 &  0 & 0 & -0.975 & 0 \\
    & 6 &  0 & 0 & -0.984 & 0 \\
    & 7 & 0 & 0 & -0.986 & 0  \\
    & 8 & 0 & 0 & -0.987 & 0  \\
    \hline
    2000 & 4 & 3.44e-5 & 1.63e-3 & -1.131 & 0.201 \\
    & 5 & -8.23e-5 & 1.69e-3 & -1.089 & 0.191 \\
    & 6 & -1.71e-4 & 1.76e-3 & -1.100 & 0.198 \\
    & 7 & -1.20e-5 & 1.66e-3 & -1.103 & 0.201 \\
    & 8 & -9.17e-5 & 1.61e-3 & -1.106 & 0.205 \\
    \hline
    5000 & 4 & 3.60e-5 & 9.82e-4 & -1.117 & 0.195 \\
    & 5 & 4.94e-6 & 1.04e-3 & -1.092 & 0.210  \\
    & 6 & -1.98e-5 & 1.09e-3 & -1.078 & 0.223  \\
    & 7 & -2.24e-6 & 1.09e-3 & -1.068 & 0.230  \\
    & 8 & -4.13e-5 & 1.03e-3 & -1.076 & 0.235  \\
    \hline
  \end{tabular}
  \caption{ Impinging jet:
    Effect of element order on the computed forces  on the walls.
  }
  \label{tab:for_order}
\end{table}


\begin{table}
  \centering
  \begin{tabular}{llllll}
    \hline
    $Re$ & $\Delta t$ & $\overline{f}_x$ & $f'_x$ & $\overline{f}_y$ & $f'_y$ \\
    300 & 0.0005 & 0 & 0 & -0.983 & 0 \\
    & 0.001 & 0 & 0 & -0.984 & 0  \\
    & 0.005 & 0 & 0 & -0.986 & 0  \\
    & 0.01 & 0 & 0 & -0.986 & 0 \\
    & 0.05 & -2.37e-3 & 7.61e-7 & -0.270 & 3.82e-3  \\
    & 0.1 &  -2.69e-4 & 8.74e-5 & -0.183 & 1.17e-3  \\
    & 0.5 &  7.65e-5 & 4.27e-6 & -0.098 & 5.09e-4  \\
    \hline
    2000 & 0.0005 & -8.93e-5 & 1.73e-3 & -1.095 & 0.197 \\
    & 0.001 & -1.71e-4 & 1.76e-3 & -1.100 & 0.198 \\
    & 0.005 & -2.58e-5 & 1.77e-3 & -1.069 & 0.244 \\
    & 0.01 & -6.06e-5 & 3.54e-3 & -0.576 & 0.391 \\
    & 0.05 & 5.93e-5 & 2.02e-3 & -0.114 & 0.0624 \\
    & 0.1 & -9.34e-4 & 1.20e-6 & -0.0609 & 0.0144 \\
    \hline
    5000 & 0.0005 & 1.84e-5 & 1.14e-3 & -1.087 & 0.223   \\
    & 0.001 & -1.98e-5 & 1.09e-3 & -1.078 & 0.223 \\
    & 0.005 & -3.38e-5 & 4.09e-3 & -0.742 & 0.578 \\
    & 0.01 & -2.05e-4 & 3.97e-3 & -0.384 & 0.459 \\
    & 0.05 & 4.05e-5 & 1.47e-3 & -0.0829 & 0.0924 \\
    & 0.1 & -2.25e-5 & 9.16e-4 & -0.0562 & 0.0406 \\
    \hline
  \end{tabular}
  \caption{ Impinging jet:
    Effect of $\Delta t$ on the computed forces on the walls.
  }
  \label{tab:jet_ldt}
\end{table}

We can obtain the statistical quantities such as the mean and
rms forces the jet exerts on the domain walls based on
the force signals like those shown in Figure \ref{fig:jet_for}.
To investigate the  mesh resolution effect,
we have performed simulations using a range of element orders.
In Table \ref{tab:for_order} we list the time-averaged
mean and rms forces on the wall computed using different element
orders at three Reynolds numbers $Re=300$, $2000$ and $5000$.
The flow at $Re=300$ is steady, and so shown in the table
are the steady-state forces and no time-averaging is performed.
It can be observed that with element orders $6$ and larger
there is little difference in the obtained mean and rms
force values, demonstrating a convergence with respect to
the mesh resolution. The majority of subsequent simulations
for this problem are performed with an element order $6$.

The current method produces stable simulation results for the
impinging jet problem, with various time step sizes ranging from
small to large values.
This is demonstrated by Table~\ref{tab:jet_ldt},
in which the mean and rms forces  obtained with
a range of $\Delta t$ values have been shown
for the Reynolds numbers $Re=300$, $2000$ and $5000$.
The deterioration in accuracy of the obtained results
when $\Delta t$ becomes large
can also be observed here,
similar to the observation from Section \ref{sec:cyl}.
Differences between the mean/rms forces corresponding
to large (or fairly large) $\Delta t$ values and those
corresponding to small $\Delta t$ are evident,
indicating a deterioration or loss of accuracy
when $\Delta t$ becomes too large.

\begin{figure}
  \centerline{
    \includegraphics[width=3.1in]{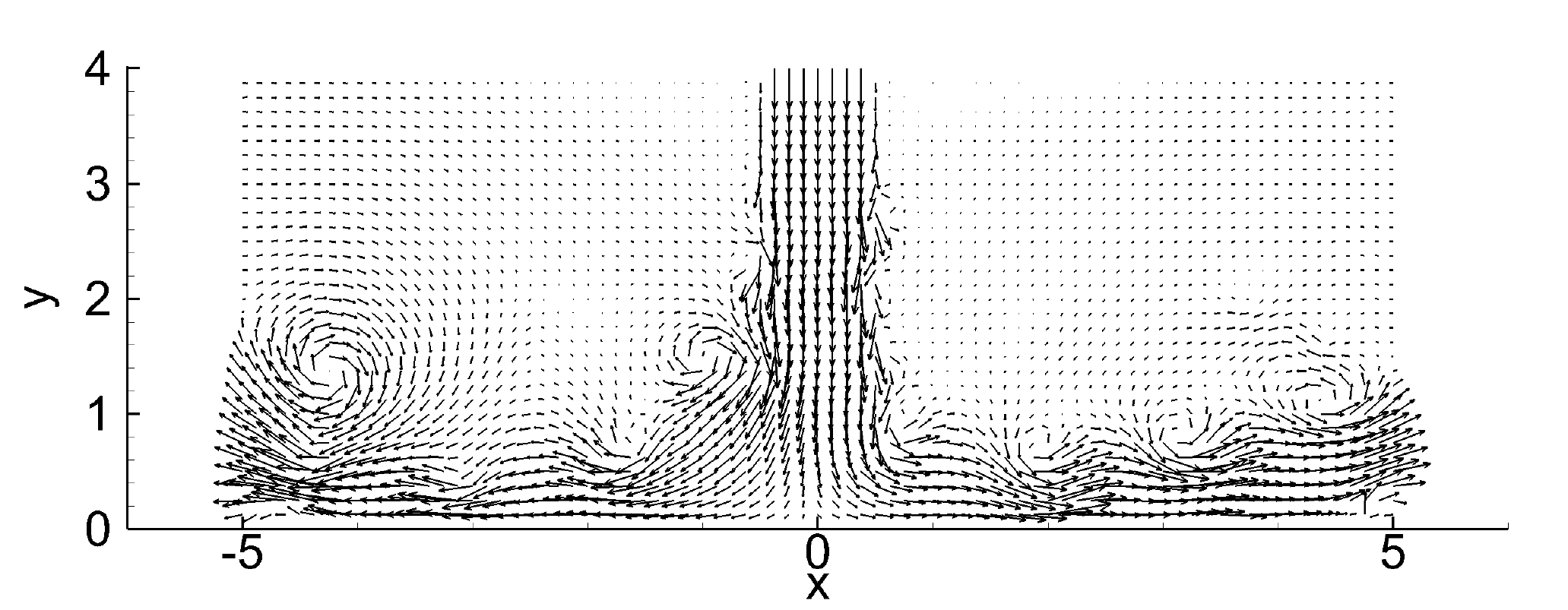}(a)
    \includegraphics[width=3.1in]{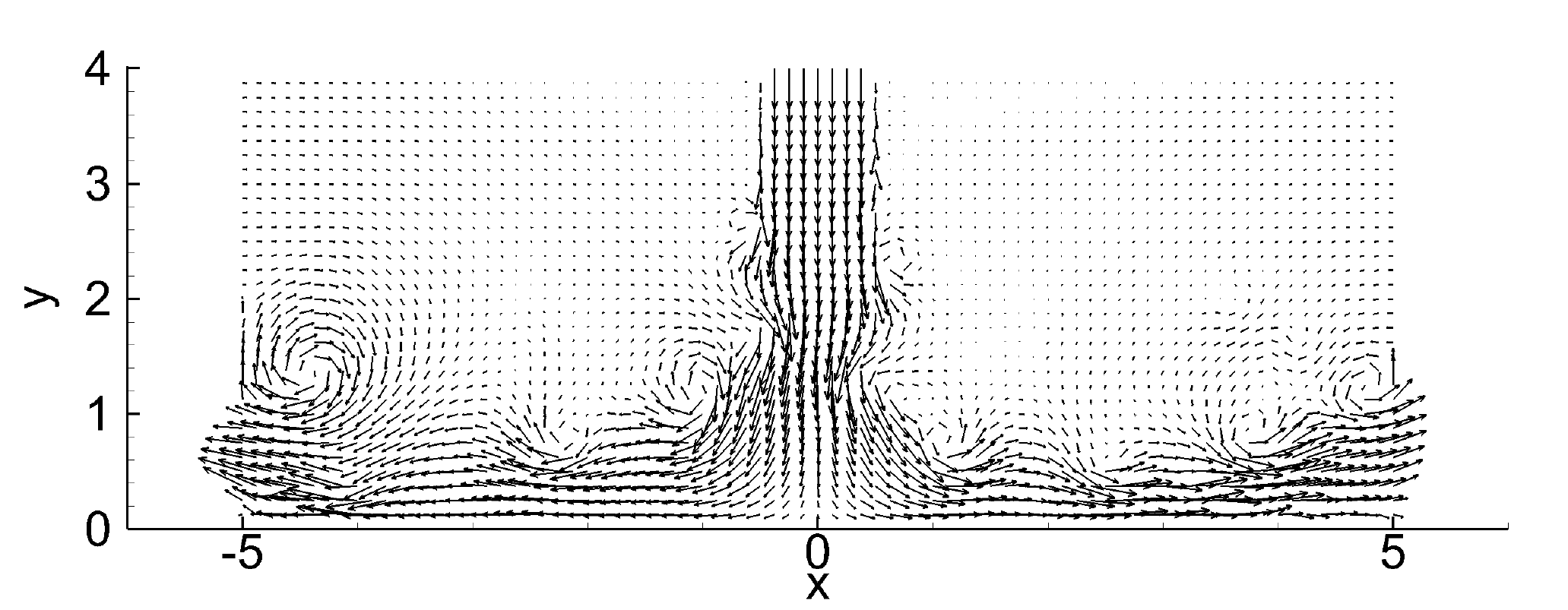}(b)
  }
  \centerline{
    \includegraphics[width=3.1in]{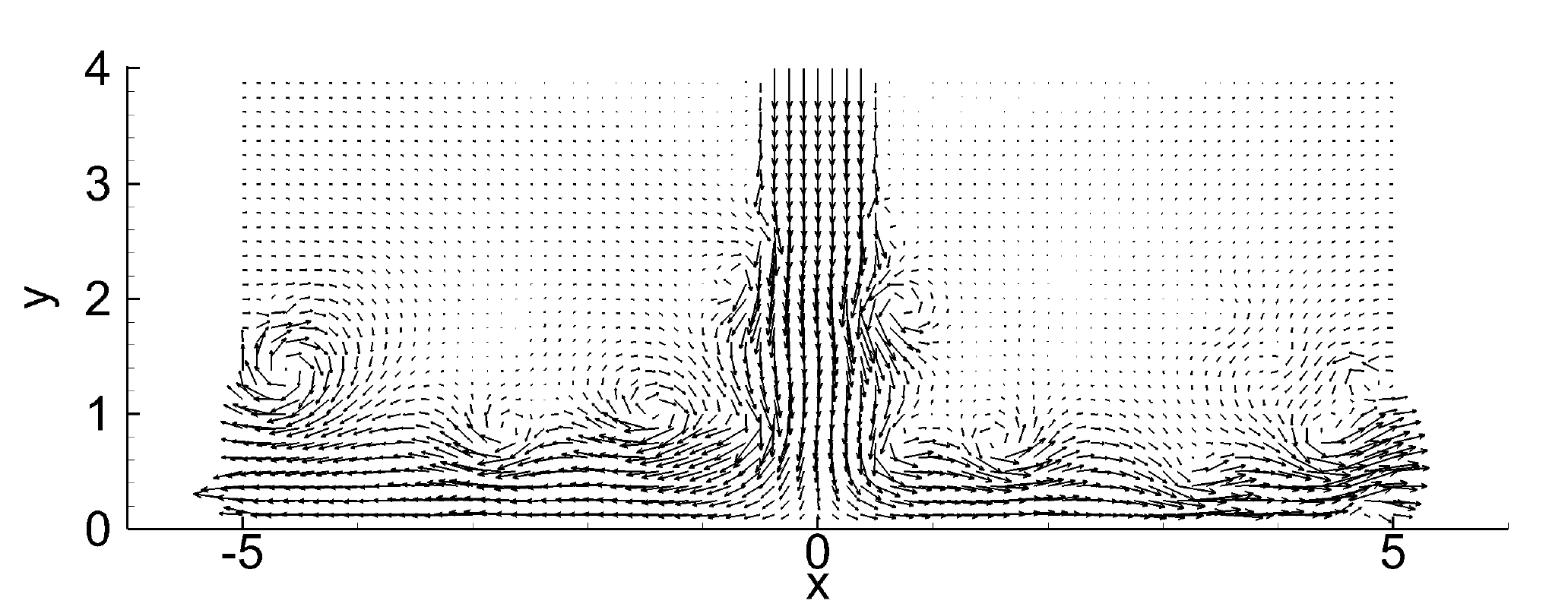}(c)
    \includegraphics[width=3.1in]{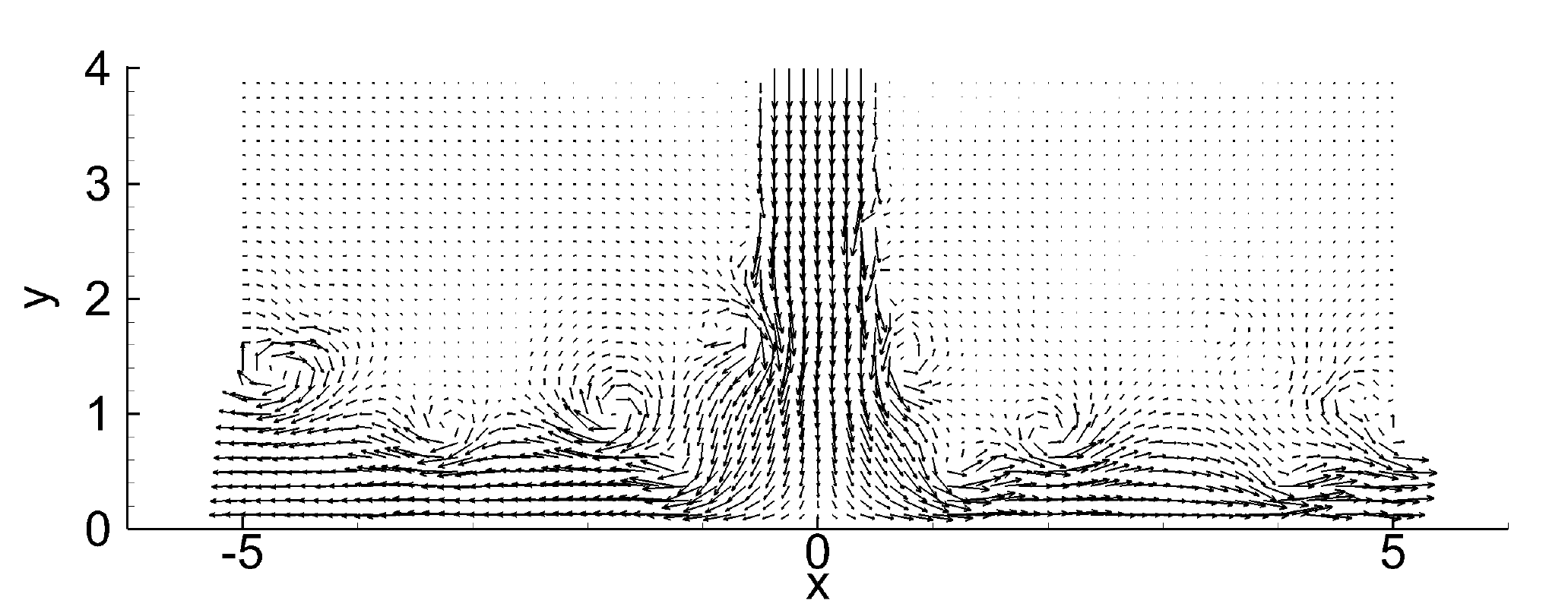}(d)
  }
  \centerline{
    \includegraphics[width=3.1in]{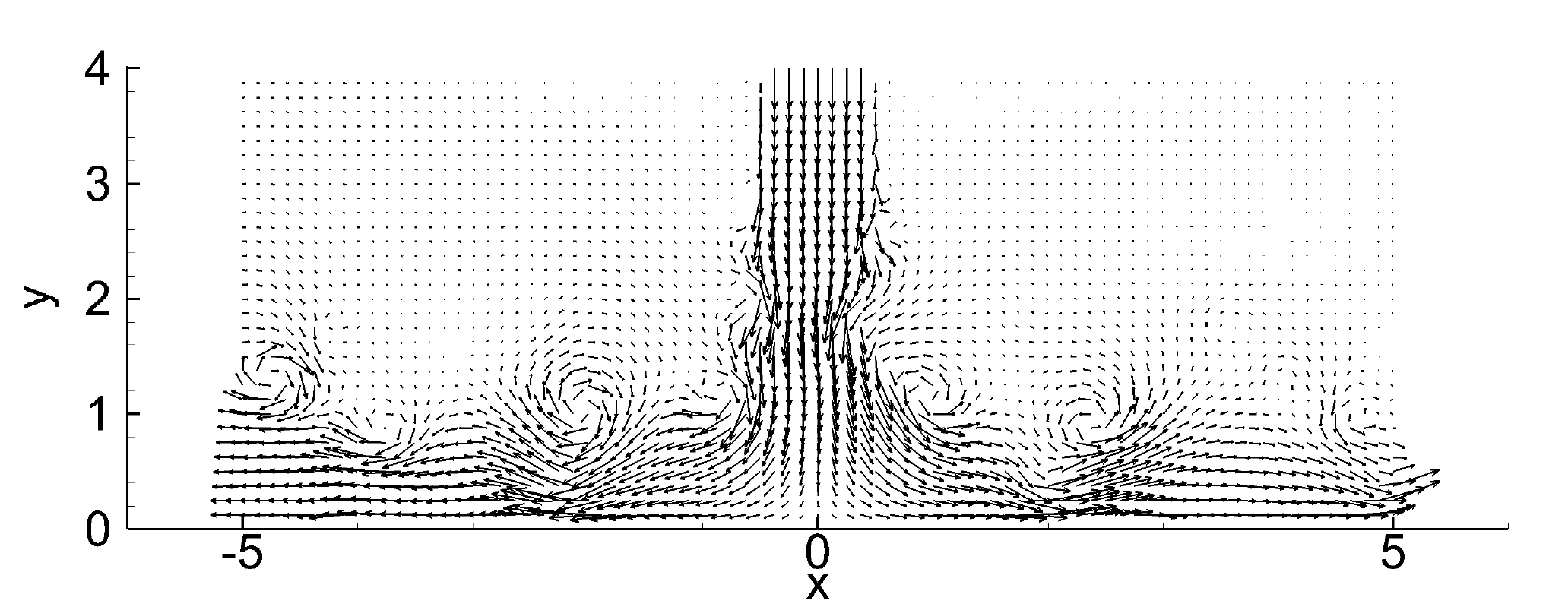}(e)
    \includegraphics[width=3.1in]{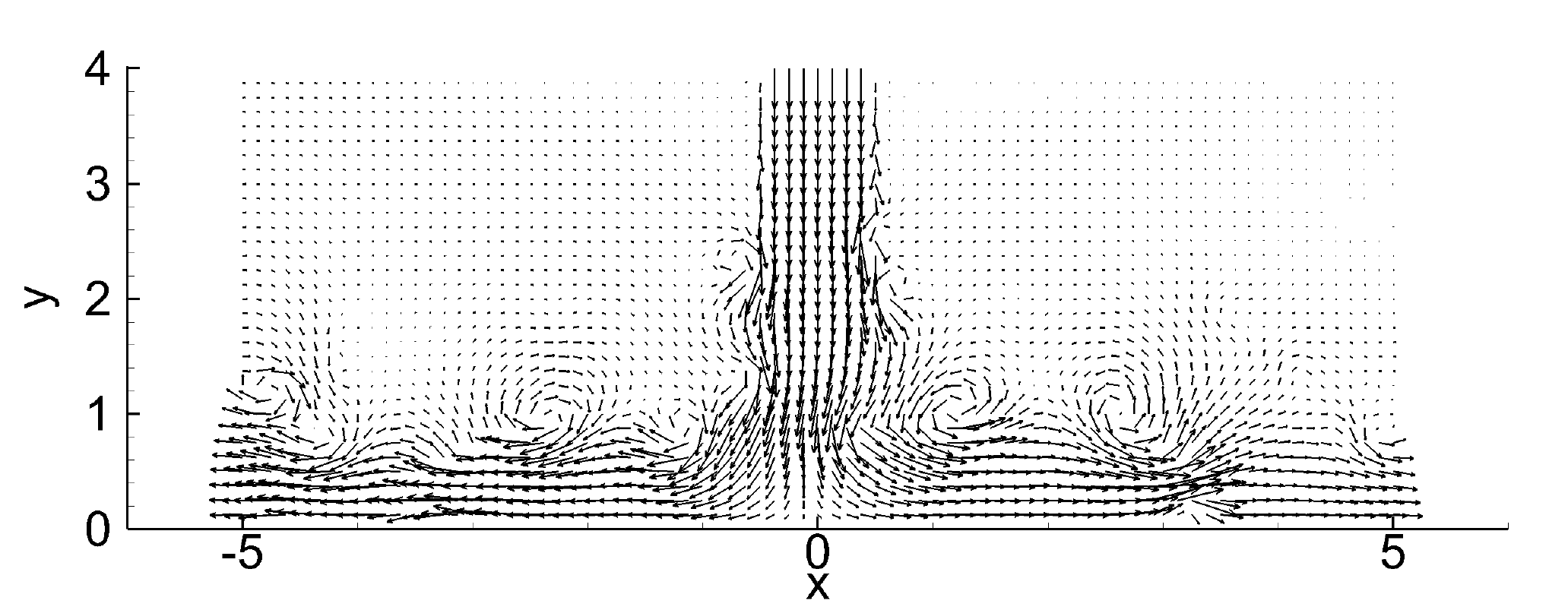}(f)
  }
  \centerline{
    \includegraphics[width=3.1in]{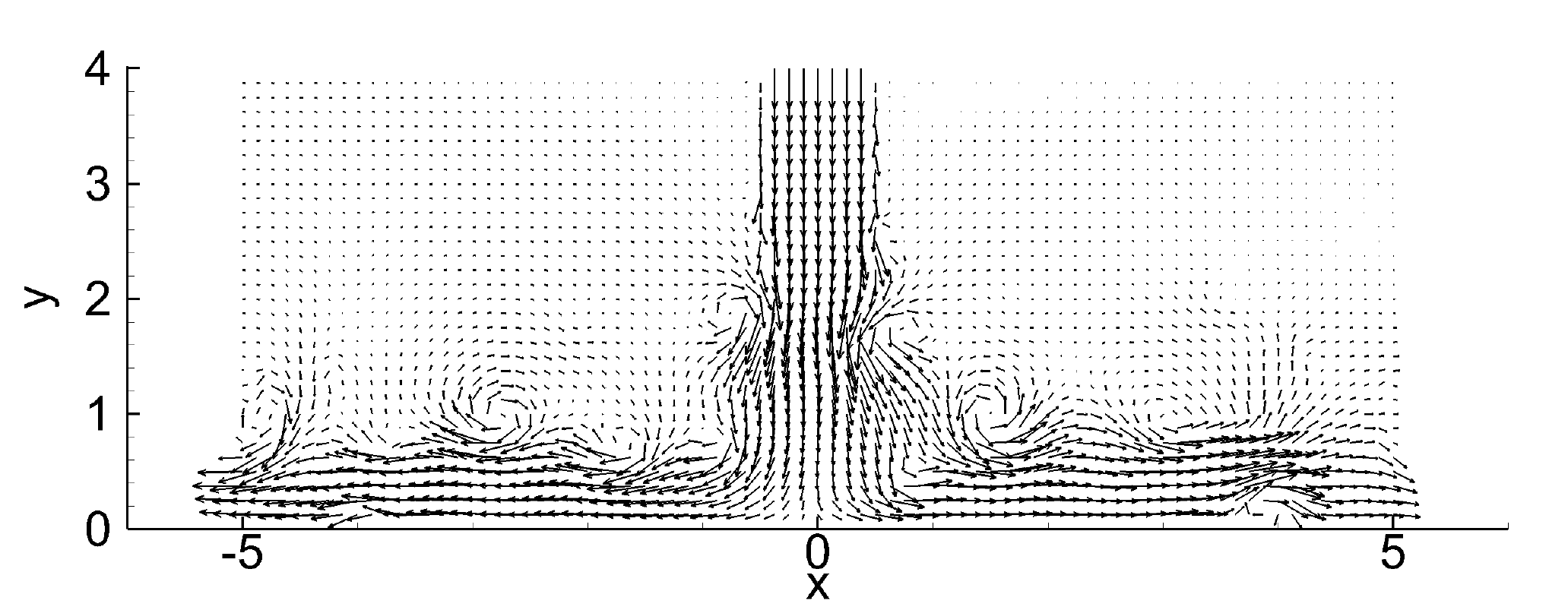}(g)
    \includegraphics[width=3.1in]{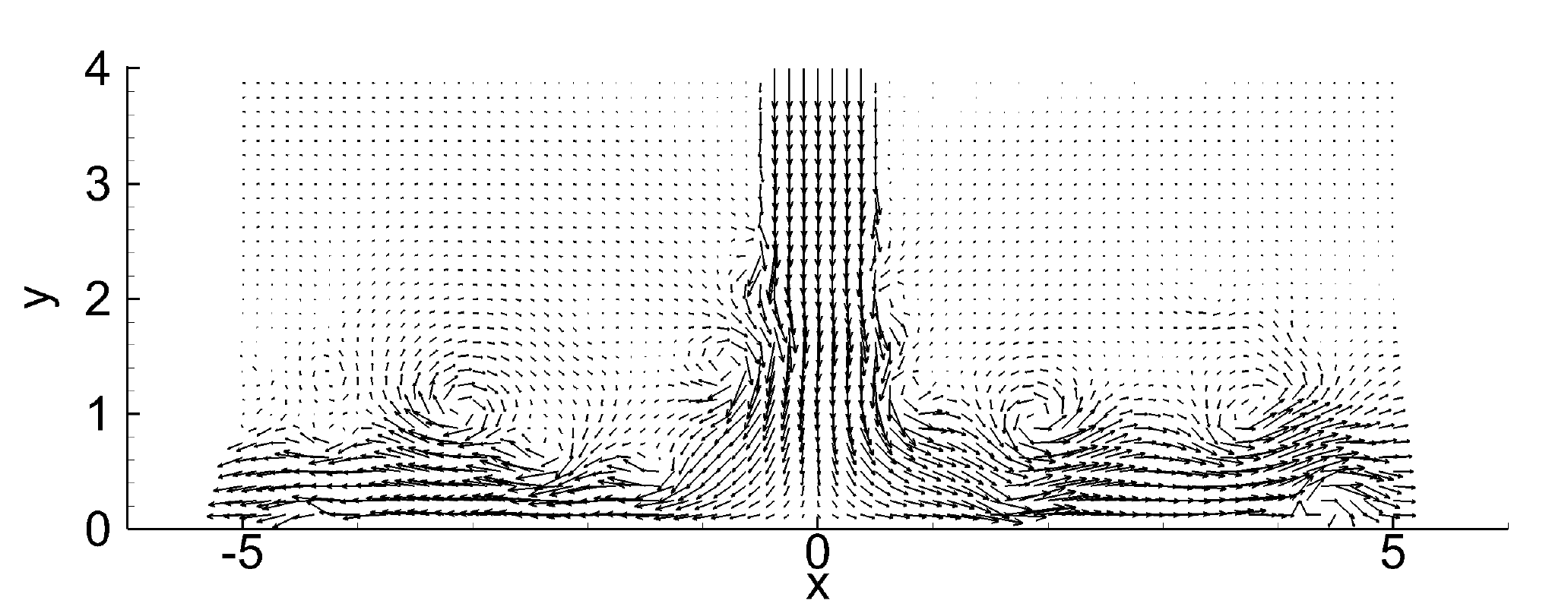}(h)
  }
  \caption{
    Impinging jet (Re=5000): temporal sequence of snapshots of
    the velocity fields at instants
    (a) $t=1569$,
    (b) $t=1570.2$,
    (c) $t=1571.4$,
    (d) $t=1572.6$,
    (e) $t=1573.8$,
    (f) $t=1575$,
    (g) $t=1576.2$,
    (h) $t=1577.4$.
    Velocity vectors are shown on a set of sparser grid points
    for clarity.
  }
  \label{fig:jet_dyn}
\end{figure}


Figure \ref{fig:jet_dyn} illustrates the dynamical features of
the impinging jet problem with a temporal sequence of snapshots
of the velocity fields at $Re=5000$.
These results correspond to zero external pressure heads ($p_{01}=p_{02}=0$)
on the open boundaries. They are computed with an
element order $7$, a time step size $\Delta t=0.001$,
and the open boundary condition \eqref{equ:obc}
with $\mbs H(\mbs n,\mbs u)$ given by \eqref{equ:H_gen}
having parameter values $(\theta,\alpha_1,\alpha_2)=(0,1,1)$.
A stable region 
immediately downstream of the jet inlet can be observed.
This region shrinks  with increasing
Reynolds number. At $Re=5000$ this stable region appears
to be shorter than a jet diameter (Figure \ref{fig:jet_dyn}).
Downstream of this stable region,
the vertical jet experiences the Kelvin-Helmholtz instability
and the shear layers roll up to form vortices along
the profile of the jet stream (Figures~\ref{fig:jet_dyn}(b)-(d)).
These vortices persist along the outgoing horizontal
streams, forming a train of vortices
in the domain (Figures~\ref{fig:jet_dyn}(c)-(h)).
These vortices are ultimately discharged from
the domain through the left and the right open
boundaries (see Figures~\ref{fig:jet_dyn}(d)-(h)). 
The presence of backflows and the passage of strong vortices on
the open boundaries make the impinging jet problem
very challenging to simulate. The energy-stable open
boundary condition~\eqref{equ:obc}, and those developed
in e.g.~\cite{DongS2015,Dong2015clesobc,DongKC2014,NiYD2019},
are critical to dealing with such open boundaries
and the successful simulation of this problem.



\begin{figure}
  \centering
  \includegraphics[width=3.5in]{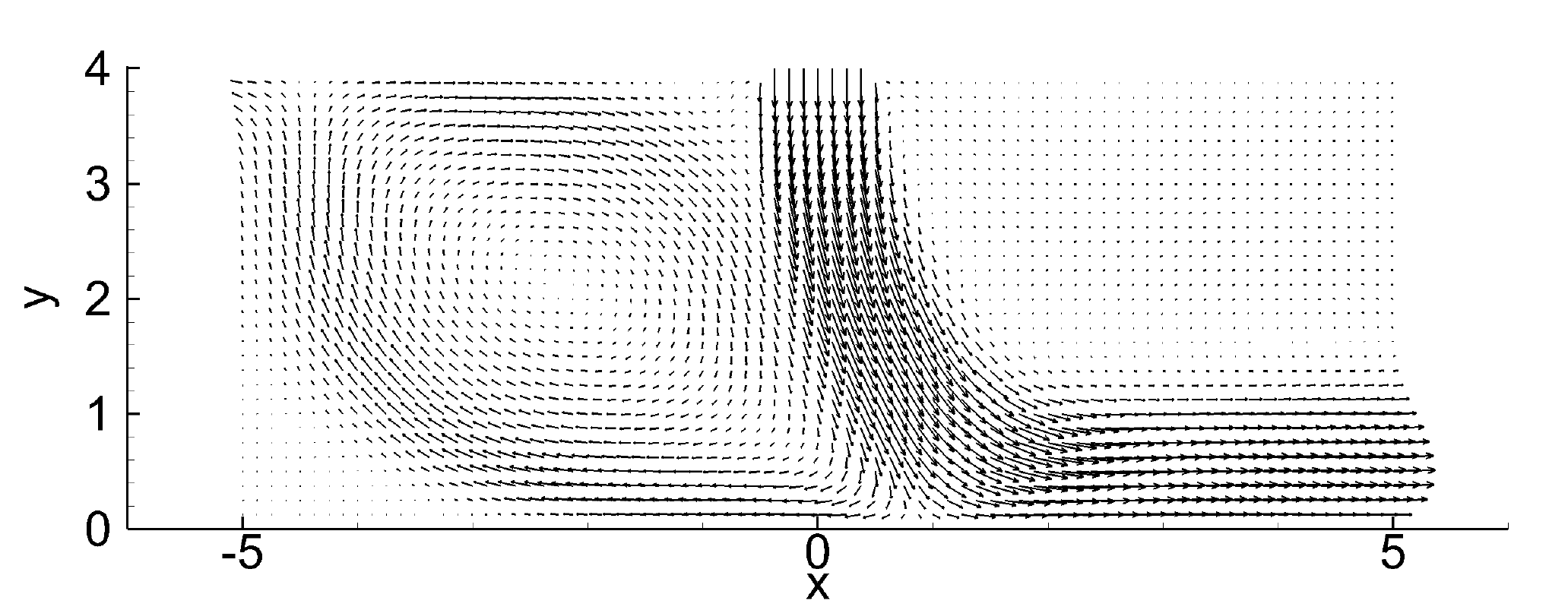}(a)
  \includegraphics[width=3.5in]{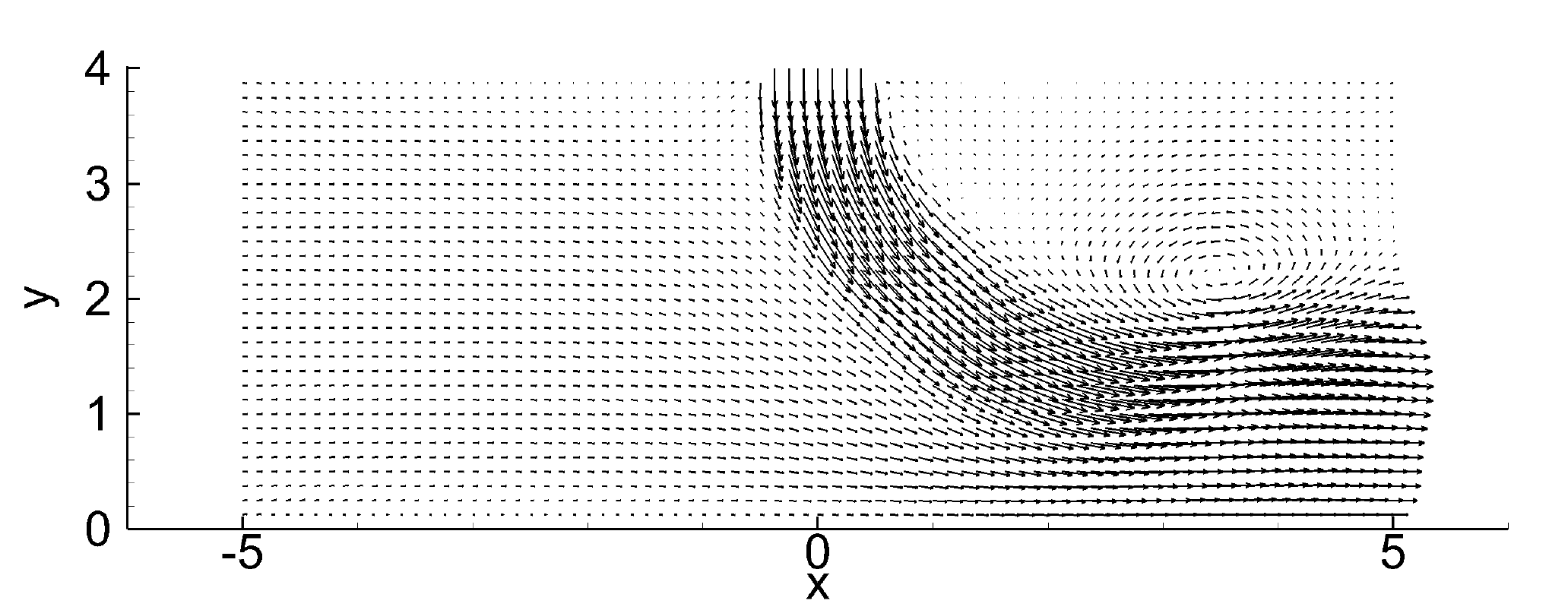}(b)
  \caption{
    Impinging jet (Re=300): effect of external pressure heads
    on the flow field. External pressure
    force on the right boundary is fixed at $p_{02}=0$, and
    on the left boundary it corresponds to
    (a) $p_{01}=0.2$ and (b) $p_{01}=0.4$.
  }
  \label{fig:jet_b300}
\end{figure}


\begin{figure}
  \centering
  \includegraphics[width=3.5in]{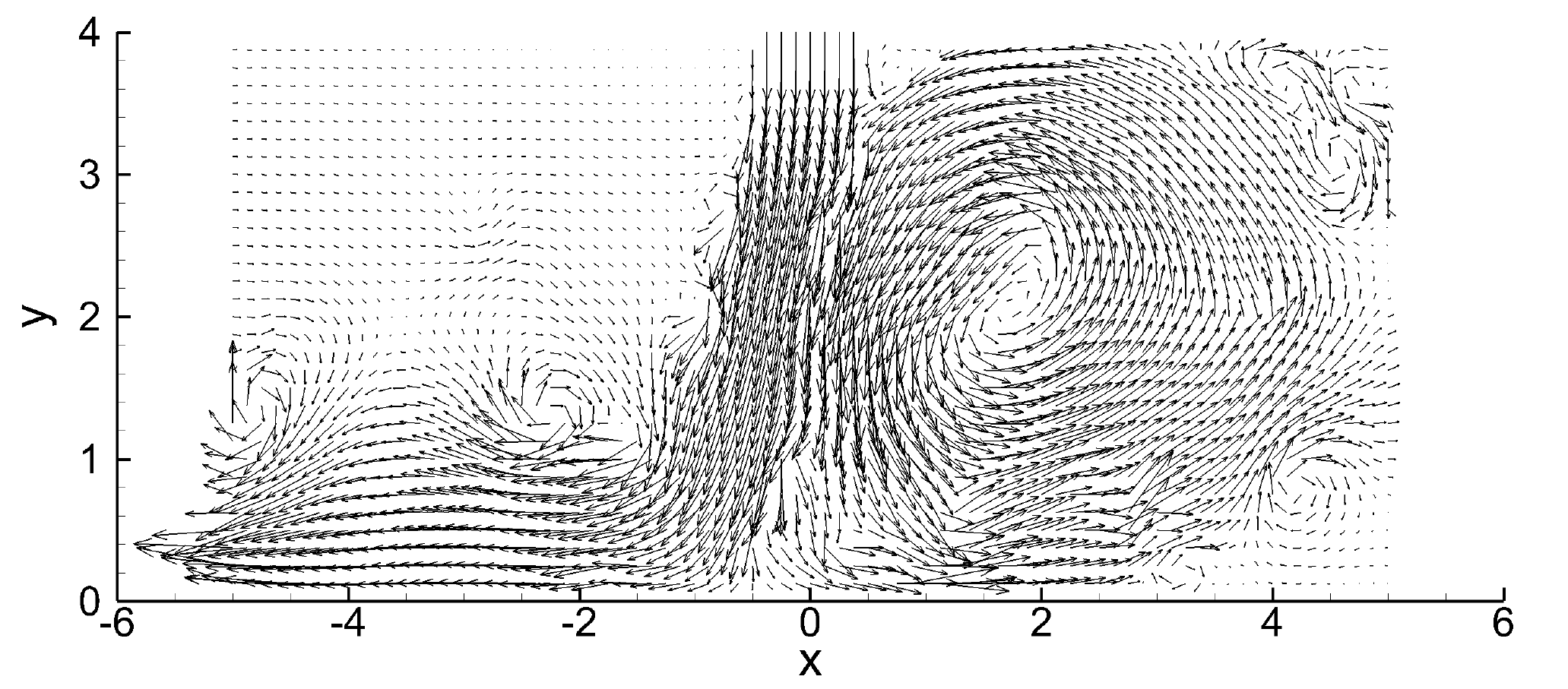}(a)
  \includegraphics[width=3.5in]{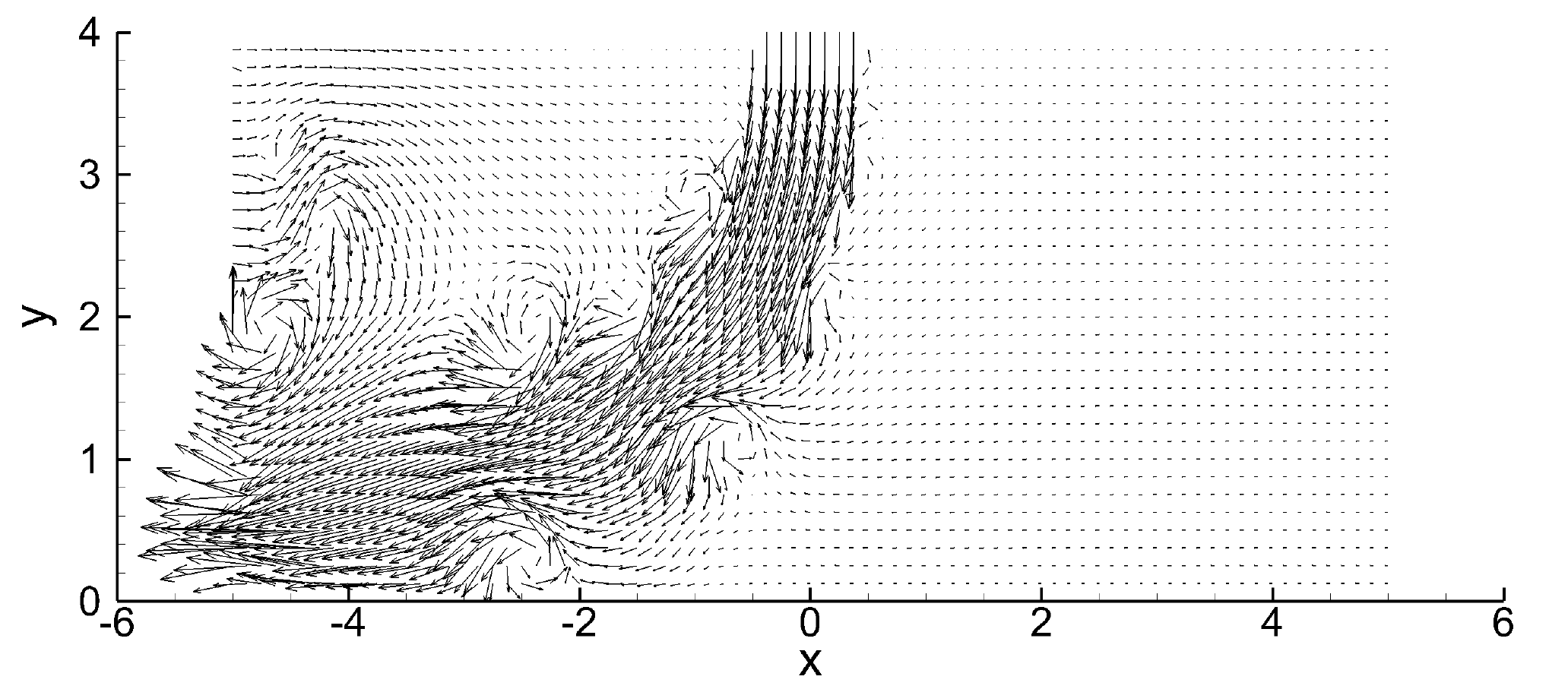}(b)
  \includegraphics[width=3.5in]{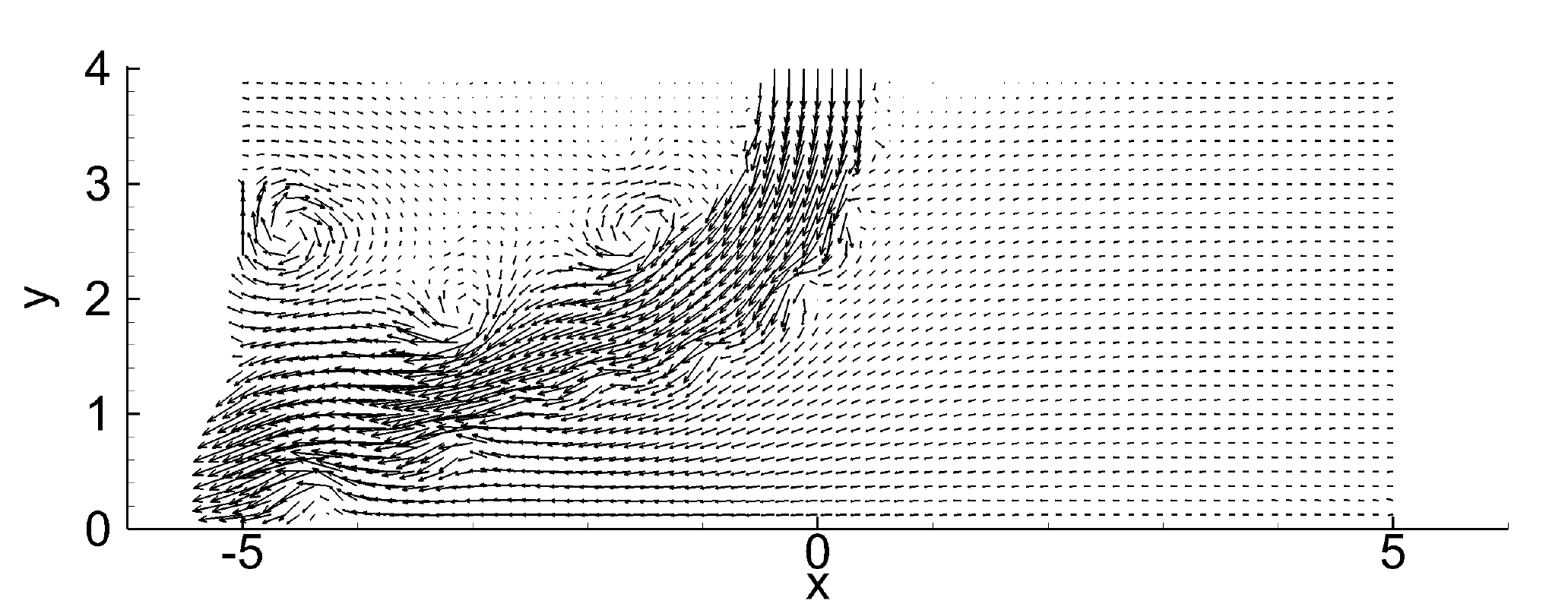}(c)
  \caption{
    Impinging jet (Re=5000): effect of external pressure heads
     on the flow field. External pressure
    force on the left boundary is fixed at $p_{01}=0$, and
    on the right boundary it corresponds to
    (a) $p_{02}=0.2$, (b) $p_{02}=0.3$, and (c) $p_{02}=0.4$.
  }
  \label{fig:jet_b5k}
\end{figure}


Let us next look into the effect
of non-zero external pressure heads ($p_{01}$ and $p_{02}$)
on the impinging jet problem.
Depending on the relative values of 
$p_{01}$ and $p_{02}$, the flow loses symmetry and
the pressure difference can induce a horizontal flow,
which if sufficiently strong will bend the jet
toward one side.
Figures \ref{fig:jet_b300} and \ref{fig:jet_b5k}
demonstrates these scenarios with Reynolds numbers
$Re=300$ and $Re=5000$, respectively.
In Figure \ref{fig:jet_b300}, on the right open
boundary a zero pressure head ($p_{02}=0$) is imposed,
and the external pressure on the left open boundary ($p_{01}$) is
varied. The two plots in Figure \ref{fig:jet_b300}
correspond to $p_{01}=0.2$ and $p_{01}=0.4$, respectively.
In Figure \ref{fig:jet_b5k}, on the left open
boundary  a zero pressure head ($p_{01}=0$) is imposed,
while the external pressure head on the right open boundary ($p_{02}$)
is varied. The plots in Figure \ref{fig:jet_b5k}
correspond to $p_{02}=0.2$, $0.3$ and $0.4$ in the simulations,
respectively.
We indeed observe that the jet can be bent toward
the right (Figure \ref{fig:jet_b300}) or the left
(Figure \ref{fig:jet_b5k}) because of the external
pressure difference in the horizontal direction.
When this pressure difference is not strong
enough to overcome the upcoming horizontal jet stream,
the horizontal stream may be deflected to form
a large vortex on one side
of the domain (see Figures \ref{fig:jet_b300}(a) and
\ref{fig:jet_b5k}(a)).
With a strong enough pressure difference,
a horizontal flow can be established in the domain
and the jet is pushed toward one side
(see Figure \ref{fig:jet_b300}(b) and Figures \ref{fig:jet_b5k}(b)-(c)).

\begin{figure}
  \centerline{
    \includegraphics[width=3.1in]{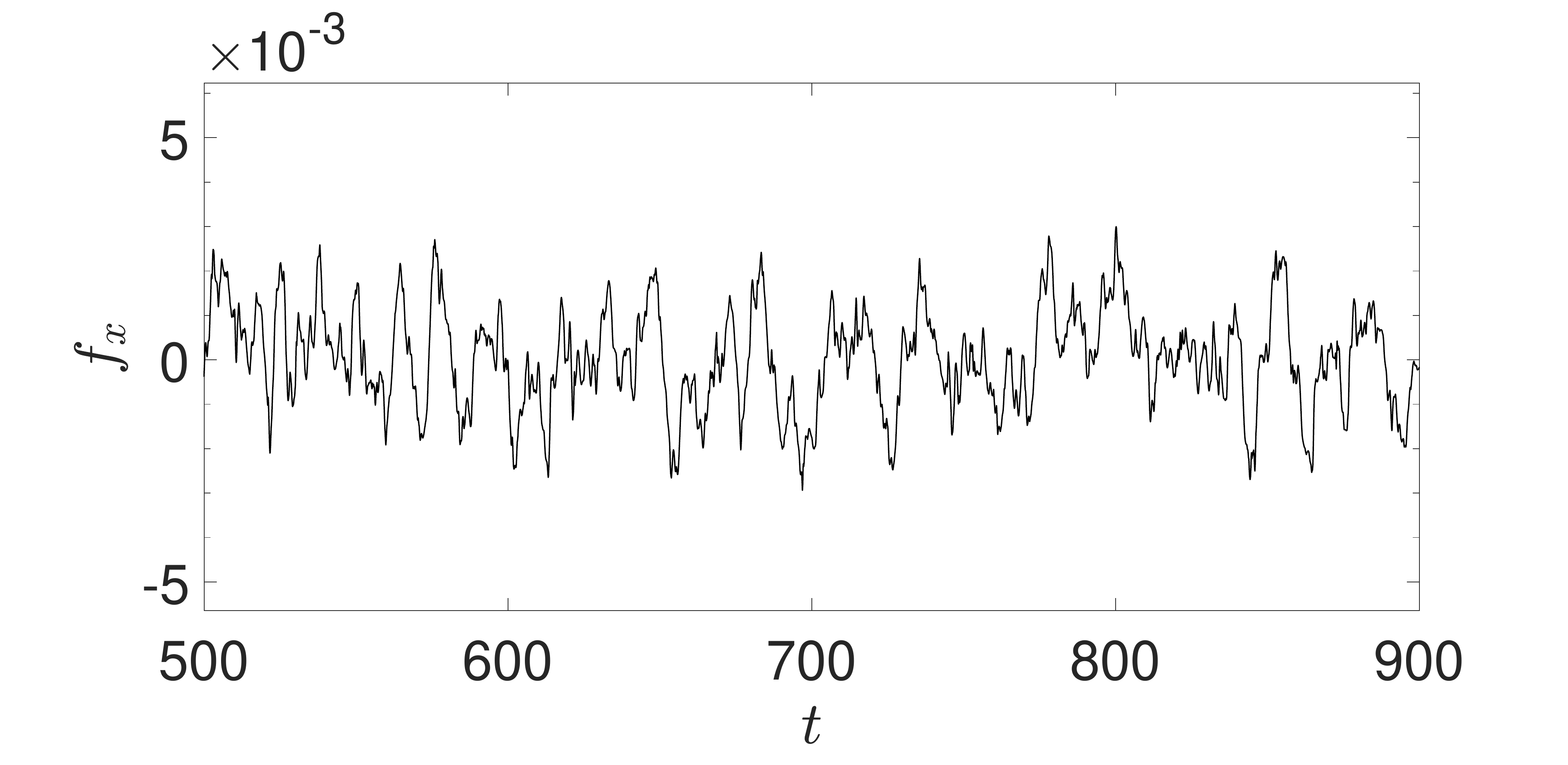}(a)
    \includegraphics[width=3.1in]{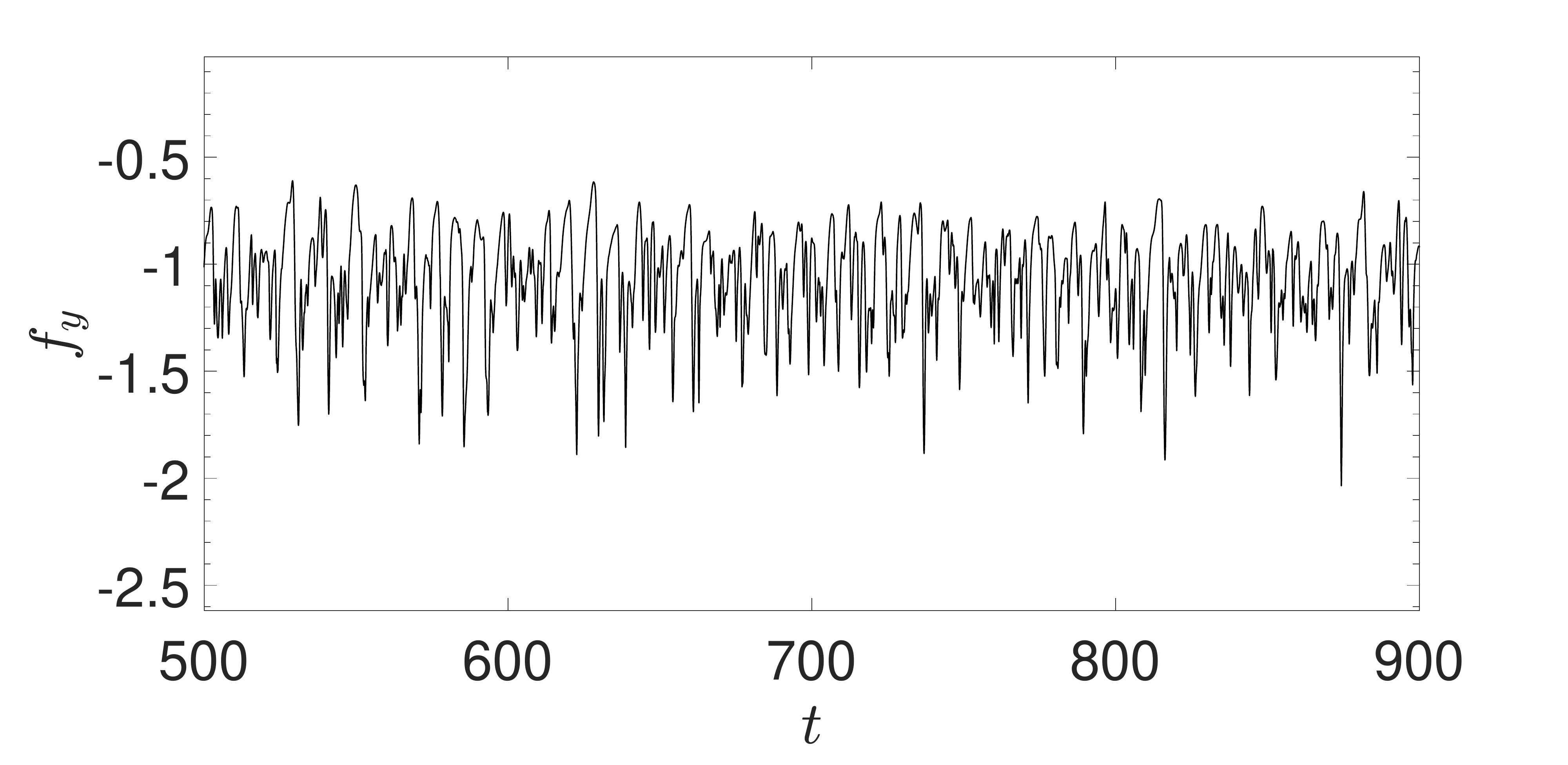}(b)
  }
  \centerline{
    \includegraphics[width=3.1in]{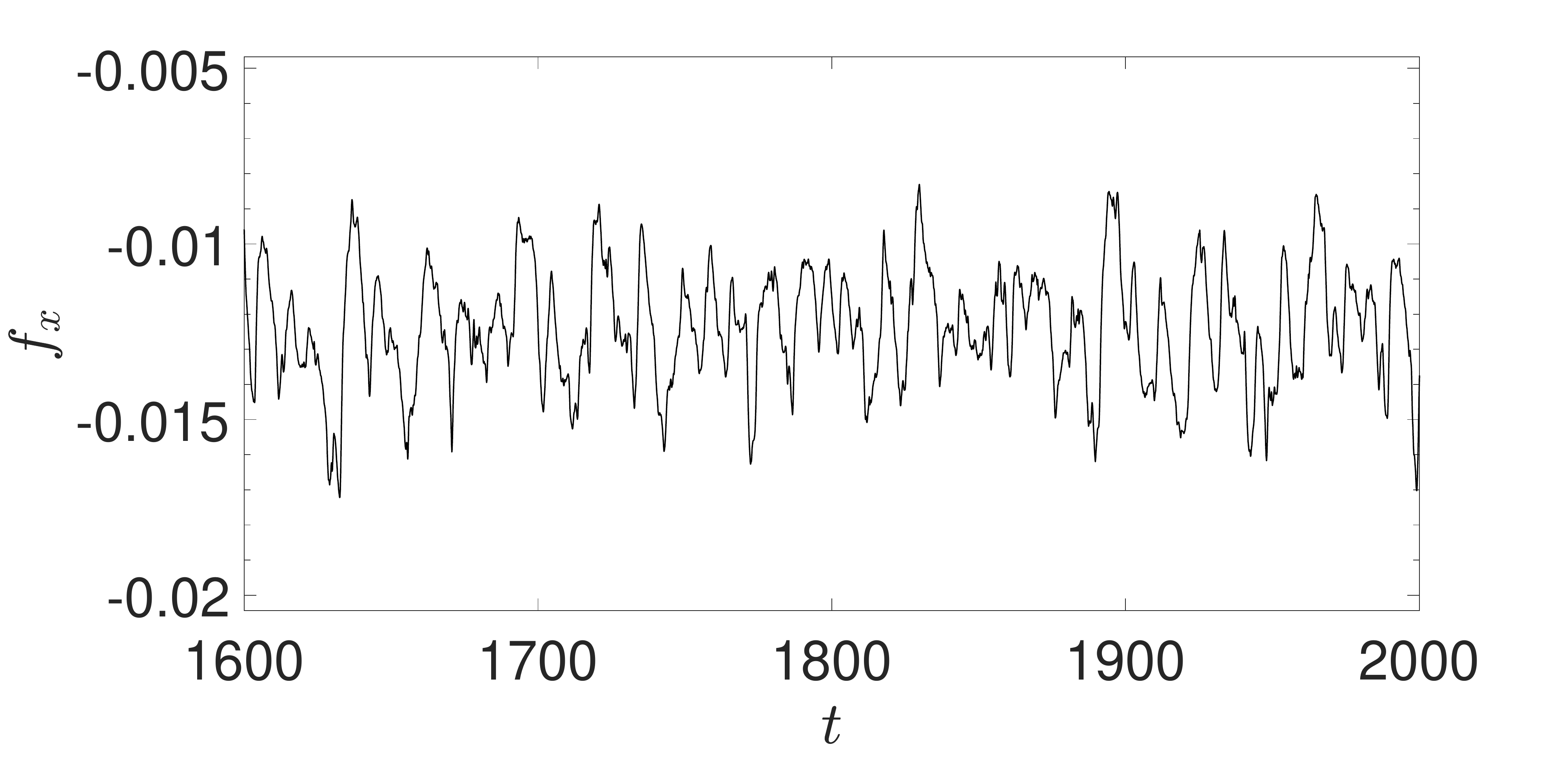}(c)
    \includegraphics[width=3.1in]{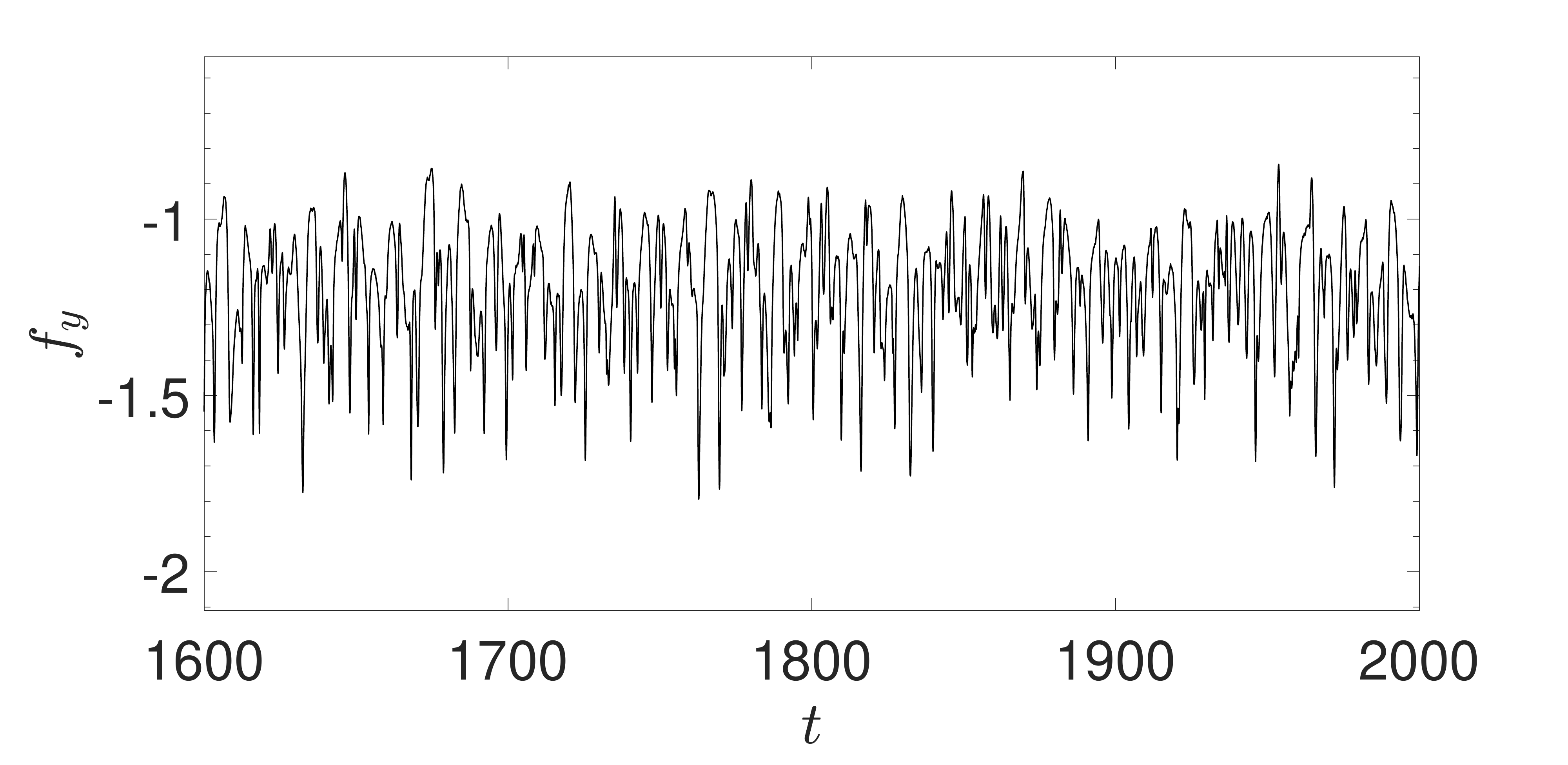}(d)
  }
  \caption{
    Impinging jet (Re=5000): time histories of the x-component (plots
    (a) and (c)) and y-component (plots (b) and (d)) of the force on the wall.
    Plots of (a) and (b) correspond to zero external pressure force
    $p_{01}=p_{02}=0$, and those in (c) and (d) correspond to
    external pressure forces $p_{01}=0$ and $p_{02}=0.3$.
  }
  \label{fig:jet_f5k}
\end{figure}

Non-zero external pressure heads can also cause the force exerting on the wall to
differ markedly when compared with the case of zero external pressure
heads. Figure \ref{fig:jet_f5k} shows time histories of the two components
of the total force on the walls
at $Re=5000$ corresponding to a zero pressure head ($p_{01}=p_{02}=0$, plots
(a) and (b)) and a non-zero pressure head ($p_{01}=0$ and $p_{02}=0.3$,
(c) and (d)). We observe that the external pressure
difference induces a mean  horizontal force on the
walls, while with zero external pressure heads the mean horizontal
force  is essentially zero.


\section{Concluding Remarks}
\label{sec:summary}

%

In the current work we have developed an unconditionally energy-stable
scheme for simulating incompressible flows on domains with
outflow/open boundaries. This scheme combines the
generalized Positive Auxiliary Variable (gPAV) approach and
a rotational velocity-correction type strategy.
The incompressible
Navier-Stokes equations, the dynamic equation
for the auxiliary variable, and the energy-stable open
boundary conditions have been reformulated based on the gPAV idea.
The discrete unconditional energy stability of the proposed
scheme has been proven.
%
Within each time step, the scheme requires the
computation of two velocity fields and two pressure fields
in a fully de-coupled fashion, by solving several
individual linear equations involving  constant and time-independent
coefficient matrices.
The auxiliary variable, being a scalar number rather than
a field function, is given by a well-defined
explicit formula, and its computed values are
guaranteed to be positive.
%
It should be noted that no nonlinear solver is involved in the current
method for either the field functions or the auxiliary variable,
and the linear algebraic systems to be solved 
involve only constant coefficient matrices that can be pre-computed.
Therefore, the current  scheme
is computationally very attractive and competitive.


Extensive numerical experiments have been provided with
a number of flow problems involving outflow/open
boundaries. In particular the flow regimes with 
backflow instability have been simulated, in which strong vortices
or backflows can occur at the outflow/open boundaries.
These numerical tests demonstrate the stability of the
proposed scheme with various time step sizes ranging
from small to large values.
At the same time these tests also show a deterioration in accuracy of
the simulation results when the time step size becomes too large.
%
These observations suggest that the simulation result using a
large time step size should only serve as a reference solution,
which cannot be fully trusted unless appropriate convergence tests
with respect to the simulation parameters (such as
the time step size) are performed.
The use of an unconditionally energy-stable scheme such as
the one presented herein (and any other numerical scheme for that matter)
is no substitute for the convergence tests in actual
production simulations.


It is worth comparing the scheme developed here with
that from the recent work~\cite{LinYD2019},
as both schemes employ an auxiliary variable
in the algorithmic construction.
In Remark \ref{rem:rem_1} we have commented on this matter
in some detail. Here we would like to emphasize two points:
\begin{itemize}

\item
  While an auxiliary variable is used in both schemes,
  the reformulated  system and its numerical treatment
  are completely different in the current scheme
  compared with that of~\cite{LinYD2019}.
  In the current method
  it is guaranteed that the solution for the auxiliary variable
  exists, that it is given by an  explicit formula, and
  that its computed value is positive.
  In contrast, in \cite{LinYD2019} the auxiliary variable is obtained
  by solving a nonlinear algebraic equation. Consequently, the solution for
  the auxiliary variable  cannot be guaranteed to exist in~\cite{LinYD2019}. Even when
  it exists, the computed value is not guaranteed to be positive,
  even though this variable should physically be.
  Some nonlinear algebraic solver such as the Newton's method is required
  in \cite{LinYD2019}.
  In the current method, on the other hand, no nonlinear solver is
  involved, for either the field functions or the auxiliary variable.

\item
  In the algorithmic formulation of the current scheme,
  the pressure and the velocity are de-coupled (barring the
  auxiliary variable) in a way that mirrors a rotational velocity-correction
  strategy. The discrete unconditional energy stability has been
  proven with this de-coupled formulation.
  In contrast, in the algorithmic formulation of~\cite{LinYD2019}
  the pressure and the velocity are fully coupled, and
  the discrete energy stability can only be proven in this fully
  coupled setting therein.  
    
\end{itemize}


Another salient feature of the current scheme lies in
the use of the function $g(\frac{R^2}{E})$
(defined in~\eqref{equ:def_g}), rather than
$\frac{R^2}{E}$, when reformulating the incompressible
Navier-Stokes equation (see equation~\eqref{equ:nse_1}).
This construction can improve the accuracy 
when small time step sizes are used in simulations.
While $\frac{R^2}{E}$ should physically equal  the
unit value on the continuum level, its numerically-computed
value is rarely exactly the unit value.
Numerical experiments indicate that, with small time step sizes,
the computed values for $\frac{R^2}{E}$ are very close
to $1.0$, but typically slightly larger than $1.0$ by a minuscule
amount (e.g.~$10^{-6}$). This can introduce
an error, while small, if $\frac{R^2}{E}$ itself is used
when reformulating the Navier-Stokes equation in~\eqref{equ:nse_1}.
The use of the function $g(\frac{R^2}{E})$ 
can get rid of this error and improve the accuracy
of simulations with small time step sizes.



Regarding the computational cost, because the current method
requires the solution of two copies of the flow field
variables (velocity and pressure), the amount of operations
per time step in the current method is approximately
twice that with a typical semi-implicit  scheme
(see e.g.~\cite{Dong2015clesobc}), which is only conditionally stable.


\section*{Acknowledgement}
This work was partially supported by
NSF (DMS-1522537) and a scholarship from the China
Scholarship Council (CSC, 201806080040).

\bibliographystyle{plain}
\bibliography{obc,mypub,nse,sem,contact_line,interface,cyl,multiphase}

\end{document}